\documentclass[a4paper,11pt]{article}
\pdfoutput=1
\usepackage{jheppub}
\usepackage{amsmath}
\usepackage{amssymb}
\usepackage{color}
\usepackage{graphicx}
\usepackage{hyperref}
\usepackage{xspace,slashed}
\usepackage[utf8]{inputenc}
\usepackage{subcaption}
\usepackage{stackengine}

\makeatletter
\g@addto@macro\bfseries{\boldmath}
\makeatother

\newcommand\brabar{\scalebox{.3}{(}\raisebox{-1.7pt}{-}\scalebox{.3}{)}}

\newcommand{\pt}{p_\perp}
\newcommand{\yjet}{y_\text{jet}}

\definecolor{mygreen}{rgb}{0.0, 0.8, 0.0}
\newcommand{\order}[1]{\mathcal{O}\left(#1\right)}
\newcommand{\as}{\alpha_s}

\newcommand{\GeV}{\,\text{GeV}}
\newcommand{\MeV}{\,\text{MeV}}

\newcommand{\Lundaff}{Department of Astronomy and Theoretical Physics, Lund University, S\"olvegatan 14A, 223 62 Lund, Sweden}

\newcommand{\OXaff}{Rudolf Peierls Centre for Theoretical Physics,
  Clarendon Laboratory, Parks Road, Oxford OX1 3PU, UK}
\newcommand{\ASCaff}{All Souls College, Oxford OX1 4AL, UK}
\newcommand{\CNRSaff}{CNRS, UMR 7589, LPTHE, F-75005, Paris, France}
\newcommand{\CERNaff}{CERN, Theoretical Physics Department, CH-1211
  Geneva 23, Switzerland} 

\newcommand{\IPhTAff}{IPhT, Universit\'{e} Paris-Saclay, CNRS UMR 3681,
  CEA Saclay, F-91191 Gif-sur-Yvette, France}

\title{Calculating the primary Lund Jet Plane density}

\author[a]{Andrew Lifson,}
\affiliation[a]{\Lundaff}
\author[b,c,*]{Gavin P.\ Salam,\note[*]{On leave from \CNRSaff\ and \CERNaff}}
\affiliation[b]{\OXaff}
\affiliation[c]{\ASCaff}
\author[d]{Gr\'egory Soyez}
\affiliation[e]{\IPhTAff}

\abstract{
The Lund-jet plane has recently been proposed as a powerful jet
substructure tool with a broad range of applications.
In this paper, we provide an all-order single logarithmic calculation
of the primary Lund-plane density in Quantum Chromodynamics, including
contributions from the running of the coupling, collinear effects for
the leading parton, and soft logarithms that account for large-angle
and clustering effects.
We also identify a new source of clustering logarithms close to the
boundary of the jet, deferring their resummation to future work.
We then match our all-order results to exact next-to-leading order
predictions.
For phenomenological applications, we supplement our perturbative
calculation with a Monte Carlo estimate of non-perturbative
corrections.
The precision of our final predictions for the Lund-plane density is
$5{-}7\%$ at high transverse momenta, worsening to about $20\%$ at the
lower edge of the perturbative region, corresponding to transverse
momenta of about $5\GeV$.
We compare our results to a recent measurement
by the ATLAS collaboration at the Large-Hadron Collider, revealing
good agreement across the perturbative domain, i.e.\ down to about 5~GeV.
}

\begin{document}

\maketitle 
%======================================================================
\section{Introduction}\label{sec:intr}

In the exploration of the fundamental interactions and particles at
high-energy colliders, jets are among the most abundantly produced and
widely used probes.
Over the past decade, it has become apparent that considerable
valuable information is carried by the internal structure of the jets,
especially at high transverse momenta (see
e.g.~\cite{Marzani:2019hun,Larkoski:2017jix,Asquith:2018igt} for
recent reviews).
That information is increasingly being used for distinguishing
hadronic decays of boosted electroweak particles from quark or
gluon-induced jets~(e.g.~\cite{Sirunyan:2019der,Aad:2019wdr,Aaboud:2019aii,Aaboud:2018ngk,Sirunyan:2019jbg,Sirunyan:2019vgt}),
for distinguishing quark and gluon-induced jets from each
other~(e.g.~\cite{Gras:2017jty,Frye:2017yrw,Metodiev:2018ftz,Larkoski:2019nwj}),
and for studying the modification of jets that propagate through the
medium produced in heavy-ion
collisions~(e.g.~\cite{Andrews:2018jcm,Sirunyan:2017bsd,Sirunyan:2018gct,Acharya:2019djg,Mehtar-Tani:2016aco,Chien:2016led,Chang:2017gkt,Milhano:2017nzm,Caucal:2019uvr,Casalderrey-Solana:2019ubu}).

A huge variety of observables has been
explored~(e.g.~\cite{Butterworth:2008iy,Krohn:2009th,Thaler:2010tr,Larkoski:2013eya,Dasgupta:2013ihk,Larkoski:2014wba,Larkoski:2014pca,Salam:2016yht,Komiske:2017aww,Dreyer:2018nbf}) for
studying jet substructure, supplemented in recent years by a range of
machine-learning
approaches
(e.g.~\cite{Cogan:2014oua,deOliveira:2015xxd,Komiske:2016rsd,Louppe:2017ipp,Egan:2017ojy,Andreassen:2018apy,Datta:2017lxt,Komiske:2017aww,Komiske:2018cqr,CMS:2019gpd,Kasieczka:2019dbj,Kasieczka:2018lwf,Qu:2019gqs}). 
With such a diverse range of observables, it has become challenging to
obtain a detailed understanding of the specific jet features probed by
each one. 
At the same time approaches have emerged in which one designs an
infinite set of observables which, taken as a whole, can encode
complete information about a jet.
Specific examples are energy-flow polynomials~\cite{Komiske:2017aww}
and the proposal~\cite{Dreyer:2018nbf} (see also \cite{Andrews:2018jcm}) to determine a
full Lund diagram~\cite{Andersson:1988gp} for each jet.
As well as encoding complete information about the radiation in a jet,
both of these approaches provide observables that can be directly
measured and that also perform well as inputs to machine-learning.
Here we concentrate on Lund diagrams.

Lund diagrams~\cite{Andersson:1988gp} are two-dimensional
representations of the phase-space for radiation in jets,
which have long been used to help understand Monte-Carlo event
generators and all-order logarithmic resummations in QCD.
The phase-space for a single emission involves three degrees of
freedom, and Lund diagrams highlight the logarithmic distribution of
two of those degrees of freedom, typically chosen to be the emission's
transverse momentum ($k_t$) and angle ($\Delta$).
In leading-order QCD, the logarithms of both variables are uniformly
distributed.

A core idea introduced in Refs.~\cite{Dreyer:2018nbf} and \cite{Andrews:2018jcm}
and briefly reviewed in section~\ref{sec:primary-lund},
is to use a Cambridge/Aachen declustering sequence to represent a
jet's internal structure as a series of points in the two-dimensional
Lund plane, with the option of concentrating on ``primary'' emissions,
those that can be viewed as emitted by the jet's main hard prong.
The location of a given point immediately indicates whether it is in a
perturbative or non-perturbative region, whether it is mainly
final-state radiation or a mix with initial-state radiation,
underlying event, etc.
The set of points obtained for a single jet can be used as an input to
multi-variate tagging methods~\cite{Dreyer:2018nbf}, or can be used to
construct other specialised observables~\cite{Dasgupta:2020fwr}.
Given an ensemble of many jets, one can also determine the average
density of points in each region of the (primary) Lund plane,
$\rho(\Delta,k_t)$.
This average density is of interest in fundamental measurements of QCD
radiation~\cite{Aad:2020zcn,Cunqueiro:2018jbh,Zardoshti:2020cwl}, both
in perturbative and 
non-perturbative regions, and in studies of modifications of jet
structure in heavy-ion collisions~\cite{Andrews:2018jcm}. 

The purpose of this article is to carry out a baseline calculation of
the all-order perturbative structure of the primary Lund-plane
density, identifying the key physical aspects that are relevant for
understanding the density, and providing a prediction that accounts
for all single-logarithmic corrections
$\alpha_s^n \ln^{m}\!\Delta\, \ln^{n-m} k_t$ multiplying the
leading-order, $\order{\alpha_s}$, result for the density.
The relevant contributions are discussed in
section~\ref{sec:all-orders} and include running-coupling effects,
collinear flavour-changing effects and various effects of soft
radiation at commensurate angles (which we compute only in the
large-$N_c$ approximation).\footnote{We only consider jets initiated
  by massless partons, though a similar calculation for jets initiated
  by a massive parton would also of interest given the sensitivity to
  dead-cone effects~\cite{Cunqueiro:2018jbh} and an associated
  recent measurement~\cite{Zardoshti:2020cwl}.}
In section~\ref{sec:fixed-order} we match the all-order results to a
next-to-leading order (NLO) calculation using the {\tt NLOJet++}
program~\cite{Nagy:2003tz}, in section~\ref{sec:np-effects} we address the
question of non-perturbative corrections and in
section~\ref{sec:results} we combine the different results into a set
of final predictions that we compare to recent experimental
measurements from the ATLAS collaboration~\cite{Aad:2020zcn}.

%======================================================================
\section{The primary Lund plane density and basic setups}\label{sec:primary-lund}

Let us assume we have a jet with transverse momentum $\pt$,
obtained from a given jet algorithm such as
the anti-$k_t$ algorithm~\cite{Cacciari:2008gp}. We first
re-cluster the constituents of the jet using the Cambridge/Aachen~(C/A)
algorithm~\cite{Dokshitzer:1997in,Wobisch:1998wt} as often used in jet
substructure techniques. We then iteratively repeat the following
steps, starting with $j$ defined as the full (re-clustered) jet:
\begin{enumerate}
\item Undo the last step of clustering: $j\to j_1+j_2$, taking $j_1$
  to be the {\em harder branch}, i.e.\ $p_{\perp
    1}>p_{\perp 2}$.
\item Record the properties of the branching ${\cal {T}}\equiv\{k_t,
  \Delta, z, ...\}$ defined as
  \begin{subequations}\label{eq:branching-variables}
    \begin{align}
      \label{eq:branching-variables-a}
      \Delta &\equiv \Delta_{12}=\sqrt{(y_1-y_2)^2+(\phi_1-\phi_2)^2} ,\\
      \label{eq:branching-variables-b}
      k_t & \equiv p_{\perp 2} \Delta_{12}, \qquad \qquad
      z \equiv \frac{p_{\perp 2}}{p_{\perp 1}\!+\!p_{\perp 2}}.
    \end{align}
  \end{subequations}
  where $y$ and $\phi$ denote the rapidity and azimuthal angle of a
  particle, specifically $y = \frac12\ln\frac{E+p_z}{E-p_z}$.
\item Redefine $j \leftarrow j_1$ and iterate (i.e.\ iterate following
  the harder branch)
\end{enumerate}
The iteration stops when $j$ can no longer be de-clustered, giving an ordered list of tuples:
\begin{equation}
  {\cal {L}}_\text{primary} = \left[{\cal {T}}^{(1)}, \dots, {\cal
      {T}}^{(i)}, \dots, {\cal {T}}^{(n)} \right].
\end{equation}
Additional variables can be added to each tuple, for example an
azimuthal angle.
One can also choose to follow softer branchings at each step, which
would lead to exploration of secondary, tertiary, etc. Lund planes.
Neither of these aspects is relevant for the discussion presented
here.

The primary Lund plane density is then defined as the density of
emissions in the (logarithmic) $\Delta$, $k_t$
plane:\footnote{Throughout this paper, we use a subscript
  ``$\perp$'' to denote transverse momenta with respect to the beam,
  and a subscript ``$t$'' to denote the transverse momenta of
  emissions relative to their emitter.}
\begin{equation}\label{eq:primary-density}
  \rho(\Delta, k_t) = \frac{1}{N_\text{jets}}
  \frac{dn_\text{emissions}}{d\ln 1/\Delta\,d\ln k_t}.
\end{equation}
Alternatively, one can introduce a primary Lund plane density
in the $\Delta$, $z$ plane:
\begin{equation}\label{eq:primary-density-z}
  \tilde\rho(\Delta, z) = \frac{1}{N_\text{jets}}
  \frac{dn_\text{emissions}}{d\ln 1/\Delta\,d\ln 1/z},
\end{equation}
as measured by the ATLAS collaboration.

Note that integrating the primary Lund plane density over
$\ln 1/\Delta$ and $\ln k_t$ (or $\ln 1/z$) gives the average number
of primary emissions per jet.
If the integration is performed with an additional
Soft-Drop~\cite{Larkoski:2014wba,Dasgupta:2013ihk} condition, one
obtains the Iterated Soft-Drop multiplicity~\cite{Frye:2017yrw} (see
also chapter~III of~\cite{Amoroso:2020lgh}).

In practice, we will focus on two kinematic configurations:
\begin{itemize}
\item {\bf High-$\pt$ setup.}
  This is close to the original proposal
  from~\cite{Dreyer:2018nbf}. We cluster jets with the anti-$k_t$
  algorithm with $R=1$, keep all jets with $\pt\ge 2$~TeV. The primary
  Lund-plane density $\rho(\Delta, k_t)$ is then reconstructed
  according to the procedure described above.
\item {\bf ATLAS setup.}
  This is similar to the ATLAS measurement presented
  in~\cite{Aad:2020zcn}. Jets are reconstructed with the anti-$k_t$
  algorithm with a radius $R=0.4$. The two largest-$\pt$ jets with
  $|\eta|<2.1$ are kept (with $\eta$ the pseudo-rapidity, defined as
  $\eta = - \ln \tan \frac\theta2$). One then
  imposes that the leading jet has a $\pt$ of at least 675 GeV and
  that the $\pt$ of the second jet is at least $\tfrac{2}{3}$ of the $\pt$ of
  the leading jet. For each of the two jets, we construct the Lund
  plane $\tilde\rho(\Delta, z)$ as follows: we take all the particles
  within a radius $R=0.4$ of the jet axis, recluster them with the C/A
  algorithm with $R=0.4$ and apply the de-clustering procedure
  highlighted above. 
  
  In practice the ATLAS measurement only includes charged tracks with
  $\pt$ above 500~MeV within a distance
  $\sqrt{(\eta-\eta_\text{jet})^2+(\phi-\phi_\text{jet})^2}=0.4$ of the
  jet axis\footnote{This distance uses pseudo-rapidity $\eta$ instead of
    rapidity $y$.
    The two variables are identical for massless objects.
    For individual experimental objects, it is the pseudorapidity that
    is measured.
    However for any object that is massive, rapidity is to be strongly
    favoured~\cite{Schegelsky:2010xi,Gallicchio:2018elx} because
    rapidity differences are invariant under longitudinal boosts,
    while pseudorapidity differences are not.
    Every stage of jet clustering creates massive objects, even
    starting from massless ones.
    The pseudorapidity of a jet displays various pathologies: for
    example, a jet consisting of two massless particles with identical
    pseudorapidities $\eta_1 = \eta_2$ has a rapidity
    $y_\text{jet} = \eta_1 = \eta_2$, while the jet's pseudorapidity
    is different $\eta_\text{jet} \ne \eta_1$, by an amount that
    depends non-trivially on the kinematics of the jet.
    Therefore even if the inputs to the initial jet clustering are selected
    based on their pseudo-rapidity, we recommend using rapidity
    for all subsequent operations.}
  in their reconstruction of the Lund
  plane. We will treat this as a non-perturbative correction (going
  from a full-particle measurement to a measurement based on tracks
  above 500~MeV).
  We note that the use of charged tracks makes this measurement
  collinear unsafe since, for example, arbitrarily collinear
  branchings can affect the relative fractions of charged and neutral
  particles in each branch, and consequently the definition of the
  harder branch in the de-clustering procedure.
  Numerically, this effect is small, as we shall verify later.
\end{itemize}
In all cases, the initial jet clustering is done using
FastJet~\cite{Cacciari:2005hq,Cacciari:2011ma} and the Lund plane is
constructed using the code available with
\texttt{fastjet-contrib}~\cite{fastjet-contrib}.

%======================================================================
\section{All-order calculation}\label{sec:all-orders}

The average Lund plane density measures an effective intensity of
radiation per unit logarithm of $k_t$ and of angle.
As such at LO, in the simultaneously soft and collinear limit, i.e.\
away from the large-angle and the collinear edges of the plane, it is
given by
\begin{equation}\label{eq:rho-fc}
  \rho_{\text{LO},i}^\text{soft-coll.}(\Delta,k_t) = \frac{2\alpha_s C_i}{\pi},
\end{equation}
where $C_i$ is the Casimir of the hard parton of flavour $i$
initiating the jet, $C_i=C_A$ for a gluon-initiated jet and $C_F$ for
a quark-initiated jet.

Beyond leading order, each additional factor of $\alpha_s$ can be
associated with up to one logarithm of either $\Delta$ or of
$\pt /k_t$.
As a result, at any given order, say $\as^{n+1}$, the logarithmically
dominant terms have the structure
$\as^{n+1} \ln^m \Delta \ln^{n-m} \frac{\pt}{k_t}$ with
$0 \le m\le n$.
Our goal is to calculate this complete set of single-logarithmic
contributions to $\rho(\Delta, k_t)$, i.e.\ for all $n$ and $m$,
including the full (non-logarithmic) $\Delta$ dependence for terms
with $m=0$ and the full $k_t$ dependence for terms with $m=n$.
In the case of $\tilde \rho(\Delta,z)$, we will equivalently aim to
account for all terms $\as^{n+1} \ln^m \Delta \ln^{n-m} z$.

The logarithms have several physical origins.
These are
(i) running coupling corrections, enhanced by logarithms of the
transverse momentum $k_t$;
(ii) hard-collinear logarithms of the emission angle $\Delta$ which
can induce flavour-changing effects and affect the behaviour of $\rho$
close to the $z=1$ line;
(iii) soft emissions at large angles enhanced by the logarithm of
either $k_t$ or the emission energy fraction $z$; and
(iv) Cambridge/Aachen clustering effects for emissions with
commensurate angles, enhanced by logarithms of $k_t$ or $z$.
Each of these effects is discussed separately in the following sections.

%----------------------------------------------------------------------
\subsection{Running coupling corrections}\label{sec:running-coupling}

This is by far the simplest correction: the scale of the running
coupling is simply set by the transverse momentum of the emission. We
therefore have
\begin{equation}\label{eq:rho-softcoll+rc}
  \rho_{\text{rc},i}(\Delta,k_t) = \frac{2\alpha_s(k_t) C_i}{\pi}.
\end{equation}
We use the 2-loop running coupling in the CMW scheme~\cite{Catani:1990rr}:
\begin{multline}\label{eq:running-alphas}
  \alpha_s(k_t)
   = \frac{\alpha_s}{1-2\alpha_s\beta_0\ln(\pt R/k_t)}\\
   - \frac{\beta_1\alpha_s^2}{\beta_0}
      \frac{\ln(1-2\alpha_s\beta_0\ln(\pt R/k_t))}
           {[1-2\alpha_s\beta_0\ln(\pt R/k_t)]^2}
    + \frac{K}{2\pi}\frac{\alpha_s^2}{[1-2\alpha_s\beta_0\ln(\pt R/k_t)]^2},
%    \nonumber
\end{multline}
with
\begin{equation}
  \beta_0 = \frac{11C_A-2n_f}{12\pi},
  \quad
  \beta_1 = \frac{17C_A^2-5C_An_f-3C_F}{24\pi^2},
  \quad
  K = \left(\frac{67}{18}-\frac{\pi^2}{6}\right)C_A-\frac{5}{9}n_f.
\end{equation}
The reference $\alpha_s\equiv\alpha_s(\pt R)$ is taken at the scale
$\pt R$ with $\pt$ the transverse momentum of the jet and $R$ the jet
radius. We use $n_f=5$ flavours for $k_t\ge m_b=4.78$~GeV, $n_f=4$ for
$m_b>k_t\ge m_c=1.67$~GeV and $n_f=3$ below.
Furthermore, we freeze the coupling at $k_t=1$~GeV.

At our accuracy, it would have been sufficient to use the 1-loop
running coupling (without the CMW scheme term, i.e.\ $K$).
We have instead used the 2-loop running for two main
reasons.
Firstly, several of our result depend on the structure of the hard
events under consideration (dijet events in our case). These events
will be obtained using the {\tt NLOJet++} program~\cite{Nagy:2003tz}
with underlying PDF sets that use at least a 2-loop running. It is
therefore more coherent to use a 2-loop running also in the
computation of $\rho$.
Secondly, the running coupling is numerically the largest of the
logarithmically-enhanced contributions. Including the 2-loop
corrections therefore makes sense from a purely phenomenological
perspective.

%----------------------------------------------------------------------
\subsection{Hard-collinear effects}\label{sec:flavour-logs}

Eq.~(\ref{eq:rho-softcoll+rc}) assumes that the primary branch
followed by the declustering procedure keeps the flavour of the
initial parton. In this case, all the emissions for a quark-initiated
jet (gluon-initiated jet) come with a $C_F$ ($C_A$)
colour factor.

In practice, however, hard and collinear branchings have two
effects at single-logarithmic accuracy: (i) they can change the
flavour of the harder branch via either a $q\to qg$ splitting where
the daughter gluon carries more than half the parent quark's momentum,
or from a $g\to q\bar q$ splitting; and (ii) successive collinear
branchings can reduce the transverse momentum of the leading parton
thereby creating a difference between the $z=1$ and the
$k_t=\pt\Delta$ lines in the Lund plane.
These effects are of the form $\alpha_s^{n+1}\ln^n 1/\Delta$,
associated with a series of emissions strongly ordered in angle and without
soft enhancement.

For an initial parton of flavour $i$, we use
\begin{equation}
  p\big(x,j | i, t_\text{coll}(\Delta; \Delta_0, \mu)\big),
\end{equation}
to denote the probability of having a leading parton of flavour $j$,
carrying a longitudinal fraction $x$, when the primary Lund
declustering procedure has reached an angular scale $\Delta$.
The dependence on
$\Delta$ is encoded through a ``collinear evolution time''
\begin{equation}\label{eq:tcoll}
  t_\text{coll}(\Delta; \Delta_0, \mu)
 = I_\alpha\left(\Delta p_\perp, \Delta_0 p_\perp;
    \mu\right),
%  = I_\alpha\left(\frac{\Delta}{R}\mu, \frac{\Delta_0}{R}\mu;
%     \mu\right),
\end{equation}
where $I_\alpha$ is the integration of the
coupling between two transverse momentum scales:
\begin{align}\label{eq:Ialpha}
  I_\alpha(k_t, k_{t0}; \mu)
  &  = \int_{k_t}^{k_{t0}}\frac{dq_t}{q_t}
     \frac{\alpha_s(q_t)}{\pi},
  \\
   & = \frac{1}{2\pi\beta_0}\left[
         \ln\frac{\lambda_0}{\lambda}
       - \frac{\beta_1\alpha_s}{\beta_0}\left(
            \frac{1+\ln\lambda}{\lambda}
          - \frac{1+\ln\lambda_0}{\lambda_0}
         \right)
       + \frac{K\alpha_s}{2\pi}\left(
            \frac{1}{\lambda}
          - \frac{1}{\lambda_0}
         \right)
     \right],\nonumber
\end{align}
with $\alpha_s\equiv \alpha_s(\mu)$,
$\lambda = 1+2\alpha_s\beta_0\ln(k_t/\mu)$ and
$\lambda_0 = 1+2\alpha_s\beta_0\ln(k_{t0}/\mu)$.
The expression on the second line corresponds to a fixed number of
flavours.
The parameter $\Delta_0$ in~(\ref{eq:tcoll}) is the large-angle scale
at which the collinear evolution starts and can be varied to get an
estimate of the uncertainties associated with the resummation of the
collinear logarithms.

The evolution of the probability densities $p(x,j | i, t)$ is given by
\begin{equation}\label{eq:dglap}
  \frac{dp(x,j|i,t)}{dt}
  = \int_0^1 dz \sum_{k=q,g}
  \left[
    \frac{{\cal {P}}_{jk}^{(R)}(z)}{z}\,p_k\left(\frac{x}{z},k\Big|i,t\right) -
    {\cal {P}}_{jk}^{(V)}(z)\,p_j(x,k|i,t)
  \right].
\end{equation}
The kernels ${\cal {P}}_{jk}^{(R)}(z)$ and ${\cal {P}}_{jk}^{(V)}(z)$
correspond to real and virtual emissions respectively and are
straightforwardly obtained from the DGLAP splitting
functions, imposing that the leading parton is defined following the
larger-$\pt$ branch:
\begin{subequations}
  \label{eq:calPij}
  \begin{align}
    {\cal {P}}_{qq}^{(R)}(z) & = P_{qq}(z) \Theta(z>1/2), &
    {\cal {P}}_{qq}^{(V)}(z) & = P_{qq}(z), \\
    {\cal {P}}_{gq}^{(R)}(z) & = P_{gq}(z) \Theta(z>1/2),  &
    {\cal {P}}_{gq}^{(V)}(z) & = 0, \\
    {\cal {P}}_{gg}^{(R)}(z) & = \left[P_{gg}(z)+P_{gg}(1-z)\right] \Theta(z>1/2), &
    {\cal {P}}_{gg}^{(V)}(z) & = P_{gg}(z) + P_{qg}(z), \\
    {\cal {P}}_{qg}^{(R)}(z) & = \left[P_{qg}(z)+P_{qg}(1-z)\right] \Theta(z>1/2), &
    {\cal {P}}_{qg}^{(V)}(z) & = 0.
\end{align}
\end{subequations}
where the $P_{ij}(z)$ are the normal full (real) DGLAP splitting
functions,
\begin{subequations}\label{eq:dglap-splittings}
\begin{align}
  P_{qq}(z) & =  C_F \frac{1+z^2}{1-z}\,,
  & P_{gq}(z) & = P_{qq}(1-z)\,, \\
  P_{gg}(z) & =  C_A\left[\frac{z}{1-z}+\frac{1-z}{z}+z(1-z)\right],
  & P_{qg}(z) & = n_f T_R\left[z^2+(1-z)^2\right],
\end{align}
\end{subequations}
including appropriate symmetry factors.
If the leading parton carries a longitudinal momentum $x \pt$ at an
angle $\Delta$, the splitting variables
(cf.~(\ref{eq:branching-variables})) are related through
$z = \frac{k_t}{x \pt\Delta}$.
Therefore, for a jet initiated by a hard parton of
flavour $i$, the primary Lund-plane density including collinear
effects takes the form
\begin{equation}\label{eq:rho-coll}
  \rho_{\text{coll},i}(\Delta,k_t)
  = \sum_j \int_0^1 dx\,
    p\big(x,j | i, t_\text{coll}(\Delta; \Delta_0, \mu)\big)
    \left[
      \frac{x \pt\Delta }{2C_jk_t}
      {\cal{P}}_j\left(\frac{k_t}{x \pt\Delta}\right) 
    \right]
    \rho_{\text{rc},j}(\Delta, k_t),
\end{equation}
This expression includes the effects of the flavour changes and the
distribution of the longitudinal momentum fraction of the leading
parton.
The factor $\frac{z}{2C_j}{\cal{P}}_j(z)$, with
${\cal{P}}_j(z) = \sum_k {\cal{P}}_{kj}^{(R)}(1-z)$, accounts for the
fact that close to the $z=1$ boundary, one should use the full
splitting function instead of its soft limit.

\begin{figure}
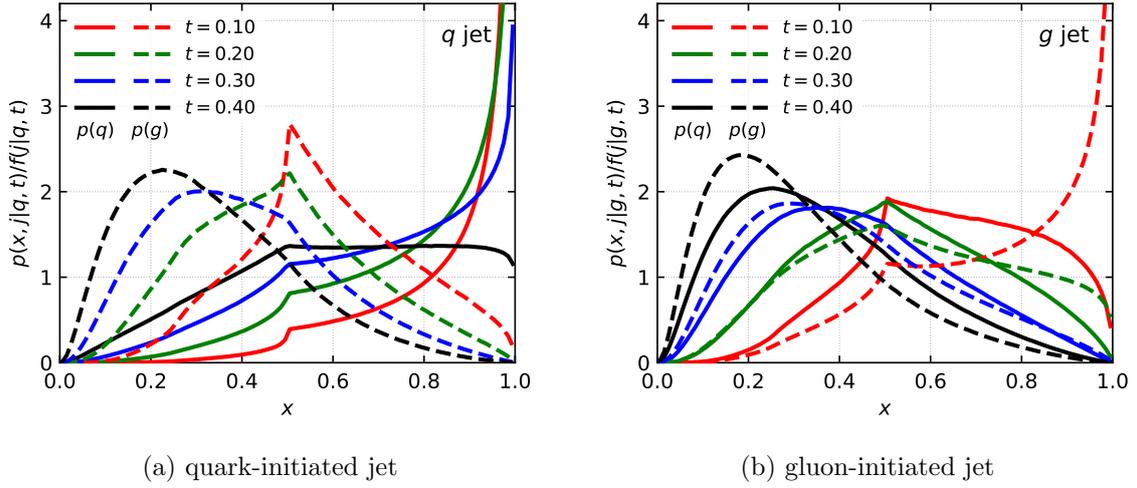

  \centering
  \begin{subfigure}[t]{0.48\textwidth}
    \includegraphics[width=\textwidth, page=7]{figs/qg-frac.pdf}
    \caption{quark-initiated jet}\label{fig:pjq-distribs}
  \end{subfigure}
  \hfill
  \begin{subfigure}[t]{0.48\textwidth}
    \includegraphics[width=\textwidth, page=8]{figs/qg-frac.pdf}
    \caption{gluon-initiated jet}\label{fig:pjg-distribs}
  \end{subfigure}
  \caption{Distributions
    $p(x,j|i,t_\text{coll})/f(j|i,t_\text{coll})$, with
    $f(j|i,t_\text{coll})$ the average fraction of partons of flavour
    $j$ after a time $t_\text{coll}$ starting from a flavour $i$ (cf.\
    Eq.~(\ref{eq:f-xbar-def})), for
    different values of $t_\text{coll}$ as a function of $x$. The left (right)
    plot correspond to quark (gluon)-initiated jets. On both plots, the
    solid (dashed) lines correspond to a quark (gluon) leading
    parton.
  }\label{fig:pji-distribs}
\end{figure}

In practice, an exact analytic calculation of
$p(x,j | i, t_\text{coll})$ is not possible. It is however
straightforward to obtain it numerically using the approach of
Ref.~\cite{Dasgupta:2014yra}.\footnote{We actually use a simplified
  version where, after a given splitting, only the harder of the two
  branches is further split.}
A sample of the resulting distributions is shown in
Fig.~\ref{fig:pji-distribs} for both quark-initiated and
gluon-initiated jets.
The distributions progressively shift towards smaller values of $x$ as
expected.

From Eq.~(\ref{eq:dglap}) one can also deduce a few analytic
properties of $p(x,j | i, t_\text{coll})$.
In particular, if one takes the $0^\text{th}$ and $1^\text{st}$ moments
of~(\ref{eq:dglap}) one obtains respectively an evolution equation for
the average fraction of quarks and gluons and the average longitudinal
momentum of the leading parton.
We therefore define
\begin{equation}\label{eq:f-xbar-def}
  f(j|i, t_\text{coll}) = \int_0^1 dx\,p(x,j|i,t_\text{coll})
  \quad\text{ and }\quad
  \bar x(j|i, t_\text{coll}) = \frac{1}{f_j(i, t_\text{coll})} \int_0^1 dx\,x\,p(x,j|i,t_\text{coll}).
\end{equation}
For both $f$ and $f\bar x$ one can write a closed equation which
admits a solution under the form of a (matrix) exponential. The
solutions to these equations are given in Appendix~\ref{sec:coll-analytic}.

\begin{figure}
  \centering
  \begin{subfigure}[t]{0.48\textwidth}
    \includegraphics[width=\textwidth, page=1]{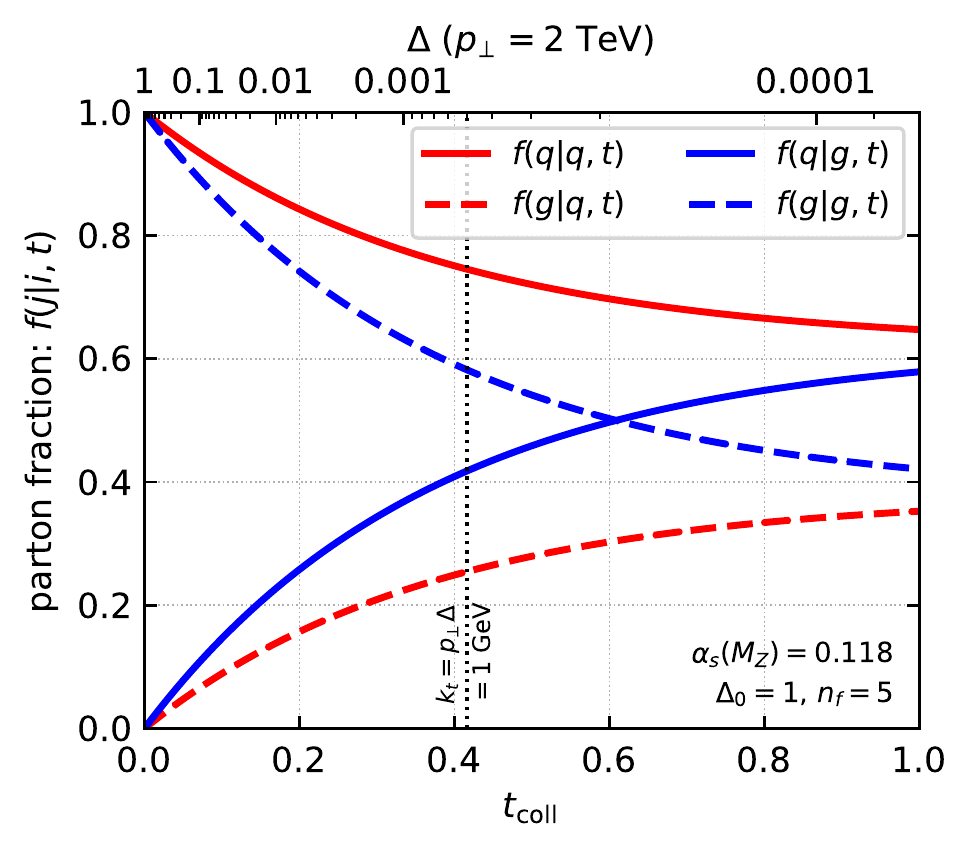}
    \caption{Average partonic fraction.}\label{fig:qg-frac}
  \end{subfigure}
  \hfill
  \begin{subfigure}[t]{0.48\textwidth}
    \includegraphics[width=\textwidth, page=2]{figs/qg-frac.pdf}
    \caption{Average leading-parton momentum fraction.}\label{fig:qg-mom-frac}
  \end{subfigure}
  \caption{Average parton fraction (left) and longitudinal momentum
    (right) as a function of the evolution time variable
    $t_\text{coll}$. The red (blue) curves correspond to a quark
    (gluon) initial parton. The solid (dashed) curves correspond to a
    quark (gluon) leading parton at time $t_\text{coll}$. The top axis
    shows the angle corresponding to $t_\text{coll}$ for a 2-TeV jet.
    The $t_\text{coll}$ axis extends significantly beyond the typical
    perturbative region so as to help illustrate the asymptotic trends.
  }\label{qg-avgs}
\end{figure}

Plots of $f(j|i,t)$ and $\bar x(j|i,t)$ are shown in
Fig.~\ref{qg-avgs} from which we can make several observations.
First, as $t_\text{coll}\to \infty$, i.e.\ $\Delta \to 0$ (modulo
Landau-pole complications), the quark and gluon fractions
tend to constants that are independent of the initial flavour of the
jet.
From~\eqref{eq:flav-solution} one finds 
\begin{equation}\label{eq:qg-asymptotic}
  f(q|\text{any},t_\text{coll}\to \infty) = \frac{s_g}{s_q+s_g}, \qquad
  f(g|\text{any},t_\text{coll}\to \infty) = \frac{s_q}{s_q+s_g},
\end{equation}
with $s_q=C_F(2\ln 2-\tfrac{5}{8})$ and $s_g=\tfrac{1}{3}n_f$.
This means about 62.1\% quarks and 37.9\% gluons for $n_f=5$, in
agreement with Fig.~\ref{fig:qg-frac}.
Furthermore, for $t_\text{coll}\to 0$, we find
\begin{equation}
  \label{eq:qg-mom-fracs}
  \lim_{t_\text{coll} \to 0} \frac{\bar x(g|q,t_\text{coll})}{t_\text{coll}}
  = \frac{w_{gq}}{s_q} \simeq 0.712,
  \qquad
  \lim_{t_\text{coll} \to 0} \frac{\bar x(q|g,t_\text{coll})}{t_\text{coll}}
  = \frac{w_{qg}}{s_g} \simeq 0.78125,
\end{equation}
where the $w_{gq}$ and $w_{qg}$ coefficients are given analytically in
Appendix~\ref{sec:coll-analytic}.
The results in Eq.~(\ref{eq:qg-mom-fracs}) correspond to the average
harder-parton momentum fractions after a single $q\to qg$ splitting
(with the gluon the harder particle) or a single $g\to q\bar q$
splitting.
The numerical values are in agreement with Fig.~\ref{fig:qg-mom-frac}.

%----------------------------------------------------------------------
\subsection{Soft emissions at large or commensurate angles}\label{sec:soft}

The average primary Lund density is subject to several classes of
effect associated with the non-trivial characteristics of soft radiation.
At $\order{\as}$, when one goes beyond the collinear limit of
Eq.~(\ref{eq:rho-fc}) and considers radiation at angles
$\Delta \sim 1$, soft gluon radiation is a coherent sum from all hard
coloured partons in the event rather than just a single parton.
At higher orders, two further effects arise:
the presence of a first soft gluon contributes to the radiation of a
subsequent second soft gluon at commensurate angles, and so forth for
higher numbers of gluons, contributing effects similar to non-global
logarithms;
and in the presence of two or more gluons at commensurate angles, one
must account for the way in which jet clustering determines whether
a given gluon is classified as a primary or a secondary Lund emission.
These two effects are present both for large and (perhaps more surprisingly) small $\Delta$.

In this section we will consider all of these effects, using the
large-$N_C$ limit so as to retain simple colour algebra.
After considering how we decompose events into separate colour flows,
we shall in section~\ref{sec:soft-largeR} examine the impact of
different colour flows on the large-angle part of the Lund plane at
$\order{\as}$.
Then in section~\ref{sec:soft-clust} we shall consider the case of
double (energy-ordered) soft gluon emission and derive the structure
of the non-global and clustering logarithms in the small-angle limit
at order $\as^2$.
Finally in section~\ref{sec:soft-resum} we will discuss how we
generalise these results to resum all sources of single soft
logarithms to all orders.

We start by recalling that in the large-$N_C$ limit, a given
Born-level process can be expressed as a sum over several partonic
channels where each of them is a weighted sum of different colour
flows:
\begin{equation}\label{eq:rho-soft}
  \rho_\text{soft}(\Delta, k_t)
  = \sum_{c\in\text{channel}}\sum_{f\in\text{flows}} w_{c,f}\, \rho_\text{soft}^{(c,f)}(\Delta,k_t).
\end{equation}
In this context, for each colour flow, one can view the Born-level
process as a superposition of colour dipoles.
Let us consider a dijet process with two incoming partons,
$p_{1,2}$, and two outgoing partons, a jet parton $p_j$ and a recoiling parton
$p_r$.
The relative weights of the different partonic channels can be obtained
from the $2\to 2$ squared matrix elements, e.g.\ using \texttt{NLOJet++}.
Then, a $q_\text{in}q_\text{in}'\to q_\text{out}q_\text{out}'$ channel
would have, in the large-$N_c$ limit, a single colour flow with weight
$w=1$ corresponding to dipoles $[(q_\text{in}q_\text{out}')+(q_\text{in}'q_\text{out})]$. Similarly, a $g_\text{in}q_\text{in}\to g_\text{out}q_\text{out}$
channel would have two colour flows:
\begin{subequations}\label{eq:flows-gq2gq}
\begin{align}
  [(g_\text{in}g_\text{out}) + (q_\text{in}g_\text{out}) + (g_\text{in}q_\text{out})]&\quad\text{ with weight }\quad\frac{s^2}{s^2+u^2},\\
  [(g_\text{in}g_\text{out}) + (g_\text{out}q_\text{out}) + (g_\text{in}q_\text{in})]&\quad\text{ with weight }\quad\frac{u^2}{s^2+u^2},
\end{align}
\end{subequations}
with
$s,t,u$ the usual Mandelstam variables.
The complete set of colour flows, dipole superpositions and weights
can, for example, be deduced from Ref.~\cite{Ellis:1986bv}.

In the next sections, it will be helpful to separate
$\rho_\text{soft}$ in different contributions according to the
Born-level flavour of the jet:
\begin{equation}\label{eq:rho-soft-flv}
  \rho_\text{soft}(\Delta, k_t)
  = \sum_{i\in\text{jet flavours}} f_i\: \rho_{\text{soft},i}(\Delta,k_t),
\end{equation}
where $f_i$ denotes the relative fraction of quark and gluon jets.
This separation in jet flavours can be straightforwardly obtained
from~(\ref{eq:rho-soft}).

We can expand $\rho_{\text{soft},i}$ as a series in $\alpha_s$:
\begin{equation}
  \rho_{\text{soft},i}(\Delta, k_t) =\sum_{n=1}^\infty \alpha_s^n \rho_{\text{soft},i}^{(n)}(\Delta, k_t).
\end{equation}
In the first subsection below, we will show that
$\rho_{\text{soft},i}^{(1)}$ deviates from~(\ref{eq:rho-fc}) by
corrections that are power-suppressed in $\Delta$.
At our single-logarithmic accuracy, we have
\begin{equation}
  \alpha_s^n\rho_{\text{soft},i}^{(n)}(\Delta, k_t) \propto 
  \alpha_s^n\ln^{n-1}\left(\frac{\pt\Delta}{k_t}\right).
\end{equation}
These soft logarithms are either due to non-global configurations
or to clustering logarithms associated with the Cambridge/Aachen
reclustering used to construct the primary Lund-plane density.
It is interesting to note that these clustering logarithms are present
at arbitrarily small angles, which is somehow uncommon (though
addressed also in \cite{Kang:2019prh}).
We show how they appear at order $\alpha_s^2$ in
section~\ref{sec:soft-clust} and provide an all-order resummation in section~\ref{sec:soft-resum}.

\subsubsection{Soft emissions at large angles: fixed-order study}\label{sec:soft-largeR}

For definiteness, let us consider the case of two incoming partons,
$\ell_{1,2}$, and two outgoing partons, $\ell_{j,r}$, with the following
kinematics:\footnote{4-momenta are written as $(p_x, p_y, p_z, E)$.}
\begin{subequations}\label{eq:incoming-outgoing}
\begin{align}
  \ell_1^\mu & \equiv \frac{Q}{2} (0,0,1,1),\\
  \ell_2^\mu & \equiv \frac{Q}{2} (0,0,-1,1),\\
  \ell_j^\mu & \equiv \pt (1,0, \sinh\yjet,\cosh\yjet),\\
  \ell_r^\mu & \equiv \pt (-1,0,-\sinh\yjet,\cosh\yjet),
\end{align}
\end{subequations}
where $Q = \pt \cosh\yjet$ (with the jet transverse momentum
$p_\perp\equiv \ell_\perp$, while its rapidity is equal to $\yjet$).
We consider the jet of radius $R$ around $\ell_j$, while $\ell_r$
corresponds to the recoiling hard jet.

In the large-$N_c$ approximation we have to consider soft gluon
emission from any of 6 possible colour dipoles: one incoming-incoming, two
jet-incoming, two recoil-incoming and one jet-recoil. For a given dipole
with legs $\ell_a$ and $\ell_b$, the contribution from a soft emission
of momentum
$k^\mu\equiv k_\perp(\cos\phi,\sin\phi,\sinh y,\cosh y)$ takes
the form
\begin{equation}
  \alpha_s\rho_\text{soft,$ab$}^{(1)}(\Delta, k_t)
  = \int_0^{\pt}\! k_\perp dk_\perp\!\int dy \int_0^{2\pi}\frac{d\phi}{2\pi}
  \frac{\alpha_s N_c}{2\pi}
  \frac{\ell_a\cdot \ell_b}{(\ell_a\cdot k) (k\cdot\ell_b)}
  \,\Delta \delta(\Delta-\Delta R) 
  \,k_t \delta(k_t-k_\perp\Delta R) ,
\end{equation}
with $\Delta R=\sqrt{(y-\yjet)^2+\phi^2}$.
For each dipole configuration, we can set $y=\yjet+\Delta R\cos\psi$
and $\phi=\Delta R\sin\psi$. The $k_\perp$ and $\Delta R$
integrations can then be evaluated trivially, leaving the integration over
$\psi$. This integration usually cannot be computed exactly so
we instead perform a series expansion in $\Delta^2$:
\begin{subequations}\label{eq:soft-em-largeR-alphas}
  \begin{align}
    \alpha_s\rho_{\text{soft},12}^{(1)}
    & = \frac{\alpha_s N_c}{\pi} \Delta^2, \\
    \alpha_s\rho_{\text{soft},1j}^{(1)}
    & = \frac{\alpha_s N_c}{\pi} \left[1 + \frac{\Delta^2}{4} +
      \frac{\Delta^4}{144} + {\cal {O}}(\Delta^8) \right],\\
    \alpha_s\rho_{\text{soft},jr}^{(1)}
    & = \frac{\alpha_s N_c}{\pi} \left[1
      +\frac{\tanh^2\yjet}{4} \Delta^2+\frac{(\cosh^2\yjet-3)^2}{144
      \cosh^4\yjet} \Delta^4 + \frac{\tanh^2\yjet}{64\cosh^4\yjet}\Delta^6 + {\cal {O}}(\Delta^8) \right],\\
    \alpha_s\rho_{\text{soft},1r}^{(1)}
    & = \frac{\alpha_s N_c}{\pi} \left[
      \frac{e^{2\yjet}}{4\cosh^2\yjet} \Delta^2
      +\frac{1}{16\cosh^4\yjet} \Delta^4
      +\frac{\tanh^2\yjet}{64\cosh^4\yjet} \Delta^6
      + {\cal {O}}(\Delta^8) \right],      
  \end{align}
\end{subequations}
where additionally $\rho_{\text{soft},2j}^{(1)}=\rho_{\text{soft},1j}^{(1)}$
and $\rho_{\text{soft},2r}^{(1)}$ is obtained from
$\rho_{\text{soft},1r}^{(1)}$ via the replacement $\yjet\to
-\yjet$.

In the collinear limit, $\Delta\ll 1$,
$\rho_{\text{soft},2j}^{(1)}$,
$\rho_{\text{soft},1j}^{(1)}$ and
$\rho_{\text{soft},jr}^{(1)}$ all tend to a constant, with
corrections taking the form of power corrections in $\Delta^2$, as
expected.
The other dipoles are suppressed by a factor $\Delta^2$.

While running-coupling and collinear flavour-changing effects only
depend on the flavour of the jet, the corrections due to soft emissions
at large angles involve the structure of the whole event.
The relative weight of each dipole depends on the channel and colour
flow under consideration, cf.~(\ref{eq:rho-soft}).

%----------------------------------------------------------------------
\subsubsection{Soft emissions and Cambridge/Aachen clustering:
  fixed-order study}\label{sec:soft-clust}

Say we want to extend the calculation from
section~\ref{sec:soft-largeR} to order $\alpha_s^2$. The same
calculation as above would have to be repeated with two soft
emissions, $k_1$ and $k_2$, strongly ordered in energy ($k_{\perp
  1}\gg k_{\perp 2}$).
Measuring the emission $k_2$ and integrating out $k_1$ yields a
contribution to the primary Lund plane density of the form
\begin{equation}\label{eq:alphas2-soft-expectation}
  \alpha_s^2\rho_{\text{soft}}^{(2)} \propto \alpha_s^2
  \ln\frac{k_t}{\pt\Delta}.
\end{equation}
In the limit $\Delta \to 0$ the prefactor is simple, and we calculate
it here.

For concreteness, we illustrate the case of a quark-induced jet.
We denote by $\theta_i$ the angle between $k_i$ and the quark and
by $\theta_{12}$ the angle between $k_1$ and $k_2$ and work in a
limit where all angles are small.
In contrast to section~\ref{sec:soft-largeR}, we now use a frame
where the jet is perpendicular to the beam.
In conjunction with the small-angle limit, this ensures that angles
and rapidity-azimuth distances are equivalent, as are energies and
transverse momenta (with respect to the beam).

Let us first consider three simple nested-collinear limits.
When $\theta_1 \gg \theta_2$, gluon $k_2$ is emitted with colour factor
$C_F$ and declustered as a primary emission, i.e.\ the LO $2\alpha_s
C_R/\pi$ emission intensity for gluon $k_2$ is unaffected by the presence
of gluon $k_1$.
The situation is similar when $\theta_1 \ll \theta_2$.
When $\theta_{12} \ll \theta_1$, gluon $k_2$ is emitted with colour
factor $C_A$ and declustered as a secondary emission, i.e.\ on gluon
$k_1$'s Lund leaf.
The only non-trivial situation is when the angles $\theta_1$,
$\theta_2$ and $\theta_{12}$ are commensurate,
$\theta_1\sim\theta_2\sim\theta_{12}\ll 1$.
In this region, one needs to account for the non-trivial matrix
element for the emission of two gluons at commensurate angles and for
the effects of the Cambridge/Aachen clustering used to construct the
Lund plane.
Together, these induce an
$\mathcal{O}(\alpha_s^2 \ln k_t/(\pt\Delta))$ correction to the
$2\alpha_s C_R/\pi$ behaviour, where the logarithm is associated with
the integral over the transverse momentum of gluon $k_1$.

This contribution to the primary Lund density for a quark-induced jet
can be written
\begin{align}\label{eq:clust-initial-expression}
  \alpha_s^2\rho_\text{soft}^{(2)}
   = & \left(\frac{\alpha_s}{\pi^2}\right)^2
  \int_0^{\pt}\frac{dk_{\perp 1}}{k_{\perp 1}}
  \int_0^{k_{\perp 1}}\frac{dk_{\perp 2}}{k_{\perp 2}}
  \int d^2\theta_1 \frac{C_F}{\theta_1^2}
  \int d^2\theta_2\;
  \Delta \, \delta(\Delta-\theta_2) \, k_t \, \delta(k_t-k_{\perp 2}\Delta) \nonumber\\
  & \left\{
    \left[
      \left(C_F-\frac{C_A}{2}\right)\frac{1}{\theta_2^2}
      + \frac{C_A}{2}\frac{1}{\theta_{12}^2}
      + \frac{C_A}{2}\frac{\theta_1^2}{\theta_2^2\theta_{12}^2}
    \right]
    \left[
      1-\Theta(\theta_{12}<\theta_1)\Theta(\theta_{12}<\theta_2)
    \right]
  - \frac{C_F}{\theta_2^2}
  \right\}, 
\end{align}
where $k_{\perp i}$ is the transverse momentum of $k_i$ relative to the beam.
In this expression the first (second) term of the curly bracket
corresponds to a real (virtual) emission $k_1$.
For the real emission, we have two factors: the first square bracket
corresponds to the matrix element for the emission of the soft gluon
$k_2$ and the second square bracket imposes that the gluon $k_2$
is reconstructed as a primary emission, i.e.\ is not clustered with
the emission $k_1$.

Eq.~(\ref{eq:clust-initial-expression}) genuinely encodes clustering
effects.
If we first consider the $C_F^2$ contribution, naively associated with
two emissions from the hard quark, the virtual term
partially cancels the real contribution, leaving a negative
contribution with a factor
$\Theta(\theta_{12}<\theta_1)\Theta(\theta_{12}<\theta_2)$, i.e.\
where emission $k_2$ clusters with emission $k_1$. Obviously,
this contribution disappears in the collinear limit
$\theta_2\ll\theta_1,\theta_{12}$ as expected.
Focusing now on the $C_F C_A$ term, naively associated with secondary
$k_2$ emission, the only contribution comes from the situation
where the emission is not clustered with its emitter, $k_1$, which
vanishes in the collinear limit $\theta_{21}\ll\theta_1$.

The $k_{\perp 1}$ integration in
Eq.~(\ref{eq:clust-initial-expression}) has a logarithmic
enhancement from strong energy ordering
$k_{\perp 2}\ll k_{\perp 1}\ll \pt$, leading to a contribution proportional to
$\alpha_s^2\ln\frac{k_t}{\pt\Delta}$, as anticipated in
Eq.~\eqref{eq:alphas2-soft-expectation}, and no collinear
divergence. The integrand is suppressed in the limits $\theta_1\to 0$
and $\theta_1\to\infty$ and only receives a contribution from
$\theta_1\sim\theta_2$.\footnote{A consequence of this is that, at our
  logarithmic accuracy, we can safely set the upper bound on the
  $\theta_1$ integration to infinity.}

The integration over $k_{\perp 1}$, $k_{\perp 2}$, $\theta_2$ and one of
the azimuthal angles ($\varphi_1$ or $\varphi_2$) can be trivially
performed, leaving an integration over $\theta_1$ and an azimuthal
angle $\varphi$.
One finds\footnote{The numerical pre-factor is analytically found to
  be $\frac{-i\pi}{36}-\frac{i}{\pi}\big[{\rm Li}_2\big(\frac{1+i\sqrt{3}}{2}\big)+\frac{1}{2}{\rm Li}_2\big(\frac{1-i\sqrt{3}}{2}\big)+\frac{5}{2}{\rm Li}_2\big(\frac{-1+i\sqrt{3}}{2}\big)+{\rm Li}_2\big(\frac{-i}{\sqrt{3}}\big)-{\rm Li}_2\big(\frac{i}{\sqrt{3}}\big)+{\rm Li}_2\big(\frac{3+i\sqrt{3}}{6}\big)-{\rm Li}_2\big(\frac{3-i\sqrt{3}}{6}\big)\big]$.}
\begin{equation}\label{eq:clustering-alphas2-quark}
  \alpha_s^2\rho_\text{soft}^{(2)}
  = 0.323066\,\left(\frac{2\alpha_s}{\pi}\right)^2 C_F (C_F-C_A)
  \ln\frac{k_t}{\pt \Delta} \qquad \qquad \text{[quark]}.
\end{equation}
The calculation for a gluon jet can be obtained by replacing $C_F \to
C_A$ in Eq.~(\ref{eq:clust-initial-expression}).
That replacement carries through directly to
Eq.~(\ref{eq:clustering-alphas2-quark}), giving
\begin{equation}\label{eq:clustering-alphas2-gluon}
  \alpha_s^2\rho_\text{soft}^{(2)}
  = 0 \qquad \qquad  \text{[gluon]}.
\end{equation}
Thus for a purely gluonic theory, the energy-ordered double-soft
emission pattern and the C/A clustering combine in such a way that
there is no $\alpha_s^2 \ln k_t/(\pt \Delta)$ correction to the Lund
density when $\Delta$ is small. 

The above results are valid at small angles.
Two additional classes of effect arise at large angles.
Firstly, the clustering effects become sensitive to the coherent structure
of the radiation from the complete hard event. This relates to the
discussion in section~\ref{sec:soft-largeR}.
Secondly, if one identifies the jet with the anti-$k_t$ algorithm and
reclusters its constituents  with the C/A algorithm, there is an
interplay between the two clusterings.
This leads to another source of logarithmic enhancement
\begin{equation}
  \alpha_s^n
  \ln^m\left(\frac{p_\perp R}{k_t}\right)
  \ln^p\left(\frac{R}{R-\Delta}\right),\qquad
  \text{with }m\le n-1,\; p\le n-1.
\end{equation}
The $\ln(\frac{R}{R-\Delta})$ structure appears when a first emission,
close to but outside the anti-$k_t$ jet boundary, splits collinearly
such that one of its offspring is inside the boundary.\footnote{
  It is related to the $\Delta\eta \ln\Delta\eta$ term observed in
  Eq.~(3.13) of~\cite{Dasgupta:2002bw}.}
The all-order resummation of these {\em boundary logarithms} is beyond
the scope of this paper. They are however briefly discussed in
Appendix~\ref{sec:boundary-logs}.
Note that if the original jet is identified with the C/A algorithm,
these boundary logarithms are absent.

%----------------------------------------------------------------------
\subsubsection{Soft emissions: all-order treatment}\label{sec:soft-resum}

The treatment of soft single logarithms to all orders requires us to
consider configurations with arbitrarily many energy-ordered gluons at
commensurate angles~\cite{Dasgupta:2001sh}, for which analytic
approaches exist only in specific limits~\cite{Banfi:2002hw}.
The technique we adopt is similar to that originally proposed for the
resummation of non-global logarithms in~\cite{Dasgupta:2001sh} and the
related clustering logarithms~\cite{Appleby:2003sj,Banfi:2005gj}.
We rely on a large-$N_C$ approximation (with subleading colour
corrections up to order $\as^2$ in the collinear limit).
Techniques that exist to resum non-global logarithms at full $N_C$ are so
far applicable only for a limited set of
observables~\cite{Hatta:2013iba,Hagiwara:2015bia}.

Compared to the typical treatment of non-global and clustering
logarithms we have one extra difficulty and one simplification.
The difficulty has the following origin.
Since non-global logarithms stem from emissions at commensurate angles
one can usually impose an angular cut-off at an angle
$\theta_\text{min}$ that is small compared to the physical angle one
probes ($\pi/2$, a jet radius, the rapidity width of a slice, ...).
This helps limit the particle multiplicity and associated
computational cost of the calculations.
In the case of the primary Lund plane, we instead have to probe a
large range of angles, meaning that potential angular cut-offs have to
be taken small/large enough to cover this extended phase-space,
resulting in increased computational demands.\footnote{%
  Similar non-global and clustering logarithms have been studied down
  to small angles in Ref.~\cite{Kang:2019prh} for the study of the
  SoftDrop grooming radius at NLL accuracy.
  There, the authors relied on the fact that the behaviour of the
  clustering logarithms becomes independent of $\Delta$ at
  small-enough $\Delta$.
  In this paper, we decided not to rely on this behaviour so as to
  also reach a good level of numerical precision for the approach to
  this asymptotic regime within our single-logarithmic accuracy, and
  to do so over the relatively large energy range needed to cover the
  full Lund plane.}
The simplification relative to normal non-global logarithm
calculations relates to the fact that the Lund plane density doesn't
involve any Sudakov suppression (unlike say a hemisphere mass).
That Sudakov suppression, specifically the part associated with
primary emissions, can lead to low computational efficiency unless
dedicated subtraction techniques are applies.
In the case of the Lund plane density one can simply generate all the
emissions without separating primary emissions from the other ones.

The basic approach to simulating soft emissions is to directly order
them in energy, as done in~\cite{Dasgupta:2001sh}. In this case we just
need to generate the full angular structure of the emissions and only
retain their energy ordering. If gluon $k_j$ is emitted after $k_i$, then
it has a much smaller energy.
In this case clustering with the anti-$k_t$ algorithm is equivalent to
keeping the particles in a radius $R$ around the hard parton. For C/A
clustering, all the necessary distances are available from the angular
structure of the event and the recombination of two particles is
equivalent to replacing them with the harder one.

For our ultimate primary Lund plane predictions, we want to focus on
the phase-space above a certain $k_t$ cut, below which the
non-perturbative effects dominate.
For a given minimum relative transverse momentum $k_{t,\text{min}}$
the minimum accessible angle is
$\Delta_\text{min}=k_{t,\text{min}}/\pt$. We therefore have to take an
angular cut-off for the event simulations that is sufficiently
smaller that $\Delta_\text{min}$.
If we generate emissions down to an energy $E_\text{min}$, many of
these emissions will have a $k_t$ much smaller than
$k_{t,\text{min}}$. For example for $\Delta=\Delta_\text{min}$,
emissions would be generated down to
$k_t=E_\text{min}\Delta_\text{min}\ll k_{t,\text{min}}$.
This is not a problem per se, except for the fact that this approach
generates many more emissions than absolutely necessary, which ends up
being computationally challenging.

For this reason, we have adopted a different approach, more
traditional in parton-shower event generators, namely we generate the
emissions ordered in $k_t$. Say we work with a fixed coupling.
The event is described as a collection of dipoles, each with a hard
scale corresponding to their invariant mass $Q$.
For the initial condition, we decompose the Born-level event as a sum
over all possible dipole configurations.
At any given stage of the event generation, corresponding to a given
scale $k_t=k_{ti}$, we should be able to generate the next emission
at a scale $k_{t,i+1}<k_{ti}$.
If a dipole $(p_1,p_2)$ of invariant mass $Q_{12}$ splits, this is
done by first generating $k_{t,i+1}$ according to the following
Sudakov factor (which includes collinear radiation at each end of the
dipole):
\begin{equation}
  \exp\left[-\int_{\ln k_{t,i+1}}^{\ln k_{ti}}d\ln q_t\int_{\ln
      \ell_t/Q_{12}}^{\ln Q_{12}/\ell_t} d\eta \frac{2\alpha_s N_c}{\pi}\right]
  = \exp\left[-\frac{\alpha_s N_c}{\pi} \ln
    \frac{k_{ti}}{k_{t,i+1}} \ln \frac{Q_{12}^2}{k_{ti}\,k_{t,i+1}}\right].
\end{equation}
One then decides the rapidity $\eta$ of the emission, uniformly
distributed between $\ln k_{t,i+1}/Q_{12}$ and
$\ln Q_{12}/k_{t,i+1}$ as well as an azimuthal angle
$\phi$. The 4-momentum of the new emission is thus reconstructed as
\begin{equation}
  k_{i+1}^\mu = \xi_1 p_1^\mu + \xi_2 p_2^\mu + k_{t,i+1}^\mu
\end{equation}
with
\begin{equation}\label{eq:splitting-kinematic-variables}
  \xi_1 = \frac{k_{t.i+1}}{Q_{12}}e^{\eta},\qquad
  \xi_2 = \frac{k_{t.i+1}}{Q_{12}}e^{-\eta},\qquad
  k_{t,i+1}^\mu = k_{t.i+1} (\cos\phi\, n_1^\mu + \sin\phi\, n_2^\mu)
\end{equation}
and $n_{1,2}$ two unit vectors orthogonal to $p_1$ and $p_2$.

Note that when a dipole $(p_1,p_2)$ splits into two new dipoles
$(p_1,k_{i+1})$, $(k_{i+1},p_2)$, the energy scales of the two new
dipoles can be straightforwardly obtained using
\begin{align}
  Q_{1k}^2 & = (p_1+k_{i+1})^2 = \xi_2\,Q_{12}^2 = Q_{12}\, k_{t,i+1}\,e^{-\eta},\\
  Q_{k2}^2 & = (p_2+k_{i+1})^2 = \xi_1\,Q_{12}^2 = Q_{12}\, k_{t,i+1}\,e^{+\eta}
\end{align}
In order to reduce the inclusion of uncontrolled corrections beyond
our intended resummation of single logarithms, we can perform another
simplification. 
Recall that we are aiming to resum the energy
logarithms due to clustering effects. Instead of computing the energy
of each particle explicitly from its 4-momentum, we can directly
project this momentum along the direction of the initial dipole.
Let us denote by $(p, \bar p)$ the hard-scattering momenta of
a given initial colour dipole.
For each emission,
we want to find the contributions $z$ and $\bar z$ of its momentum
fractions along the $p$ and $\bar p$ directions respectively.
For the emission of a new particle $k$ from a $(p_1, p_2)$ dipole, we
have
\begin{align}
  p_1^\mu & = z_1 p^\mu + \bar z_1 \bar p^\mu + p_{t,1}^\mu \\
  p_2^\mu & = z_2 p^\mu + \bar z_2 \bar p^\mu + p_{t,2}^\mu 
\end{align}
The new emission $k$  therefore has a projection $z_k$,
$\bar z_k$ along the $p$, $\bar p$ directions given by
\begin{align}
  z_k & = \xi_1 z_1 + \xi_2 z_2
        \approx \frac{k_{t}}{Q_{12}}
                \max(z_1 e^\eta, z_2 e^{-\eta}),\\
  \bar z_k & = \xi_1 \bar z_1 + \xi_2 \bar z_2
        \approx \frac{k_{t}}{Q_{12}}
                \max(\bar z_1 e^\eta, \bar z_2 e^{-\eta}),
\end{align}
where we have used Eq.~(\ref{eq:splitting-kinematic-variables}) and
replaced the sum over the two contributions by its maximum at our
accuracy.

Iterating the above procedure for emissions ordered in $k_t$
produces an event where each particle has a 4-momentum as well as
longitudinal fractions $z$ and $\bar z$ along their initial $(p,\bar p)$
dipole. 
To reconstruct the primary Lund plane density we then proceed as
follows: the anti-$k_t$ jet of radius $R$ is made of all the emissions
within a radius $R$ of an initial hard parton. These particles can
then be clustered using the C/A jet algorithm. This clustering uses
the exact 4-momenta of the jets and a
winner-takes-all-like~\cite{Larkoski:2014uqa} recombination scheme
where the recombined particle is taken as the one of the two
recombining particles with the largest \stackon[.1pt]{$z$}{\brabar}
momentum along the jet direction.\footnote{The
  usage of the \stackon[.1pt]{$z$}{\brabar} momentum fractions
  guarantees that in the collinear limit only the
  logarithmically-enhanced contributions are kept. At large angles,
  and, in particular, for initial dipoles which do not involve the jet
  momentum, one could generate subleading corrections as well.
  In practice however, our fixed-order tests indicate that these
  subleading corrections are very small, if present at all.}
When we reconstruct the primary Lund plane density, the $\Delta$
coordinate is again taken from the exact angular kinematics, the $z$
variable from the event \stackon[.1pt]{$z$}{\brabar} and hence $k_t$
is obtained as $p_{t,\text{hard}} z \Delta$ with $p_{t,\text{hard}}$
the jet transverse momentum (w.r.t.\ the colliding beams) of the
initial hard parton.

The above discussion is strictly speaking valid only in a
fixed-coupling approximation.
To account for running-coupling effects, we use the following
procedure.
For given Born-level kinematics (see e.g.\
(\ref{eq:incoming-outgoing})) and a given colour flow corresponding to
a given initial set of dipoles, we generate a Monte Carlo event using
a fixed $\alpha_s$.\footnote{In practice, we take the coupling at the
  scale $p_\perp R$.}
Following the procedure outlined in the previous
paragraph we obtain the coordinates $\Delta$ and $z$ of the primary
Lund declusterings. From the $z$ coordinate, we then determine an
emission ``time'' $t_{\text{soft},\text{fc}}$ defined as $t_{\text{soft},\text{fc}}=\alpha_s\ln 1/z$. This procedure
yields a resummed density $\rho_\text{soft}(\Delta, t_\text{soft,fc})$.
To include running-coupling corrections at a given $k_t$ and $\Delta$,
we simply use $\rho_\text{soft}(\Delta, t_\text{soft})$ with a
$t_\text{soft}$ defined to include running-coupling effects:
\begin{equation}\label{eq:rho-soft-resummed}
  \rho_\text{soft}\big(\Delta, t_\text{soft}(k_t, \pt\Delta; \mu)\big),
\end{equation}
with (cf.\ Eq.~(\ref{eq:tcoll}))
\begin{equation}\label{eq:tsoft}
  t_\text{soft}(k_t, \pt\Delta;\mu)
   = \int_{k_t}^{\pt\Delta}
   \frac{dq_{t1}}{q_{t1}}\frac{\alpha_s(q_{t1})}{\pi}
   \equiv I_\alpha(k_t, \pt\Delta;\mu).
\end{equation}
Additionally, this approach can be straightforwardly extended to
generate results at fixed-order. This will be useful to compare to the
results derived in section~\ref{sec:soft-largeR}
and~\ref{sec:soft-clust} as well as for matching with exact
fixed-order results in section~\ref{sec:fixed-order}.

\begin{figure}
  \centering
  \begin{subfigure}[t]{0.48\textwidth}
    \includegraphics[width=0.98\textwidth]{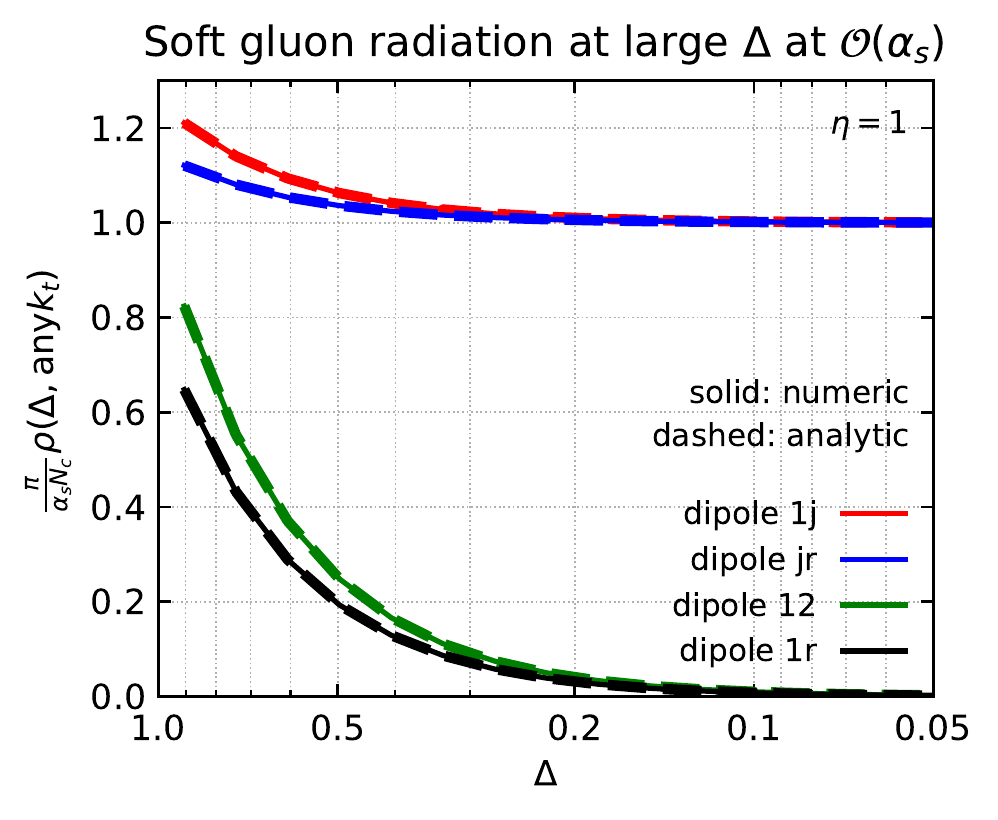}
    \caption{Effects of soft gluon radiation at large angles for
      different dipole configurations (including or not the
      jet).}\label{fig:largeR-soft-alphas}
  \end{subfigure}
  \hfill
  \begin{subfigure}[t]{0.48\textwidth}
    \includegraphics[width=1.015\textwidth]{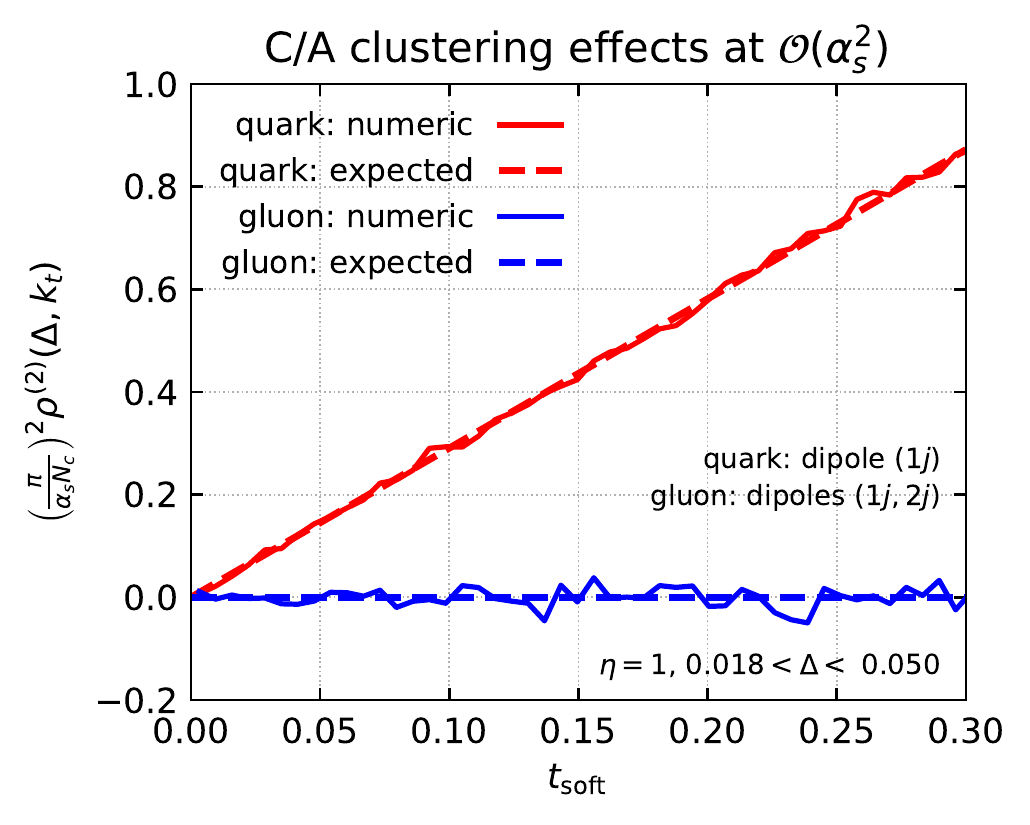}
    \caption{Effects of Cambridge/Aachen clustering at order
      $\alpha_s^2$ in the collinear limit. Results are shown for two
      dipole configurations corresponding to a quark and gluon jet
      respectively.}\label{fig:clustering-alphas2}
  \end{subfigure}
  \caption{Comparisons at fixed order between our numerical results
    for soft gluon radiation (solid lines) and analytic predictions
    from sections~\ref{sec:soft-largeR} and~\ref{sec:soft-clust}
    (dashed lines).
  }\label{fig:soft-gluon-fixed-order}
\end{figure}

As a validation of our numerical approach, we first compare the output
of the numerical approach to the analytic results for soft gluon
radiation at fixed order.
The predictions for soft radiation at large angle at
${\cal {O}}(\alpha_s)$, obtained in section~\ref{sec:soft-largeR}, are
compared to our numerical results in Fig.~\ref{fig:largeR-soft-alphas}.
The comparison is done assuming the large-$N_c$ limit and is
independent of $\ln k_t$ (modulo the soft approximation of the
kinematic constraint, $k_t/(p_\perp\Delta)\le 1$).
The figure shows excellent agreement with the analytic results from
Eq.~(\ref{eq:soft-em-largeR-alphas}).

With Fig.~\ref{fig:clustering-alphas2}, we study the numerical results
for C/A clustering effects at ${\cal {O}}(\alpha_s^2)$,
Eqs.~\eqref{eq:clustering-alphas2-quark}
and~\eqref{eq:clustering-alphas2-gluon}.
For a gluon jet, we need to consider two dipoles.
Since our calculation is done in the collinear limit, we have
considered a range of small values of $\Delta$.
The linear rise with $\ln z$, with the expected analytic coefficient,
is clearly visible for quark jets, together with no effects at this
order for gluon jets.

\begin{figure}
  \centering
  \begin{subfigure}[t]{0.48\textwidth}
    \includegraphics[width=\textwidth, page=1]{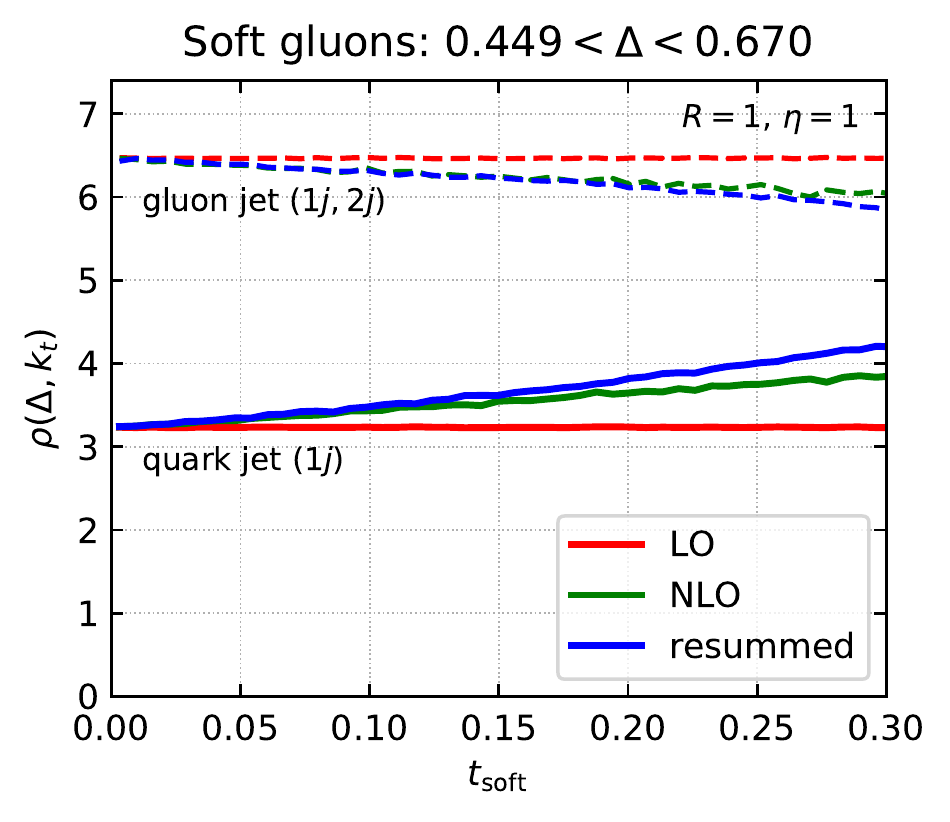}
    \caption{Large angle, $0.449<\Delta<0.670$}\label{fig:soft-resum-large}
  \end{subfigure}
  \hfill
  \begin{subfigure}[t]{0.48\textwidth}
    \includegraphics[width=\textwidth, page=2]{figs/all-orders.pdf}
    \caption{Small angle, $0.061<\Delta<0.091$}\label{fig:soft-resum-small}
  \end{subfigure}
  \caption{All-order resummation of logarithms from soft-gluon
    emissions. For both plots, we give the results for a
    quark configuration (solid lines) and a gluon configuration (dashed
    lines). LO (${\cal {O}}(\alpha_s)$) and NLO
    (${\cal {O}}(\alpha_s^2)$) results are shown for
    comparison.
    For the quark configuration we have used a $(1j)$ dipole
    connecting the incoming particle $\ell_1$ and the outgoing ``jet''
    $\ell_j$, while for the gluon configuration, we have considered the
    superposition of 2 dipoles $(1j,2j)$, i.e.\ one dipole connecting
    the incoming $\ell_1$ with $\ell_j$ and a second dipole connecting the
    other incoming parton $\ell_2$ with $\ell_j$ (cf.\ (\ref{eq:incoming-outgoing})).
    }\label{fig:soft-gluon-resum}
\end{figure}

All-order results are shown in Fig.~\ref{fig:soft-gluon-resum} for two
different regions in angle.
We see that apart from the region of very small $z$ (large
$t_\text{soft}$), the resummation has a relatively small effect
compared to the NLO result.
A feature that is particularly intriguing is that in the collinear
region, $\Delta \ll 1$, Fig.~\ref{fig:soft-resum-small}, the result
appears to be independent of $t_\text{soft}$.
Recall that for gluon jets, at order $\alpha_s^2$ the soft logarithmic
term was identically zero for small $\Delta$,
Eq.~(\ref{eq:clustering-alphas2-gluon}).
Fig.~\ref{fig:soft-resum-small} leads us to wonder whether the soft
single logarithmic terms remain zero for gluon jets at all orders, or
whether they are non-zero but simply too small to observe in our
calculation.
Note however that at large angles, Fig.~\ref{fig:soft-resum-large},
there is a clear $t_\text{soft}$ dependence both at $\as^2$ and
beyond, i.e.\ the soft single logarithmic coefficients are non-zero.

%----------------------------------------------------------------------
\subsection{Full resummed result}\label{sec:resum-final}

Our final resummed predictions include all the effects discussed in
this section: the running of the strong coupling, collinear effects ---
flavour changes, splitting functions and the momentum of the leading
parton --- as well as soft-gluon emissions to all orders including
large-angle contributions and clustering effects for emissions at
commensurate angles:
\begin{multline}\label{eq:master}
  \rho_\text{resum}(\Delta, k_t | \pt)
  =
  \sum_{i,j=q,g} f_i(\mu_F)
  \int_0^1 dx\, p\big(x,j| i,t_\text{coll}(\Delta_0,\Delta;\mu_R)\big)
  \frac{\alpha_s(\xi_Kk_t)}{\pi}
  \\
  \left(
    \frac{\mathcal{P}_{j}(z\equiv k_t/(x \pt\Delta))}{2C_j / z}
  \right)
  \rho_{\text{soft},j}\big(\Delta, t_\text{soft}(x \pt \Delta, \xi_Zk_t;\mu_R) \big)
\end{multline}
In this expression, the factor $\alpha_s/\pi$ includes the 2-loop
running coupling discussed in section~\ref{sec:running-coupling}.
The scales $\mu_R=\xi_Rp_\perp R$, $\mu_F=\xi_Fp_\perp R$ and the
factors $\xi_K$ and $\xi_Z$ probe the scale uncertainties and are
discussed below.
The factor $p(x,j|i,t_\text{coll})$ --- computed numerically by
solving Eq.~(\ref{eq:dglap}) with an approach similar to
Ref.~\cite{Dasgupta:2014yra} --- encodes the probability for the
leading parton to have a momentum fraction $x$ and a flavour $j$,
starting from a jet of flavour $i$ (with initial fraction $f_i$)
computed in the collinear limit as in section~\ref{sec:flavour-logs}.
Similarly, the factor $z\mathcal{P}_j/(2C_j)$ accounts for the collinear
structure associated with an observed Lund-plane emission at finite
$z$ (cf.\ e.g.\ Eq.~(\ref{eq:rho-coll})).
Finally, the factor $\rho_\text{soft}$ resums the soft logarithms at
large angles as well as C/A clustering logarithms, as described in
section~\ref{sec:soft-resum}.
In practice, $\rho_\text{soft}$ depends on the full colour structure of
an event. We have computed it by interfacing Born-level events
obtained with the \texttt{NLOJet++} program to the numerical code from
section~\ref{sec:soft-resum}. Each Born-level event is separated (at
large-$N_c$) into different (weighted) dipole configurations. The result
is binned as a function of $t_\text{soft}$, $\Delta$, and the jet $\pt$
and flavour. The different dipole configurations contributing to a
given jet flavour are summed, as only the sum is needed to combine
$\rho_\text{soft}$ with the collinear effects in
writing~(\ref{eq:master}).

\begin{figure}
  \begin{minipage}[c]{0.48\textwidth}
    \centering
    \includegraphics[width=0.9\textwidth]{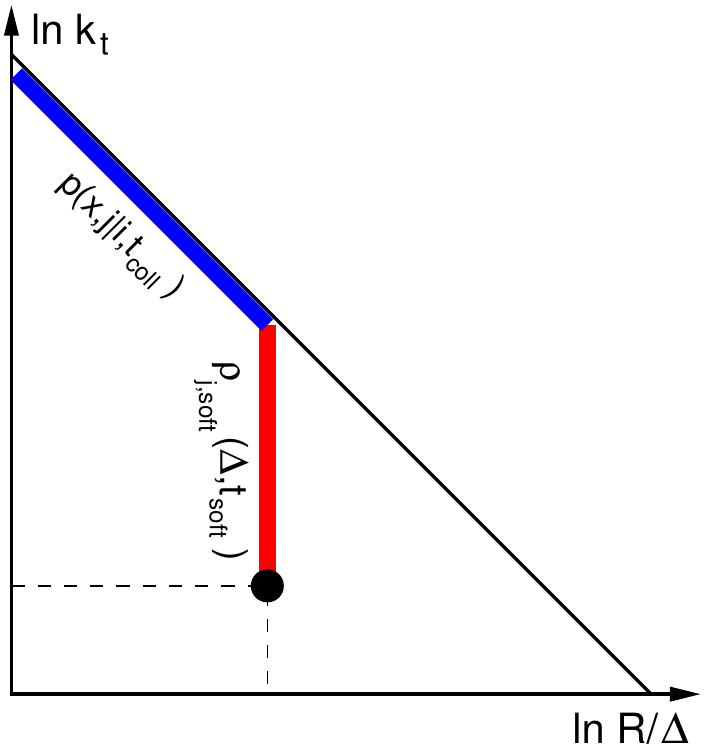}
  \end{minipage}
  \hfill
  \begin{minipage}[c]{0.45\textwidth}
    \centering
    \caption{Schematic representation of Eq.~(\ref{eq:master}) where
      the density at a point $(\Delta, k_t)$ in the Lund plane is
      obtained by first resumming collinear effects at angles larger
      than $\Delta$ and then soft gluons (at commensurate angles) down
      to a scale $k_t$.}\label{fig:master}
  \end{minipage}
\end{figure}

The structure of Eq.~(\ref{eq:master}) is illustrated in
Fig.~\ref{fig:master}, which shows that to 
obtain the density at a point $(\Delta, k_t)$ in the Lund plane (the
black dot), one first resums collinear effects down to the angle
$\Delta$ (the solid blue line) then resums the soft emissions at
commensurate angles between $x\pt\Delta$ and $k_t$ (the solid red
line).
In particular, one sees that at large angles, where the details
of the dipole configuration matter, collinear effects can be
neglected in~(\ref{eq:master}) and the sum over dipole
configurations can be performed trivially. At small angles, the
clustering logarithms resummed in $\rho_\text{soft}$ depend only on
the jet flavour.

Our results for the resummation of the soft gluons are
strictly-speaking obtained in the large-$N_c$ limit. It is however
possible to restore the full-$N_c$ behaviour up to and including
$\mathcal{O}(\alpha_s^2)$ in the collinear limit, i.e.\ in the limits
that have been discussed in sections~\ref{sec:soft-largeR}
and~\ref{sec:soft-clust}.
First, we multiply the soft density for quark jets,
$\rho_{\text{soft},q}$ by a factor $2C_F/C_A$ to guarantee the proper
result from Section~\ref{sec:soft-largeR} at order $\alpha_s$. Then,
we multiply $t_\text{soft}$ by $2(C_A-C_F)/C_A$ for quark jets, to
guarantee the proper expansion,
Eq.~(\ref{eq:clustering-alphas2-quark}), at order $\alpha_s^2$ in the
collinear limit.
At large angles the structure of subleading-$N_C$ corrections
is more complicated, and we will rely on matching with fixed-order
calculations to address these terms up to order $\as^2$.

To obtain our final predictions integrated over $\pt$, we have again
used  \texttt{NLOJet++} to obtain the jet cross-section, the quark and gluon
fractions $f_{q,g}$, the average jet $\pt$ and the average
$\alpha_s(\mu_R)$ (as well as $\rho_\text{soft}$) in a series of bins
in $\pt$. The contribution of each bin is evaluated
using~(\ref{eq:master}) at the average $\pt$ in the bin and summed
with weight proportional to the bin cross-section.

\begin{figure}
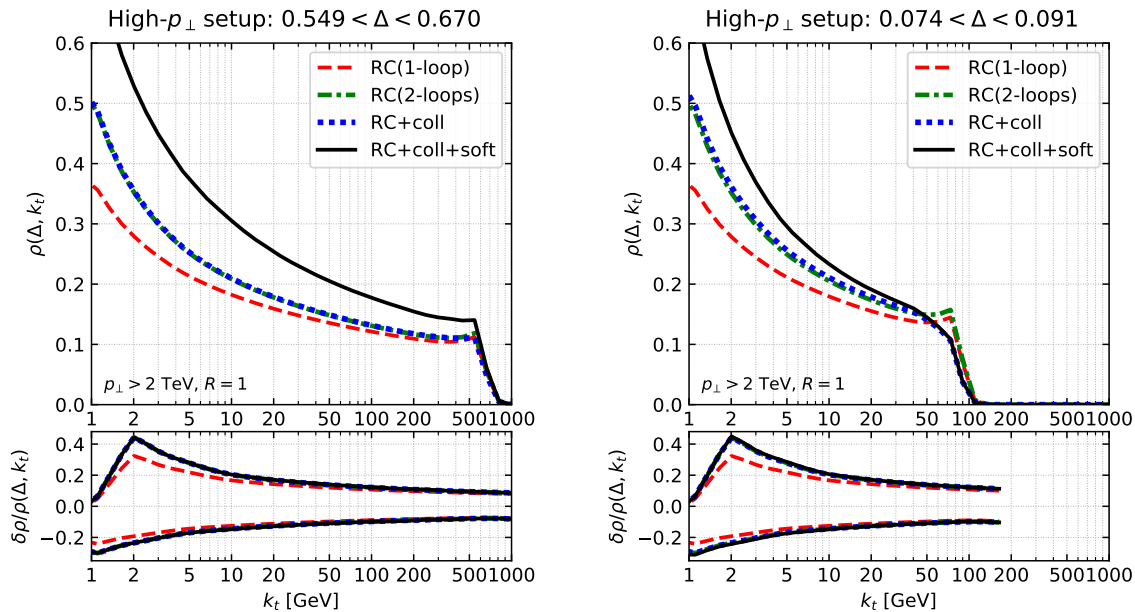

  \centering
  \begin{subfigure}[t]{0.48\textwidth}
    \includegraphics[width=\textwidth,page=2]{figs/resum.pdf}
    \caption{large angles: $0.549<\Delta<0.670$}\label{fig:resum-slice-largedelta}
  \end{subfigure}
  \hfill
  \begin{subfigure}[t]{0.48\textwidth}
    \includegraphics[width=\textwidth,page=5]{figs/resum.pdf}
    \caption{small angles: $0.074<\Delta<0.091$}\label{fig:resum-slice-smalldelta}
  \end{subfigure}
  \caption{Slices of the resummed primary Lund-plane
    density $\rho_\text{resum}(\Delta,k_t)$ at constant $\Delta$. The upper panels correspond to the
    density $\rho_\text{resum}(\Delta,k_t)$ itself while the lower panels show the
    relative scale uncertainties
    $\tfrac{\delta\rho}{\rho}(\Delta,k_t)$.
    We show results including different contributions: the dashed(red)
    line includes only 1-loop running, the (dash-dotted) green line includes
    2-loop running-coupling corrections, the (dotted) blue lines adds the
    resummation of collinear effects (flavour changes and
    leading-parton momentum), and the solid black line is our full
    resummed result, including soft-gluon resummations as
    well.}\label{fig:resum-slices-delta}
\end{figure}

\begin{figure}
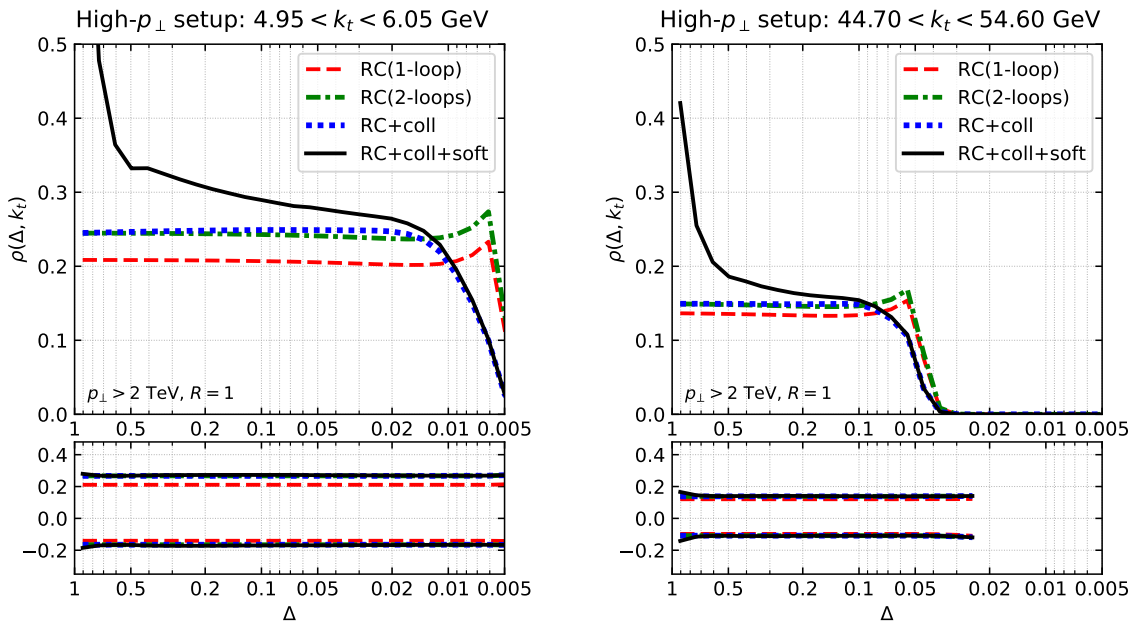

  \centering
  \begin{subfigure}[t]{0.48\textwidth}
    \includegraphics[width=\textwidth,page=10]{figs/resum.pdf}
    \caption{small $k_t$, $4.95<k_t<6.05$ GeV}\label{fig:resum-slice-smallkt}
  \end{subfigure}
  \hfill
  \begin{subfigure}[t]{0.48\textwidth}
    \includegraphics[width=\textwidth,page=12]{figs/resum.pdf}
    \caption{large $k_t$, $44.7<k_t<54.6$ GeV}\label{fig:resum-slice-largekt}
  \end{subfigure}
  \caption{Same as Fig.~\ref{fig:resum-slices-delta}, now for slices
    of constant $k_t$.}\label{fig:resum-slices-kt}
\end{figure}

Compared to section~\ref{sec:soft-resum}, the definition of
$t_\text{soft}$ from Eq.~(\ref{eq:tsoft}) has to be adjusted to ensure
$t_\text{soft}\to 0$ when $k_t \to \tfrac{1}{2}x\pt\Delta$
(i.e. $z\to \tfrac{1}{2}$). This is simply done by writing
\begin{equation}\label{eq:tsoft-scaled}
  t_\text{soft}(\xi_Zk_t, \pt\Delta;\mu)
   \equiv I_\alpha\left(\frac{\xi_Z x k_t}{x-(2-\xi_Z)\frac{k_t}{\pt\Delta}}, x \pt\Delta;\mu\right),
\end{equation}
where we have introduced a parameter $\xi_Z$
that allows us, by a standard variation of $\xi_Z$ between $1/2$ and
2, to probe the uncertainties associated with the resummation of soft
gluons.
Similarly, we estimate the renormalisation ($\mu_R=\xi_R p_\perp R$) and factorisation
($\mu_F=\xi_F p_\perp R$) scale uncertainties using the 7-point
rule~\cite{Cacciari:2003fi} around $\mu_R=\mu_F=p_\perp R$ ($\xi_R=\xi_F=1$).
The factorisation scale only influences the Born-level spectrum and
the quark/gluon fractions $f_{q,g}$.
The choice of $\mu_R$ should also be reflected in the factor
$\alpha_s/\pi$ in~(\ref{eq:master}) as well as in the definition of
$t_\text{coll}$ and $t_\text{soft}$, via the reference scale
$\mu_R=\xi_Rp_\perp R$ for
$\alpha_s$ in~(\ref{eq:running-alphas}).
Additionally, the uncertainty of the choice of scale for the argument
of $\alpha_s$ in~(\ref{eq:rho-softcoll+rc}) is taken into account by
setting the scale to $\xi_K k_t$ and varying $\xi_K$ between $1/2$ and
$2$.
This is the dominant source of uncertainty in our calculation.
The uncertainty on the collinear resummation could be estimated by
varying $\Delta_0$ in~(\ref{eq:master}). However, since the effect of
the collinear resummation is small (see e.g.\
Figs.~\ref{fig:resum-slices-delta} and~\ref{fig:resum-slices-kt}), we
have neglected this and set $\Delta_0=R$.\footnote{Varying $\Delta_0$
  would come with the additional complication that, for $\Delta_0>R$,
  collinear radiation at angles larger than the jet radius would cause
  the Born-level $\pt$ and the jet $\pt$ to differ.
  Since, in our case, $\ln(R)$ is not large, we can neglect this effect.}
To be conservative, the final perturbative uncertainty is obtained by
summing in quadrature the three individual sources of uncertainties:
the 7-point variation of $\mu_R$ (or $\xi_R$) and $\mu_F$ (or $\xi_F$), the variation of $\xi_K$
and the variation of $\xi_Z$.\footnote{Recall that $t_\text{coll}$,
  Eq.~(\ref{eq:tcoll}) and $t_\text{soft}$,
  Eq.~(\ref{eq:tsoft-scaled}) are written terms of $I_\alpha$,
  Eq.~(\ref{eq:Ialpha}) and that they all have a structure
  $\as^n L^n$, where each factor of $L$ can be one of $\ln \Delta$ or
  $\ln p_t R/k_t$.
  To probe uncertainties, we should examine variations that generate
  terms $\as^n L^{n-1}$.
  The variation of $\mu$ in Eq.~(\ref{eq:Ialpha}) does not generate
  such terms, but only terms $\as^n L^{n-2}$.
  One approach to generating terms $\as^n L^{n-1}$ is to change
  the argument of $\as$ within the integral in Eq.~(\ref{eq:Ialpha}),
  i.e.\ replacing $\as(q_t)$ with $\as(\xi q_t)$, where $\xi$
  is the scale variation factor.
  This is equivalent to replacing
  $I_\alpha(k_t,k_{t0};\mu) \to I_\alpha(\xi k_t, \xi
  k_{t0};\mu)$, i.e.\ changing both integration boundaries.
  A second approach is to change just one boundary by a factor
  $\xi$, which can be thought of as a replacement
  $L \to L \pm \ln \xi$.
  The prescription that we have adopted for $t_\text{soft}$
  corresponds to the second approach, specifically varying the lower
  boundary (which has a larger numerical impact than varying the upper
  boundary).
  Ultimately, the choice we make here is not especially critical,
  because the overall perturbative uncertainty is dominated by the
  $\xi_K$ variations.}

We present some representative results obtained with
Eq.~(\ref{eq:master}) in Figs.~\ref{fig:resum-slices-delta} (for
slices of the Lund plane in a narrow bin of $\Delta$)
and~\ref{fig:resum-slices-kt} (for slices in a narrow bin of $k_t$).
In each plot, we show results obtained using
Eq.~(\ref{eq:rho-softcoll+rc}), i.e.\ just running-coupling effects, with 1-loop
(red) and 2-loop (green) running. The blue curves then add
collinear effects (i.e.\ Eq.~(\ref{eq:rho-coll})) and the black curves
add soft-gluon emissions corresponding to our full resummed results
from Eq.~(\ref{eq:master}).
The bottom panels of each plot show the corresponding scale
uncertainties.
These plots show that 2-loop running-coupling corrections are
numerically similar
in size to the resummation of the soft logarithms.
Those soft-gluon effects are most significant
at small $k_t$ and at large angle.
It is worth noting though that their effect is also visible at large
$k_t$ in Fig.~\ref{fig:resum-slice-largedelta}.
This is due to the power
corrections in $\Delta^2$ starting at order $\mathcal{\alpha}_s$,
as discussed in section~\ref{sec:soft-largeR}.

Collinear effects are small except close to the
$k_t=\tfrac{1}{2}\pt\Delta$ endpoint where the
use of the full splitting function and the probability distribution
for the momentum fraction of the leading parton have a clearly visible
effect (see Fig.~\ref{fig:resum-slices-kt} in particular).
Flavour-changing collinear effects are small but are still visible in
Fig.~\ref{fig:resum-slice-smallkt}, reflected in the difference between the
green and blue lines for $\Delta\gtrsim 0.02$.
In particular, as one goes to smaller values of $\Delta$, there is an
increase in the fraction of jets whose leading parton is a gluon.
This flavour-changing effect is modest in size, in part because the
initial Born-level spectrum has a quark fraction of about $77.5\%$,
relatively close to the asymptotic fraction of $62\%$ that is visible
in Fig.~\ref{fig:qg-frac} (cf.\ Eq.~(\ref{eq:qg-asymptotic})).

The perturbative scale uncertainties are about 10\% at large $k_t$,
slowly growing to $\sim 15$\% at $k_t\sim 20$~GeV and to $\sim 30$\%
at $k_t=2$~GeV (averaging the upper and lower uncertainties). They are
dominated by the scale variation, $\xi_K$, in the argument of
$\alpha_s$ with an additional small contribution from the variation of
$\xi_Z$ at small $k_t$.\footnote{The kink in the upper uncertainty
  bands between 1 and 2~GeV comes from our freezing of the running
  coupling at 1~GeV.}

While all the above expressions are given for the primary
Lund-plane density $\rho(\Delta, k_t)$, they can almost
straightforwardly be adapted to $\tilde\rho(\Delta, z)$, as measured
e.g.\ by the ATLAS collaboration.
Specifically, Eq.~(\ref{eq:master})
becomes
\begin{multline}\label{eq:master-atlas}
  \tilde\rho_\text{resum}(\Delta, z | \pt)
  =
  \sum_{i,j=q,g} f_i
  \int_0^1 dx\, p\big(x,j| i,t_\text{coll}(\Delta_0,\Delta;\pt R)\big)
  \frac{\alpha_s(xz\pt\Delta)}{\pi}
  \\
  \left(
    \frac{\mathcal{P}_{j}(z)}{2C_j / z}
  \right)
  \rho_{\text{soft},j}\big(\Delta, t_\text{soft}(x \pt \Delta, xz\pt\Delta;\pt R) \big).
\end{multline}
We just note that, while keeping $k_t$ large enough in
Eq.~(\ref{eq:master}) guarantees that we stay in the perturbative
region, the integration over $x$ in~(\ref{eq:master-atlas})
potentially extends to arbitrarily small $xz\pt\Delta$ momentum
scales. This is regulated by our freezing of the running coupling at
1~GeV.
In practice, this only affects the small values of $z$ in a region
where the non-perturbative corrections dominate anyway.

In anticipation of the matching of our resummed predictions to exact
fixed-order results for $\rho(\Delta,k_t)$, we note that our all-order
equations~(\ref{eq:master}) and (\ref{eq:master-atlas}) can be expanded to fixed-order.
For Eq.~(\ref{eq:master}), at NLO we have 
\begin{subequations}\label{eq:master-expansion}
\begin{align}
  \alpha_s(k_t) & = \alpha_s + 2\alpha_s^2
                  \beta_0\,\ln\frac{\pt R}{k_t},\\
  p(x, j | i,t_\text{coll}) & = \delta_{ij}\delta(1-x)
                              + \frac{\alpha_s}{\pi}\ln\frac{\Delta_0}{\Delta}
                              \int dz
                              \left[\mathcal{P}^{(R)}_{ji}(z)\delta(z-x)-\mathcal{P}^{(V)}_{ji}(z)\delta(1-x)\right],\\
  \rho_{\text{soft},j}\big(\Delta, t_\text{soft}\big)  
                & = \alpha_s \rho_{\text{soft},j}^{(1)}(\Delta,k_t) + \alpha_s^2\rho_{\text{soft},j}^{(2)}(\Delta,k_t)
                 = \alpha_s \rho_{\text{soft},j}^{(1)}(\Delta) + \frac{\alpha_s^2}{\pi}\ln\frac{\pt\Delta}{k_t}\rho_{\text{soft},j}^{(2\text{-em})}(\Delta).
\end{align}
\end{subequations}
with $\alpha_s\equiv\alpha_s(\pt R)$. We have explicitly
written
$\alpha_s^2\rho_\text{soft}^{(2)}(\Delta,k_t)=\tfrac{\alpha_s^2}{\pi}\ln\tfrac{p_\perp\Delta}{k_t}\rho_\text{soft}^{(2\text{-em})}(\Delta)$,
i.e.\ as a logarithm times a
factor depending only on $\Delta$. The coefficients of the
$\rho_\text{soft}$ expansion can be obtained, as for
$\rho_\text{soft}$ itself, using the numerical approach from
section~\ref{sec:soft-resum} with a Born-level spectrum from
\texttt{NLOJet++}.
Inserting the elements of Eq.~(\ref{eq:master-expansion}) into
(\ref{eq:master}), we get a trivial LO contribution involving
$\rho_{\text{soft},j}^{(1)}(\Delta)$. The NLO, i.e.
$\mathcal{O}(\alpha_s^2)$, result receives 3 contributions, one from
each of the lines of Eq.~(\ref{eq:master-expansion}).
A similar fixed-order expansion can be obtained for $\tilde\rho(\Delta,z)$.

%======================================================================
\section{Matching with fixed-order}\label{sec:fixed-order}

In order to get a full coverage of the primary Lund-plane density,
including regions which are not dominated by large logarithms, it is
useful to supplement our resummation with as many orders
of the $\alpha_s$ series expansion of $\rho(\Delta, k_t)$ as are known
exactly. 

In this paper we focus on dijet events, for which we can obtain the
primary Lund-plane density using 
the {\tt NLOJet++} program,
\begin{equation}\label{eq:fixed-order-expansion}
  \rho_\text{fixed-order}(\Delta, k_t)
  = \alpha_s(\pt R) \,\rho^{(1)}(\Delta, k_t)
  + \alpha_s^2(\pt R) \, \rho^{(2)}(\Delta, k_t)
  + {\cal {O}}(\alpha_s^3).
\end{equation}
The first (LO) and second (NLO) contributions are accessible using
respectively LO and NLO 3-jet calculations~\cite{Nagy:2003tz}.

Compared to the all-order calculation discussed in
section~\ref{sec:all-orders}, the LO contribution includes the
first-order soft gluon radiation at large angles.
%^
The NLO contribution includes the first non-trivial running-coupling,
flavour-changing and clustering corrections.\footnote{In these
  fixed-order calculations the central renormalisation and
  factorisation scales have been set to $\pt R$ with $\pt$ the jet
  transverse momentum.}
We have checked numerically that there was an agreement between {\tt
  NLOJet++} and our analytic calculations for the soft-and-collinear
behaviour at $\mathcal{O}(\alpha_s)$ and for the logarithmic
dependence at $\mathcal{O}(\alpha_s^2)$, although small deviations
expected from our large-$N_c$ approximation --- used to calculate
dipole decompositions and soft logarithms beyond the collinear limit
--- are observed at large angles.
We show some explicit examples in Appendix~\ref{sec:NLO-slopes}.

Knowing both the all-order resummation and the exact fixed-order
results, we obtain a matched prediction using
\begin{equation}\label{eq:matching}
\rho(\Delta, k_t) = \frac{\rho_\text{resum}(\Delta, k_t)\,\rho_\text{NLO}(\Delta, k_t)}{\rho_\text{resum,NLO}(\Delta, k_t)},
\end{equation}
where $\rho_\text{resum,NLO}$ is the expansion to ${\cal {O}}(\alpha_s^2)$
of the resummed result~(\ref{eq:master}).
This expression is such that it reproduces the resummed calculation in
the region where large logarithms are present, and the exact NLO
result when expanded to second order in $\alpha_s$.

\begin{figure}
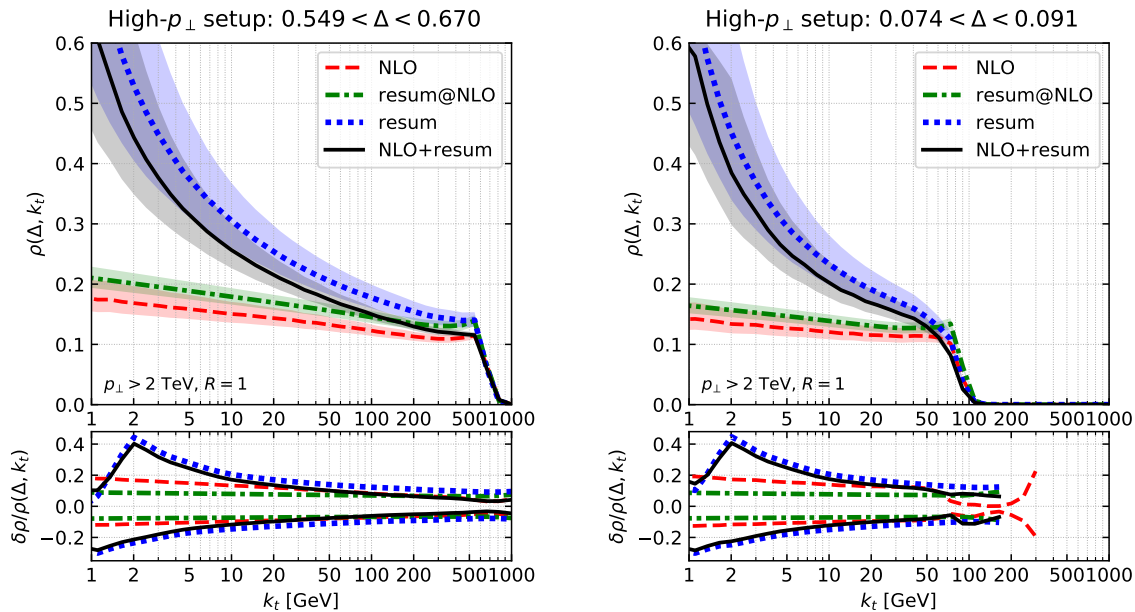

  \centering
  \begin{subfigure}[t]{0.48\textwidth}
    \includegraphics[width=\textwidth,page=2]{figs/matching.pdf}
    \caption{large angles: $0.549<\Delta<0.670$}\label{fig:matched-slice-largedelta}
  \end{subfigure}
  \hfill
  \begin{subfigure}[t]{0.48\textwidth}
    \includegraphics[width=\textwidth,page=5]{figs/matching.pdf}
    \caption{small angles: $0.074<\Delta<0.091$}\label{fig:matched-slice-smalldelta}
  \end{subfigure}
  \caption{Slices of the primary Lund-plane
    density $\rho(\Delta,k_t)$ at constant $\Delta$. The upper panels correspond to the
    density $\rho(\Delta,k_t)$ itself, while the lower panels show the
    relative scale uncertainties
    $\tfrac{\delta\rho}{\rho}(\Delta,k_t)$.
    We show results for the exact NLO calculation (red), the NLO
    expansion of the resummed results (green), the full resummed
    result (blue) and the NLO+resummation matched result
    (black).}\label{fig:matched-slices-delta}
\end{figure}

Explicit examples of matched predictions, including different levels
of approximations for the resummation, are presented in
Fig.~\ref{fig:matched-slices-delta} for the $k_t$ dependence at fixed
$\Delta$.
First, we see that the exact NLO results (red) are close to what is
obtained using the expansion of our resummed calculation (green).
Next, the resummed result (blue) shows a strong enhancement at small
$k_t$, primarily due to the running coupling, and to soft-gluon
clustering effects.
Finally, the matched result (black) smoothly interpolates between the
fixed-order result at large angle and large $k_t$ and the resummed
result at smaller angle or $k_t$.

The bands in the upper panel of Fig.~\ref{fig:matched-slices-delta} as
well as the curves in the lower panel show our theoretical
uncertainties.
One of the striking features is that the matching with NLO reduces the
uncertainties compared to the resummed result. This is valid across
the whole kinematic range and especially visible at larger $k_t$.
Final uncertainties after matching are $\sim 6\%$ at $k_t=200$~GeV,
increasing to $\sim 12\%$ at 20~GeV and $\sim 30\%$ at 2~GeV.

%======================================================================
\section{Non-perturbative effects}\label{sec:np-effects}

\begin{figure}[ht]
  \centering
  \begin{subfigure}[t]{0.48\textwidth}
    \includegraphics[width=\textwidth,page=1]{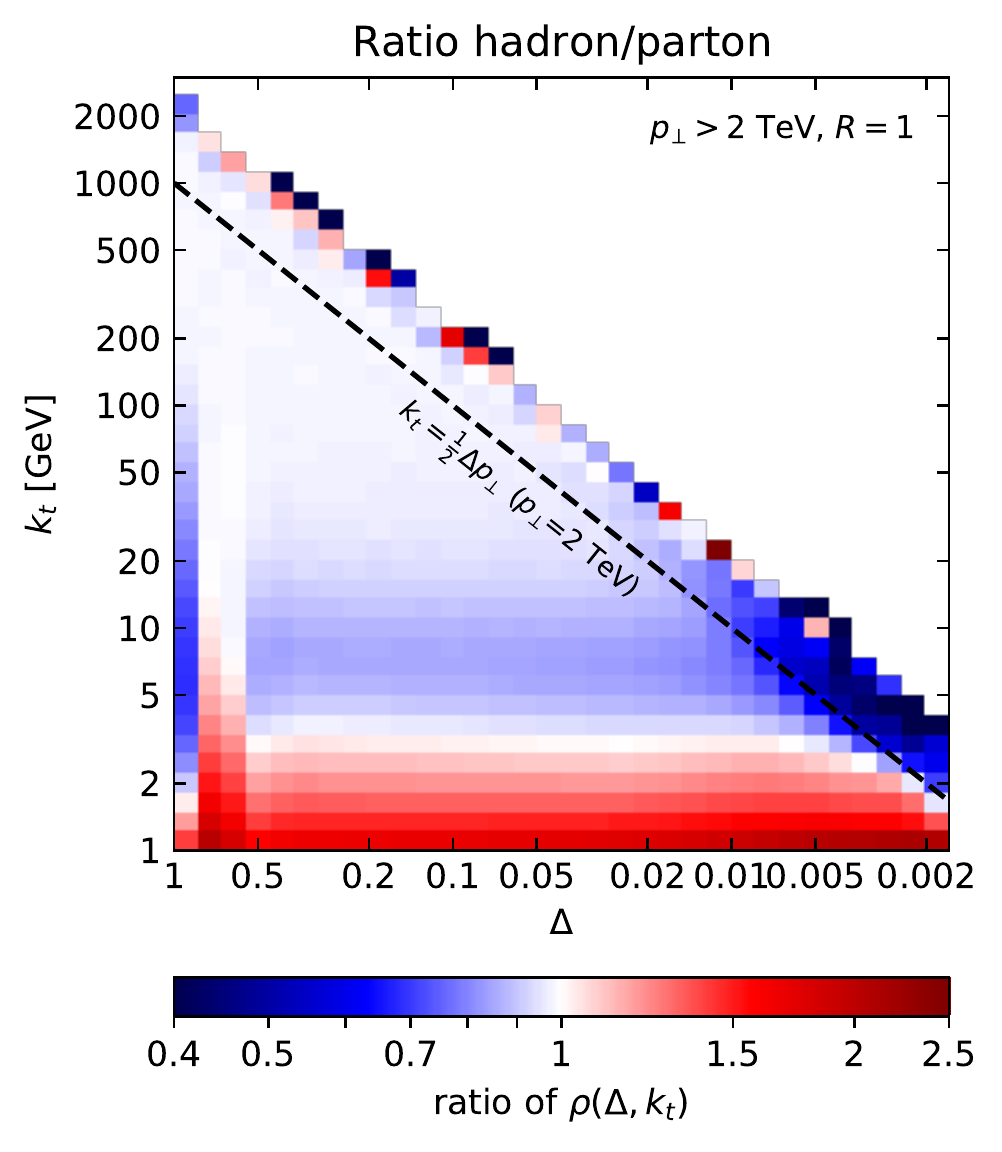}
    \caption{Hadronisation corrections}\label{fig:np-hadronisation}
  \end{subfigure}
  \hfill
  \begin{subfigure}[t]{0.48\textwidth}
    \includegraphics[width=\textwidth,page=2]{figs/np-effects-highpt.pdf}
    \caption{Underlying Event/MPI corrections}\label{fig:np-ue}
  \end{subfigure}\\
  \begin{subfigure}[t]{0.48\textwidth}
    \includegraphics[width=\textwidth,page=1]{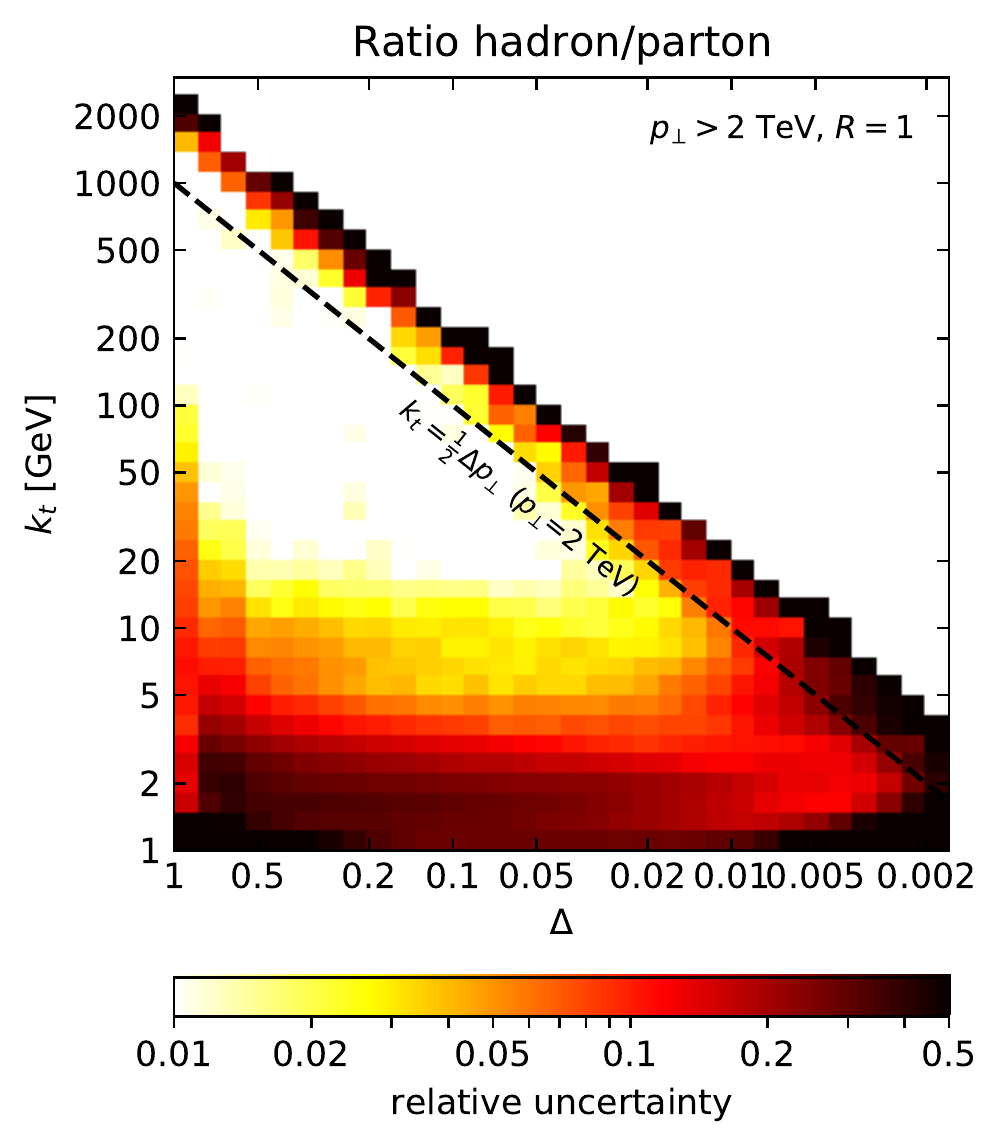}
    \caption{Hadronisation uncertainties}\label{fig:np-hadronisation-uncertainties}
  \end{subfigure}
  \hfill
  \begin{subfigure}[t]{0.48\textwidth}
    \includegraphics[width=\textwidth,page=2]{figs/np-uncertainties-highpt.pdf}
    \caption{Underlying Event/MPI uncertainties}\label{np-ue-uncertainties}
  \end{subfigure}
  \caption{Non-perturbative corrections (top) and uncertainties (bottom).
    The diagonal dashed line corresponds to the kinematic limit,
    $k_t=\tfrac{1}{2}p_\perp\Delta$, for jets with $p_\perp=2$~TeV.}\label{fig:np-corrections}
\end{figure}

Before making our final predictions it is interesting to estimate
non-perturbative corrections to the calculation we have provided so
far.
We do so using a Monte-Carlo approach.
We have studied the primary Lund-plane density using 5 different
Monte-Carlo generators/tunes: Pythia8 (v8.230)~\cite{Sjostrand:2014zea} with
the Monash~2013 tune~\cite{Skands:2014pea},
tune~4C~\cite{Corke:2010yf} and the ATLAS~2014
tune~\cite{ATL-PHYS-PUB-2014-021} (the variant with NNPDF~2.3
PDFs~\cite{Carrazza:2013axa}),
Herwig7.2.0~\cite{Corcella:2002jc,Bellm:2015jjp,Bellm:2019zci} and
Sherpa~2.2.8~\cite{Gleisberg:2008ta}.
For each generator/tune we first study the primary Lund-plane density at
parton level.
We can then switch to hadron level to study the effect of
hadronisation, include multi-parton interactions (MPI) to study the
effects of the Underlying Event, and examine the impact of using only
charged tracks as done in the ATLAS measurement~\cite{ATLAS:2019sol}.

For the central value of the non-perturbative corrections, we take the
average of the Monte-Carlo generators, excluding Herwig7.
The reason behind this exclusion is that our perturbative results are
in the same ballpark as parton-level results from Pythia8 and Sherpa
but differ significantly from parton-level Herwig7 results (see
Appendix~\ref{sec:v-mc}).
We obtain the (upper and lower) uncertainties on the non-perturbative
corrections from the envelope of the Lund-plane density ratios for the
5 Monte-Carlo generators/tunes.
To remain conservative, we keep the Herwig7 results in our
non-perturbative uncertainty estimates.

Our results are presented in Fig.~\ref{fig:np-corrections}, for our
high-$\pt$ setup separately for hadronisation and Underlying Event
corrections.
It is clearly visible that hadronisation corrections become sizeable
at low $k_t$, with a negative effect above $\sim 3$~GeV and a positive
effect below. Their effect is almost invisible for
$k_t\gtrsim 10{-}20$~GeV.
Underlying-Event corrections are instead important (and positive) at
low-to-moderate $k_t$ and large angles.
The non-perturbative uncertainties --- shown in
Fig.~\ref{fig:np-hadronisation-uncertainties}
and~\ref{np-ue-uncertainties} for hadronisation and the Underlying
Event, respectively --- are small, $\mathcal{O}(1{-}2)\%$, whenever the
overall corrections are themselves small.
At large $\Delta$, the non-perturbative corrections appear to have
additional structure and enhanced uncertainties.
This structure can be attributed to the interplay between the initial
anti-$k_t$ clustering and the C/A re-clustering as already discussed
in~\cite{Dreyer:2018nbf} and related boundary logarithms discussed in
section~\ref{sec:soft-clust} and Appendix~\ref{sec:boundary-logs}.

The diagonal dashed line in Fig.~\ref{fig:np-corrections} corresponds
to $k_t = \frac{1}{2}\pt\Delta$ for $\pt=2$~TeV. This is the kinematic
limit for the lowest-energy selected jets.
The Lund plane density quickly decreases above that line.
The large fluctuations and uncertainties observed in
Fig.~\ref{fig:np-corrections} around the dotted line are a trace of
the statistical fluctuations in our Monte Carlo samples.

%======================================================================
\section{Final predictions}\label{sec:results}

\begin{figure}
  \centering
  \begin{subfigure}[t]{0.32\textwidth}
    \includegraphics[width=\textwidth,page=7]{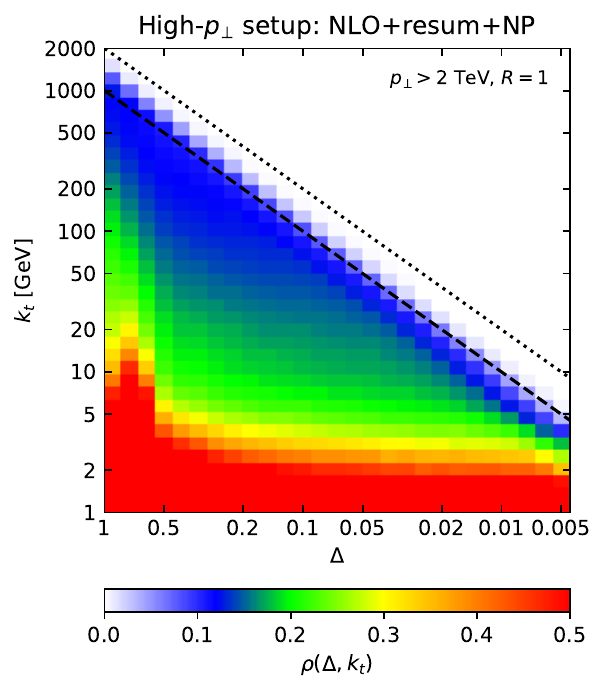}
    \caption{Lund-plane density
      $\rho(\Delta,k_t)$}\label{fig:highpt-full}
  \end{subfigure}
  \hfill
  \begin{subfigure}[t]{0.32\textwidth}
    \includegraphics[width=\textwidth,page=6]{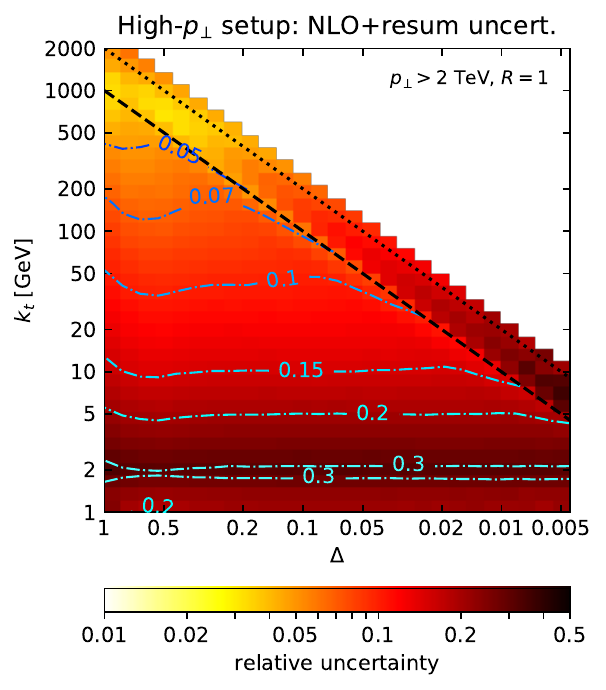}
    \caption{Perturbative uncertainty
    }\label{fig:highpt-pert-unc}
  \end{subfigure}
  \hfill
  \begin{subfigure}[t]{0.32\textwidth}
    \includegraphics[width=\textwidth,page=8]{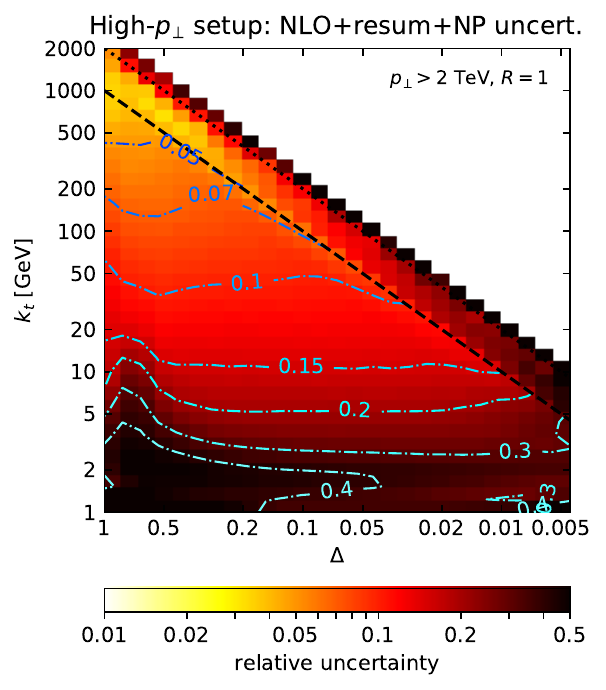}
    \caption{Full uncertainty
    }\label{fig:highpt-np-unc}
  \end{subfigure}
  \caption{Predictions for the primary Lund plane density for the
    high-$\pt$ setup (a) and associated perturbative (b) and full (c) uncertainties.
    Full uncertainties sum the perturbative  and
     non-perturbative contributions in quadrature. 
  }\label{fig:highpt}
\end{figure}

Our final predictions include both the matched perturbative
predictions, discussed in section~\ref{sec:fixed-order}, multiplied by
the non-perturbative corrections obtained in
section~\ref{sec:np-effects}.

\begin{figure}
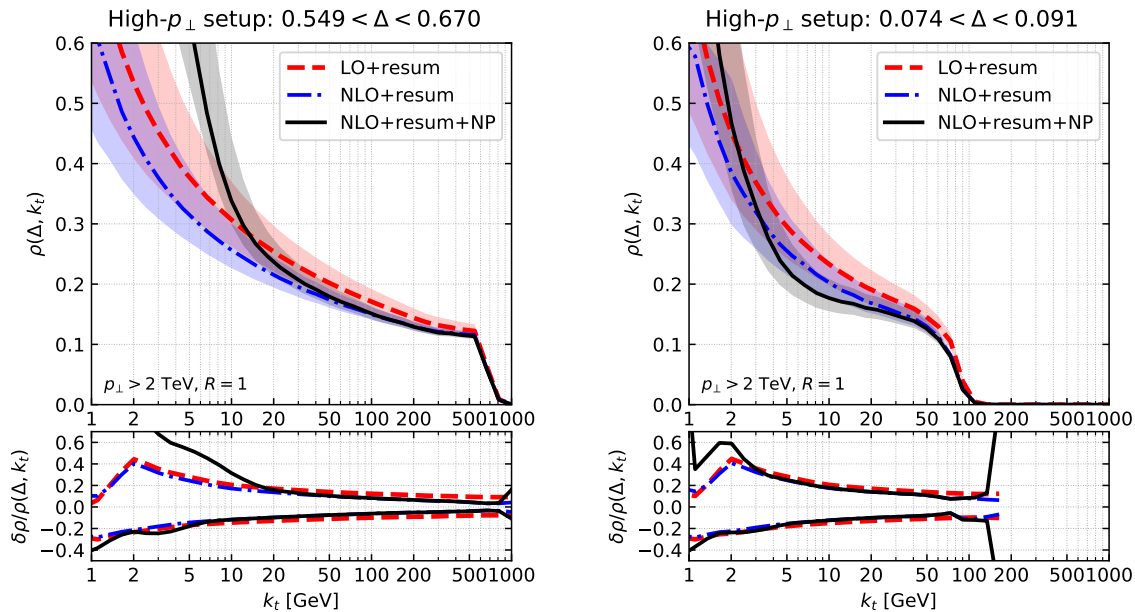

  \centering
  \begin{subfigure}[t]{0.48\textwidth}
    \includegraphics[width=\textwidth,page=2]{figs/final.pdf}
    \caption{large angles: $0.549<\Delta<0.670$}\label{fig:final-slice-largedelta}
  \end{subfigure}
  \hfill
  \begin{subfigure}[t]{0.48\textwidth}
    \includegraphics[width=\textwidth,page=5]{figs/final.pdf}
    \caption{small angles: $0.074<\Delta<0.091$}\label{fig:final-slice-smalldelta}
  \end{subfigure}
  \caption{Slices of the primary Lund-plane
    density $\rho(\Delta,k_t)$ at constant $\Delta$. The upper panels correspond to the
    density $\rho(\Delta,k_t)$ itself while the lower panels show the
    relative scale uncertainties
    $\tfrac{\delta\rho}{\rho}(\Delta,k_t)$.
    We show results for the matched result at LO (red) and NLO (blue),
    as well as NLO results including non-perturbative corrections
    (black).}\label{fig:final-slices-delta}
\end{figure}

We show in Fig.~\ref{fig:highpt-full} the resulting two-dimensional average
primary
Lund-plane density $\rho(\Delta, k_t)$,  and in
Figs.~\ref{fig:highpt-pert-unc} and~\ref{fig:highpt-np-unc} the
associated relative uncertainty at perturbative level and at the
non-perturbative level respectively.
Fig.~\ref{fig:final-slices-delta} shows slices at fixed angle
$\Delta$, which help to better visualise certain features.
The density plot, Fig.~\ref{fig:highpt-full}, shows all the expected
features: the gradual increase towards small $k_t$ due to the running
of $\alpha_s$; the extra enhancement due to soft-gluon emissions, both
at large angles and at small $k_t/\Delta$ (or equivalently $z$); the
reduction close to the kinematic limit associated with the ``energy
loss'' of the leading branch; and the increase at low $k_t$ and in the
bottom-left corner of the Lund plane due to non-perturbative effects.

The uncertainties are dominated by the perturbative component for
$k_t\gtrsim 3{-}5$~GeV, except at large angles where non-perturbative effects can have a
sizeable impact up to $k_t\sim 10{-}20$~GeV.
The total uncertainty is found to be about $20\%$ at $k_t \sim 5$~GeV
(away from the large-angle region), and decreases to $5{-}7\%$ for $k_t$
in the $200{-}500$~GeV range.
Relative to the LO+resum results, visible in Fig.~\ref{fig:final-slices-delta}, the inclusion of NLO corrections
reduces the uncertainties mainly at high $k_t$.
Even if the non-perturbative corrections have a negligible impact on
the uncertainty above $\sim 3{-}5$~GeV ($10{-}20$~GeV) at small (large)
$\Delta$, they result in a (small-but-visible) shift of the
central value up to larger values of $k_t$.

\begin{figure}
  \centering
  \includegraphics[width=0.33\textwidth,page=1]{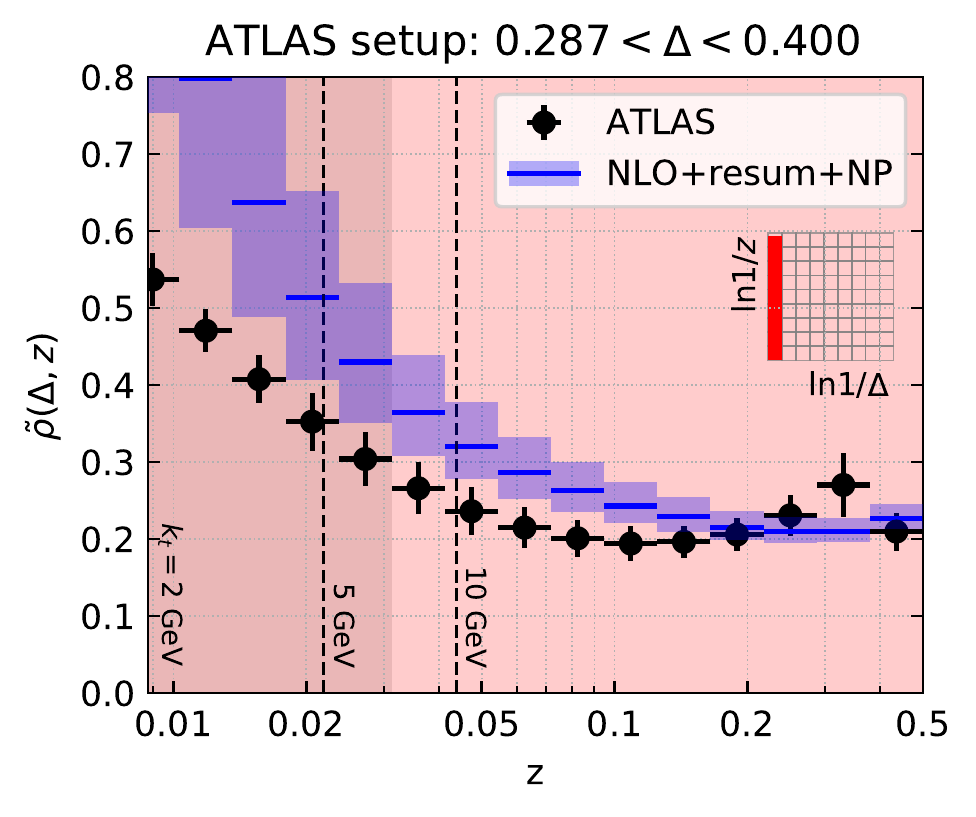}%
  \includegraphics[width=0.33\textwidth,page=2]{figs/v-atlas.pdf}%
  \includegraphics[width=0.33\textwidth,page=3]{figs/v-atlas.pdf}%
  \\%
  \includegraphics[width=0.33\textwidth,page=4]{figs/v-atlas.pdf}%
  \includegraphics[width=0.33\textwidth,page=5]{figs/v-atlas.pdf}%
  \includegraphics[width=0.33\textwidth,page=6]{figs/v-atlas.pdf}%
  \\%
  \includegraphics[width=0.33\textwidth,page=7]{figs/v-atlas.pdf}%
  \includegraphics[width=0.33\textwidth,page=8]{figs/v-atlas.pdf}%
  \includegraphics[width=0.33\textwidth,page=9]{figs/v-atlas.pdf}%
  \caption{Comparison between our calculations and the ATLAS measurement
    from Ref.~\cite{ATLAS:2019sol}, for different bins of $\Delta$.
    The dashed vertical lines, corresponding to $z=\tfrac{k_t}{\pt\Delta}$ for
    $\pt=675$~GeV and several $k_t$ values, are meant to indicate the
    transverse scales one is typically sensitive to.
    The shaded grey bands indicate bins where the relative uncertainty on the
    non-perturbative corrections is larger than 10\%.
    The shaded red regions indicate that our calculation is incomplete
    because of the missing resummation of the boundary logarithms.
  }\label{fig:atlas}
\end{figure}

\begin{figure}
  \centering
  \includegraphics[width=0.33\textwidth,page=1]{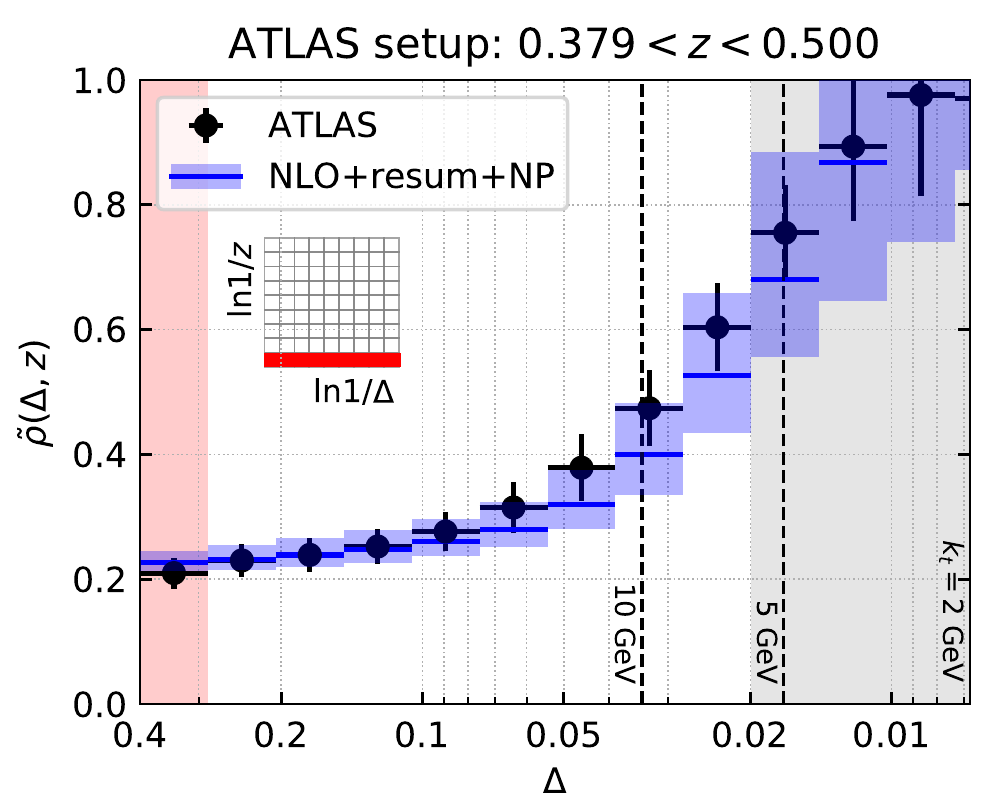}%
  \includegraphics[width=0.33\textwidth,page=2]{figs/v-atlas-zslice.pdf}%
  \includegraphics[width=0.33\textwidth,page=3]{figs/v-atlas-zslice.pdf}%
  \\%
  \includegraphics[width=0.33\textwidth,page=4]{figs/v-atlas-zslice.pdf}%
  \includegraphics[width=0.33\textwidth,page=5]{figs/v-atlas-zslice.pdf}%
  \includegraphics[width=0.33\textwidth,page=6]{figs/v-atlas-zslice.pdf}%
  \\%
  \includegraphics[width=0.33\textwidth,page=7]{figs/v-atlas-zslice.pdf}%
  \includegraphics[width=0.33\textwidth,page=8]{figs/v-atlas-zslice.pdf}%
  \includegraphics[width=0.33\textwidth,page=9]{figs/v-atlas-zslice.pdf}%
  \caption{Same as Fig.~\ref{fig:atlas} this time for slices at
    constant $\ln(1/z)$.}\label{fig:atlas-zslice}
\end{figure}

Finally, we discuss our analytic calculations supplemented with
non-perturbative corrections for $\tilde\rho(\Delta,z)$, corresponding
to the ATLAS setup.
Besides the differences discussed in section~\ref{sec:primary-lund}, we
follow the same strategy as for the high-$\pt$ setup: the resummed
prediction is obtained using Eq.~(\ref{eq:master-atlas}), matched to
{\tt {NLOJet++}} fixed-order results using Eq.~(\ref{eq:matching}) and
supplemented with non-perturbative corrections --- this time correcting
so as to correspond to a measurement performed using charged-tracks above 500~MeV --- following
the procedure outlined in section~\ref{sec:np-effects}.
Details of the non-perturbative corrections are given in Appendix
\ref{sec:atlas-np}.

We compare our results to the ATLAS data from
Ref.~\cite{ATLAS:2019sol} for slices in $\Delta$ in
Fig.~\ref{fig:atlas} and slices in $z$ in Fig.~\ref{fig:atlas-zslice}.
The vertical dashed lines correspond to the $k_t$ scales
estimated using $z=k_t/(\pt\Delta)$, i.e.\ assuming a jet at
the lower $\pt$ cut of 675~GeV and a leading parton/subjet carrying a
fraction $x=1$ of the initial jet transverse momentum.
The shaded grey bands indicate regions where the uncertainty on the
non-perturbative corrections is larger than 10\%.
Shaded red bands correspond to the regions sensitive to the boundary
logarithms discussed in section \ref{sec:soft-clust}.
We recall that we have not resummed these terms, so our calculation
should be considered incomplete in the red shaded regions.
A rough estimate of their potential size is given in
Appendix~\ref{sec:boundary-logs}.

For all unshaded bins in Figs.~\ref{fig:atlas} and
\ref{fig:atlas-zslice}, we see agreement between our predictions and
the data to within the experimental and theoretical uncertainties.
Generally speaking, the theoretical uncertainties are larger than the
experimental ones, though they are comparable at values of $z$ and
$\Delta$ that correspond to large $k_t$ values.
Recall that the theoretical uncertainties are to a large extent
dominated by the choice of scale of $\as$ in the resummation and a
higher-order resummation would therefore be beneficial to reduce the
uncertainties.

If we consider the grey shaded regions, i.e.\ those where
non-perturbative uncertainties are larger than $10\%$, the agreement
between data and theory remains good to within the total uncertainties
in most of the bins, almost all the way down to $2\GeV$.
In practice this agreement is facilitated by the non-perturbative
blow-up of the uncertainties at low $k_t$ and our predictions' central
values are systematically above the data points.
Recall, however, that our estimates of non-perturbative corrections
rely on the assumption that the parton-level event-generator results
are structurally similar to a full perturbative calculation.
This assumption is questionable at low $k_t$: for example, a parton
shower may contain a low-$k_t$ cut, with the phase-space below that
$k_t$ value being filled up by hadronisation (there is a hint of this
occurring in Appendix~\ref{sec:v-mc}, Fig.~\ref{fig:v-mc});
in contrast our perturbative calculation has no such cut, and so the
hadronisation contribution to that region, supplemented with our
perturbative contribution, could effectively lead to double counting
and so an overestimate relative to the data.
In this respect it might be interesting to develop a more analytic
understanding of expected hadronisation effects on the Lund plane
density.

One region where there is clear disagreement between our predictions
and the data is in the (red-shaded) largest angle bin
$0.287 < \Delta < 0.400$.
This disagreement is only mildly alleviated by our estimate of the
potential size of boundary logarithms, cf.\
Fig.~\ref{fig:boundary-improved} in Appendix~\ref{sec:boundary-logs}.
Several avenues could be of interest for further exploring this
region, for example a full resummation of the boundary logarithms, or
a measurement with jets whose original clustering was with the C/A
algorithm (rather than anti-$k_t$), so as to remove these boundary
logarithms altogether.
Note also that this region is potentially sensitive to
underlying-event effects, and if they are incompletely modelled in
event generators, this could also contribute to the disagreement.%
\footnote{Further contributions can come from subleading-$N_c$
  corrections, both from the colour-flow decomposition of the hard
  matrix elements and from the resummation of soft logarithms.
  Our expectation is that the former should be modest given the use of
  $R=0.4$, and that the latter would not be confined to $\Delta \sim
  R$.}

%======================================================================
\section{Conclusions and outlook}\label{sec:conclusions}

In this paper we have carried out the first calculation of all-order
logarithmic contributions to the average primary Lund plane density.
We have resummed three classes of single logarithmic terms: (i)
running-coupling effects, which are relatively straightforward and the
numerically dominant contribution over most of the Lund plane;
(ii) soft effects, which involve large-angle contributions and
clustering logarithms, both evaluated in the large-$N_C$ limit;
and (iii) collinear effects at large momentum fractions, which include
contributions that can change both the momentum and the flavour of the
leading parton.
We have also discovered a new class of logarithmic effects in jets
that arise when reclustering an anti-$k_t$ jet's constituents with the
Cambridge/Aachen algorithm.
The corresponding terms are relevant close to the large-angle boundary
of the Lund plane.
We defer their full single logarithmic resummation to future work.

For the purposes of making phenomenological predictions, we have
matched our all-order, resummed, calculation to an exact (3-jet)
calculation at next-to-leading order with the {\tt NLOJet++} program.
We then supplemented the perturbative predictions with non-perturbative
effects extracted from Monte Carlo simulations with Herwig, Pythia and
Sherpa.

The theoretical uncertainty on our perturbative predictions ranges from
$5{-}7\%$ at large $k_t$ to $\sim 20\%$ at $k_t\approx 5$~GeV.
Hadronisation and underlying-event corrections are relevant below
$20{-}30\GeV$, but in most of the Lund plane dominate the overall
uncertainty only below $k_t\sim 3{-}5$~GeV ($15\GeV$ at large angles,
where the underlying event is a significant contributor), cf.\
Figs.~\ref{fig:highpt} and \ref{fig:final-slices-delta}.

We have made our predictions for two variants of the Lund plane
definition, one using angle and absolute transverse momentum (the
default for most of this paper), and the other using angle and
relative momentum fraction in a given branching.
The latter corresponds to the choice made by the ATLAS collaboration
in their recent pioneering unfolded measurement of the primary Lund
plane density with charged tracks~\cite{Aad:2020zcn}.\footnote{The
  former has been adopted in preliminary measurements by the ALICE
  collaboration~\cite{Zardoshti:2020cwl} that probe dead-cone
  effects.}
We have compared our results to the ATLAS data, including an
additional non-perturbative correction to account for the use of
charged tracks, and found good agreement in all regions where we have
confidence in our predictions, i.e.\ the non-shaded regions
of Figs.~\ref{fig:atlas} and \ref{fig:atlas-zslice}.
This includes a broad swathe of the Lund plane, down to scales
corresponding to transverse momenta of about $5\GeV$.

Our work opens a series of questions that should be kept in mind for
future work.
First, it would be interesting to extend our calculation beyond
single-logarithmic accuracy.
We expect that this would give a considerable reduction in the
uncertainty, notably from control over the effective scale to be used
in the coupling.
Such a calculation, however, remains challenging.
Other effects like the subleading-logarithmic and subleading-$N_c$
corrections to the clustering logarithms, or NNLO fixed-order
corrections (requiring a NNLO $pp\to 3$-jet calculation) would also be
expected to bring significant improvements in certain specific regions
of the Lund plane.

It would also be of interest to understand the resummation of the
boundary logarithms that originate from the interplay between the
initial anti-$k_t$ clustering and the C/A reclustering.
Practically, however, these logarithms could be avoided by using the
C/A algorithm for both the initial clustering and the reclustering.
One observation that would also deserve better understanding is the
apparent absence of any resummation effect from clustering logarithms
in the soft-collinear part of the Lund plane for gluon-induced jets

Keeping the above theoretical limitations in mind (and possible future
improvements), one might wish to investigate whether a measurement of
the Lund plane density, which intrinsically covers a wide range of
transverse-momentum scales, could be helpful to make an extraction of
the strong coupling constant, $\alpha_s$, extending existing work on
strong coupling determinations from soft-drop
measurements~\cite{Bendavid:2018nar,Marzani:2019evv}.
In a similar spirit, one could perhaps extend the approach
of Ref.~\cite{Dokshitzer:1995zt} to develop an analytic approach to
non-perturbative corrections at small $k_t$ and potentially even use
Lund-plane measurements to determine an effective coupling constant
down to small transverse momenta.

Finally, it would be interesting to compare both analytical
predictions and measurements of the primary Lund-plane density to
recent efforts to develop parton showers with perturbative control
beyond leading double logarithmic accuracy (e.g.\
\cite{Dasgupta:2020fwr,Forshaw:2020wrq}) and leading colour (e.g.\
\cite{Forshaw:2019ver,Nagy:2019pjp}).

%----------------------------------------------------------------------
\section*{Acknowledgements}

We are grateful to Paul Caucal and Fr\'ed\'eric Dreyer, as well as to
our ATLAS colleagues (Reina Camacho, Matt Leblanc, David Miller, Ben
Nachmann and Jennifer Roloff), for many interesting discussions.
We also thank Zoltan Nagy for much-appreciated help to improve the
coverage of the phase-space for soft-and-collinear emissions with {\tt
  NLOJet++}.
A.L.\ wished to thank the IPhT for hospitality during his Master's 
internship, when the first steps of project were discussed.
This work has been supported in part by 
by the French Agence Nationale de la
Recherche, under grant ANR-15-CE31-0016 (GS),
by a Royal Society Research Professorship
(RP$\backslash$R1$\backslash$180112) (GPS),
and by the European Research Council (ERC) under the European Union’s
Horizon 2020 research and innovation programme (grant agreement No.\
788223, PanScales) (GPS, GS).

\appendix

\section{Analytic results for collinear  resummation}\label{sec:coll-analytic}

In this Appendix we give the explicit analytic solutions for the
average quark/gluon fractions $f(j|i,t_\text{coll})$ and momentum
fraction $\bar x(j|i,t_\text{coll})$ defined in
Eq.~\eqref{eq:f-xbar-def}. We find
\begin{subequations}\label{eq:flav-solution}
\begin{align}
  f(q|i,t) & = \delta_{iq}
             - \frac{s_q \delta_{iq} - s_g \delta_{ig}}{s_q+s_g}
               \left(1-e^{-(s_q+s_g)t} \right),\\
  f(g|i,t) & = \delta_{ig}
             + \frac{s_q \delta_{iq} - s_g \delta_{ig}}{s_q+s_g}
                  \left(1-e^{-(s_q+s_g)t} \right),\\
  f\bar x(q|i,t)
     & = e^{tw_+} \left\{\left[
         \cosh(t\tilde w_+) 
         +\frac{w_-}{\tilde w}\sinh(t\tilde w)
         \right] \delta_{iq}
       + \frac{w_{qg}}{\tilde w}\sinh(t\tilde w)\,
         \delta_{ig}\right\}\\
  f\bar x(g|i,t)
    & = e^{tt_+} \left\{\left[
        \cosh(t\tilde w_+) 
        -\frac{w_-}{\tilde w}\sinh(t\tilde w)
        \right] \delta_{ig}
      + \frac{t_{gq}}{\tilde w}\sinh(t\tilde w)\,
        \delta_{iq}\right\},
\end{align}
\end{subequations}
with
\begin{equation}
  w_\pm = \frac{w_{qq}\pm w_{gg}}{2},
  \qquad
  \tilde w=\sqrt{w_-^2+w_{qg}w_{gq}}.
\end{equation}
and
\begin{subequations}\label{eq:flav-chg-coefs}
\begin{align}
  s_q    & = \int_0^1 dz\, \left[{\cal{P}}_{qg}^{(R)}(z) - {\cal{P}}_{qg}^{(V)}(z)\right]
           = C_F \left(2\ln 2-\frac{5}{8}\right),\\
  s_g    & = \int_0^1 dz\, \left[{\cal{P}}_{gq}^{(R)}(z) - {\cal{P}}_{gq}^{(V)}(z)\right]
           = \frac{2 n_f T_R}{3}, \\
  w_{qq} & = \int_0^1 dz\,\left[z {\cal{P}}_{qq}^{(R)}(z) - {\cal{P}}_{qq}^{(V)}(z)\right]
           = -C_F\left(2\,\ln 2+\frac{1}{6}\right),\\
  w_{qg} & = \int_0^1 dz\,\left[z {\cal{P}}_{qg}^{(R)}(z) - {\cal{P}}_{qg}^{(V)}(z)\right]
           = \frac{25}{48}n_fT_R,\\
  w_{gq} & = \int_0^1 dz\,\left[z {\cal{P}}_{gq}^{(R)}(z) - {\cal{P}}_{gq}^{(V)}(z)\right]
           = \frac{11}{24}C_F,\\
  w_{gg} & = \int_0^1 dz\,\left[z {\cal{P}}_{gg}^{(R)}(z) - {\cal{P}}_{gg}^{(V)}(z)\right]
           = -\frac{2}{3}n_fT_R-C_A\left(2\,\ln 2-\frac{43}{96}\right).
\end{align}
\end{subequations}
Note that the coefficients $s_q$ and $s_g$ are in agreement with the
flavour-changing effects calculated in~\cite{Dasgupta:2014yra}.

\section{Boundary logarithms for $\Delta\sim R$}\label{sec:boundary-logs}

In this Appendix, we show how new logarithms of $R-\Delta$ arise from
secondary emissions at order $\alpha_s^2$. We show that these
logarithms are a consequence of the interplay between the initial
anti-$k_t$ clustering used to obtain the initial jets and the C/A
clustering used to construct the primary Lund plane.

Say we start from a dipole $(\ell_i,\ell_j)$ and have 2 emissions,
$k_1$ and $k_2$, strongly ordered in transverse momentum as discussed
in Section~\ref{sec:soft-clust}. Emission $1$ (real or virtual) is
integrated over and the softer emission 2 (real) is measured as a
contributing to $\rho^{(2)}_\text{soft}(k_t,\Delta)$.

We denote by $(k|ij)$ the geometrical pattern associated with the
radiation of gluon $k$ from the dipole $(\ell_i,\ell_j)$ (i.e.\ the transverse
momenta with respect to the beam are factored out).
We also denote by $R_i$ the distance of $i$ to the jet axis (in
rapidity-azimuth) and $R_{ij}$ the distance between $i$ and $j$.
For a parton with Casimir $C_R$, we have
\begin{align}\label{eq:starting-point}
  \alpha_s^2\rho^{(2)}_\text{soft}
  = \left(\frac{\alpha_s}{\pi}\right)^2 &
    \int_0^{\pt} \frac{dk_{\perp 1}}{k_{\perp 1}} \int_0^{k_{\perp 1}} \frac{dk_{\perp 2}}{k_{\perp 2}}
    \int dy_1 dy_2
    \int_0^{2\pi}\frac{d\phi_1}{2\pi}\frac{d\phi_2}{2\pi}
    \Delta\delta(\Delta-R_2)\,k_t\delta(k_t-k_{\perp 2}\Delta)\nonumber\\
  & C_R (1|ij)  \Bigg\{ \left[
    \frac{C_A}{2}(2|i1)+\frac{C_A}{2}(2|1j)+\left(C_R-\frac{C_A}{2}\right)(2|ij)\right]\\
  & \phantom{ C_R (1ij) }
    \left[\Theta(R_1>R) + \Theta(R_1<R)\Theta(R_1<R_{12}\text{ or }R_2<R_{12})\right]
    - C_R (2|ij)\Bigg\}.\nonumber
\end{align}
If one combines the $C_R$ contributions, performs the
$k_{\perp i}$ integrations and switches to polar coordinates for the
$y_2$, $\phi_2$ integration and uses the $\delta(\Delta-R_2)$
constraint to simplify, we get
\begin{align}\label{eq:simplified}
  \alpha_s^2\rho^{(2)}_\text{soft}
  =
  &\left(\frac{\alpha_s}{\pi}\right)^2\Delta^2\ln\left(\frac{\pt\Delta}{k_t}\right)
    \int dy_1 \int_0^{2\pi}\frac{d\phi_1}{2\pi}
    \frac{d\varphi_2}{2\pi} C_R(1|ij)\\
  & \bigg\{\frac{C_A}{2}[(2|i1)+(2|1j)-(2|ij)]
    \left[\Theta(R_1>R) + \Theta(R_1<R)\Theta(R_1<R_{12}\text{ or }R_2<R_{12})\right]\nonumber\\
  & - C_R (2|ij)\Theta(R_1<R)\Theta(R_{12}<R_1)\Theta(R_{12}<R_2)\bigg\}\nonumber
\end{align}
We can evaluate this numerically, separating the $C_RC_A$ term
in an ``inside'' contribution where $k_1$ is inside the jet (integrated in polar
coordinates around the jet axis) and an ``outside'' contribution where $k_1$
is outside the jet (integrated directly in $y_1$ and $\phi_1$).
We have done this explicitly as a check of the Monte-Carlo
implementation introduced in section~\ref{sec:soft-resum} and found
perfect agreement (in the large-$N_c$ limit).
Note that the combination of dipoles in the first square bracket of
Eq.~\eqref{eq:simplified} vanishes when $y_1\to \pm\infty$, showing
explicitly that there are no divergences collinear with the beam.

The main purpose of this Appendix is to show that the ``out'' $C_RC_A$
contribution has a collinear divergence when $\Delta \to R$. To see
this, we set $\Delta = R-\epsilon$ with $\epsilon\to 0$ (or take
$\ln(R/\Delta)\to 0$).
The collinear divergence comes from the situation where emission $k_1$
is close to emission $k_2$ (with $k_1$ outside the jet and $k_2$
inside), where the combination of dipoles can be simplified to
$4/\theta_{12}^2$.
After a few straightforward manipulations, we reach
\begin{align}
  \alpha_s^2\rho^{(2)}_\text{soft}(\Delta,k_t)
  & = \left(\frac{\alpha_s}{\pi}\right)^2 C_R C_A R^2\ln\left(\frac{\pt\Delta}{k_t}\right)
    \ln\left(\frac{R}{R-\Delta}\right)\,\int_0^{2\pi}\frac{d\varphi_2}{2\pi}(2|ij)\\
  & = \frac{2\alpha_sC_A}{\pi} \ln\left(\frac{\pt\Delta}{k_t}\right)
    \ln\left(\frac{R}{R-\Delta}\right)
    \left[\alpha_s\rho^{(1)}_\text{soft}(R,k_t) \right],\label{eq:log-div}
\end{align}
where $\alpha_s \rho^{(1)}_\text{soft}(R,k_t)$ can be taken
from Eq.~(\ref{eq:soft-em-largeR-alphas}).
This exhibits a logarithmic behaviour when $\Delta\to R$ (which is
integrable if one considers a bin in $\Delta$ between some lower bound
and $R$).\footnote{A similar enhancement was observed for narrow
  slices in Ref.~\cite{Dasgupta:2002bw}.}
We have checked that this behaviour is reproduced by the
Monte-Carlo described in section~\ref{sec:soft-resum}.

The physical origin of the collinear enhancement in~(\ref{eq:log-div})
is the interplay between the anti-$k_t$ clustering used to obtain the
initial jet and the C/A clustering used to construct the primary Lund
plane.
For a jet initially clustered with the C/A algorithm, emissions $k_1$
and $k_2$ would be clustered together and emission $k_2$ would then
not be seen as a primary emission.

Equation~(\ref{eq:log-div}) exhibits a double logarithmic
behaviour. One should also expect single-logarithmic corrections,
proportional to $\ln(R/(R-\Delta))$ without the soft enhancement. In
principle, these single-logarithmic terms should be resummed to all
orders.
We have not, so far, found
a simple prescription to achieve this resummation, which involves an
interplay between the complex structure of soft emissions at
commensurate angles and (potentially hard) collinear splittings at the
boundary of the jet.

We however give in this Appendix a simple (incomplete) prescription
from which one can gauge the potential impact of this resummation.
Coming back to our calculation at $\mathcal{O}(\alpha_s^2)$, we see
that the boundary logarithm $\ln(R/(R-\Delta))$ comes from the fact
that emissions $k_1$ and $k_2$ are collinear to each other and the
logarithm $\ln(\pt R/\Delta)$ comes from the energy ordering between
the two emissions.
One would obviously get a single-logarithmic contribution if the two
emissions were still collinear but no longer strongly ordered in
energy. This contribution, where the first emission is soft and just
outside the jet, and the second emission is collinear to the first one
and inside the jet, can be straightforwardly computed. A calculation
similar to the previous one shows that the energy logarithm is
replaced by an integration over the Altarelli-Parisi splitting
function $P(z)$ with $z$ the momentum fraction of the collinear
branching. One then gets
\begin{equation}\label{eq:log-div+Bg}
  \rho^{(2)}_{\text{soft}+B_g}
  = \left(\frac{\alpha_s}{\pi}\right)^2 R^2\left[\ln\left(\frac{\pt R}{k_t}\right)+B_g\right]
  C_R C_A
\ln\left(\frac{R}{R-\Delta}\right)\,\int_0^{2\pi}\frac{d\varphi_2}{2\pi}(2|ij),
\end{equation}
with $B_g=-\tfrac{11}{12}+\tfrac{n_f}{6C_A}$ the standard gluon
hard-collinear branching contribution obtained from integrating the
finite part of the gluon (to anything) splitting function.
This is but the first of a tower of terms enhanced by logarithms of
$R/(R-\Delta)$.
We will examine its magnitude shortly.
The full structure of the series involves other potentially
complicated effects: (a) an interplay between non-trivial clustering
logarithms and these new purely collinear effects; and (b) the way in
which the anti-$k_t$ jet clustering affects the jet axis and
subsequent identification of the set of particles (or tracks) that
gets reclustered with the C/A algorithm, specifically in presence of
hard splittings at angles comparable to the jet radius.
In addition to these subtleties, one might want to consider a number
of combinations of jet clustering: e.g.\ reclustering a full
anti-$k_t$ jet with $R_{C/A}=\infty$, reclustering it with
$R_{C/A}=R_\text{anti-$k_t$}$, reclustering only the particles within
a distance $R_\text{anti-$k_t$}$ of the anti-$k_t$ jet axis, etc.
Given that these effects concern only a single bin in $\Delta$, and
that their treatment brings many complications, we postpone their
study to future work.

\begin{figure}
  \centering
  \includegraphics[width=0.48\textwidth]{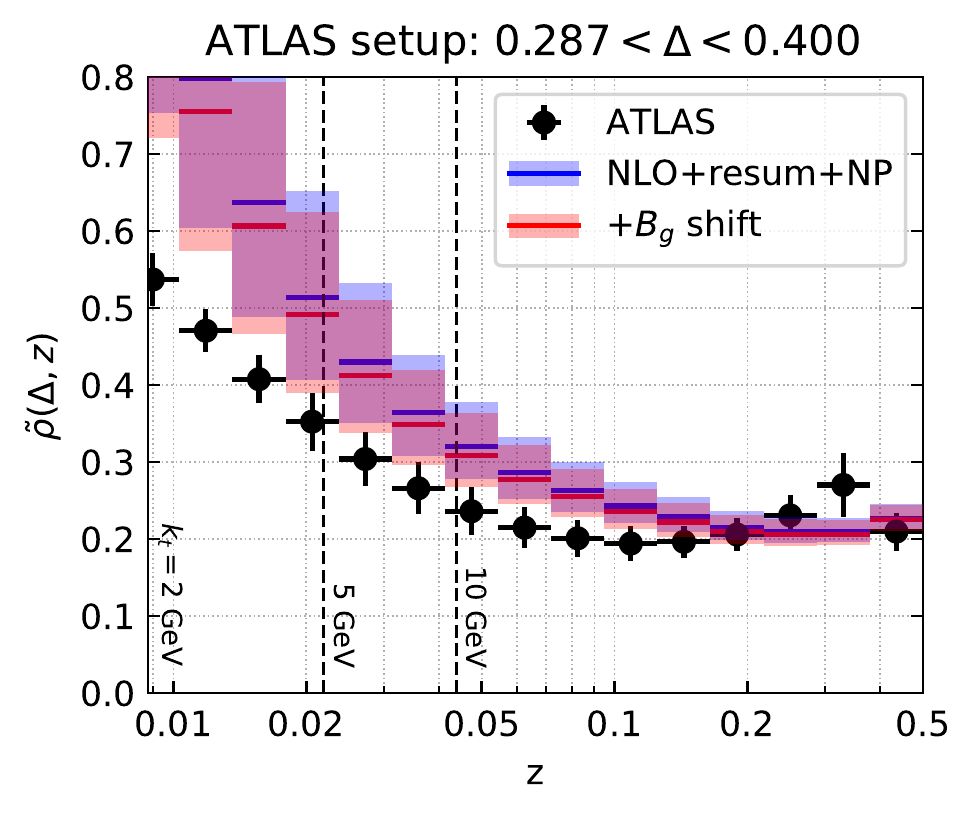}
  \caption{Estimate of the effect of a hard collinear splitting for the
    largest-$\Delta$ bin of the ATLAS data, using
    Eqs.~(\ref{eq:log-div+Bg}) and (\ref{eq:tsoft-boundary-redefinition}).
  }\label{fig:boundary-improved}
\end{figure}

We do nevertheless wish to investigate the size of the one
contribution we have outlined in Eq.~(\ref{eq:log-div+Bg}).
It can be included in the all-order resummation via the following
redefinition of $t_\text{soft}$
\begin{equation}\label{eq:tsoft-boundary-redefinition}
  t_\text{soft}
  = \int_{k_t}^{\pt\Delta}
  \frac{dq_t}{q_t}\frac{\alpha_s(q_t)}{\pi}
  \quad \longrightarrow \quad
  t_\text{soft}^{\text{(shifted)}}
  = \int_{k_te^{-B_g\Delta/R}}^{\pt\Delta}
  \frac{dq_t}{q_t}\frac{\alpha_s(q_t)}{\pi},
\end{equation}
which would only affect large values of $\Delta$ where the boundary
logarithms are present.
The effect of this (ad-hoc) prescription on the largest bin in
$\Delta$ is shown in Fig.~\ref{fig:boundary-improved}.
While the effect is relatively small (in particular, relative to our
uncertainties and to the discrepancy with the data), we see that our
results move in the right direction.
Pending a full treatment of these boundary logarithms --- left for
future work --- the bin closest to the jet edge should be treated with
caution.
We signal this limitation by shading the corresponding region in red
in our overall comparisons with the ATLAS data, Figs.~\ref{fig:atlas}
and \ref{fig:atlas-zslice}.

\section{Validation of the resummation at NLO}\label{sec:NLO-slopes}

\begin{figure}
  \centering
  \begin{subfigure}[t]{0.48\textwidth}
    \includegraphics[width=\textwidth,page=1]{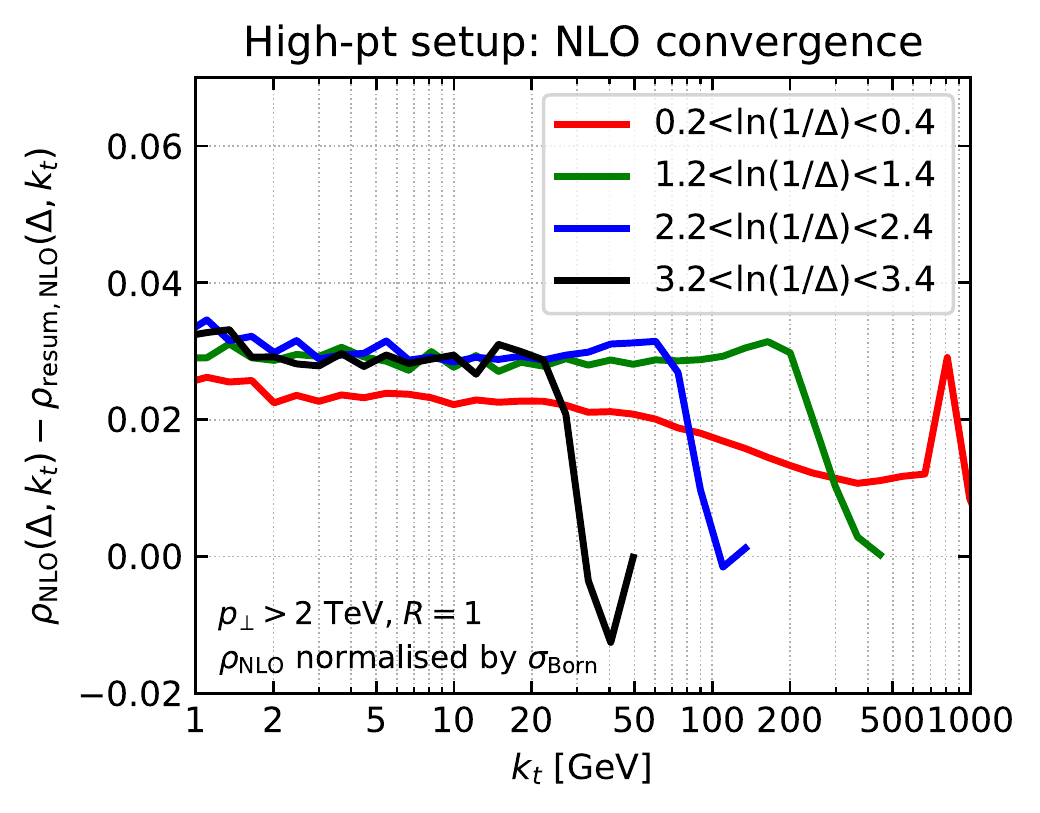}
    \caption{slices of fixed angle}\label{fig:v-nlo-slice-delta}
  \end{subfigure}
  \hfill
  \begin{subfigure}[t]{0.48\textwidth}
    \includegraphics[width=\textwidth,page=2]{figs/v-nlo.pdf}
    \caption{slices of fixed $k_t$}\label{fig:v-nlo-slice-kt}
  \end{subfigure}
  \caption{Differences between the exact NLO results and the NLO
    expansion of our resummation for different slices of the Lund
    plane.
    Control of single logarithmic terms in our resummation implies
    that these differences should tend to a constant both for small $k_t$
    and for small $\Delta$, as observed, except as concerns
    subleading-$N_C$ terms in the small-$k_t$, large $\Delta$ region.
    The exact NLO results are normalised here to
    the Born-level jet cross-section.
  }\label{fig:v-nlo-slices}
\end{figure}

As with any resummed calculation, it is important to check that its
expansion to fixed order reproduces the behaviour seen in the exact
fixed-order calculation, to within the expected accuracy of the
resummation.
In our case, this means that, at NLO, one should reproduce all
contributions of the form $\alpha_s^2\ln$, where the argument of the
logarithm is any variable in the Lund plane.
In practice, one therefore expects the difference
$\rho_\text{NLO}-\rho_{\text{resum,NLO}}$ to tend to a constant when
$k_t$ becomes small at a fixed $\Delta$, or when $\Delta\ll 1$ at a
fixed $k_t$.
Fig.~\ref{fig:v-nlo-slices} shows that this is indeed the case for
both limits.
We note that, while in the main text of the paper, the NLO Lund-plane
density has been normalised to the NLO inclusive jet cross-section,
for the purpose of Fig.~\ref{fig:v-nlo-slices} both $\rho_\text{NLO}$
and $\rho_\text{resum,NLO}$ have been normalised using the Born-level jet
cross-section.

\section{Comparison between our calculation and Monte Carlo simulations}\label{sec:v-mc}

\begin{figure}
  \centering
  \begin{subfigure}[t]{0.48\textwidth}
    \includegraphics[width=\textwidth,page=1]{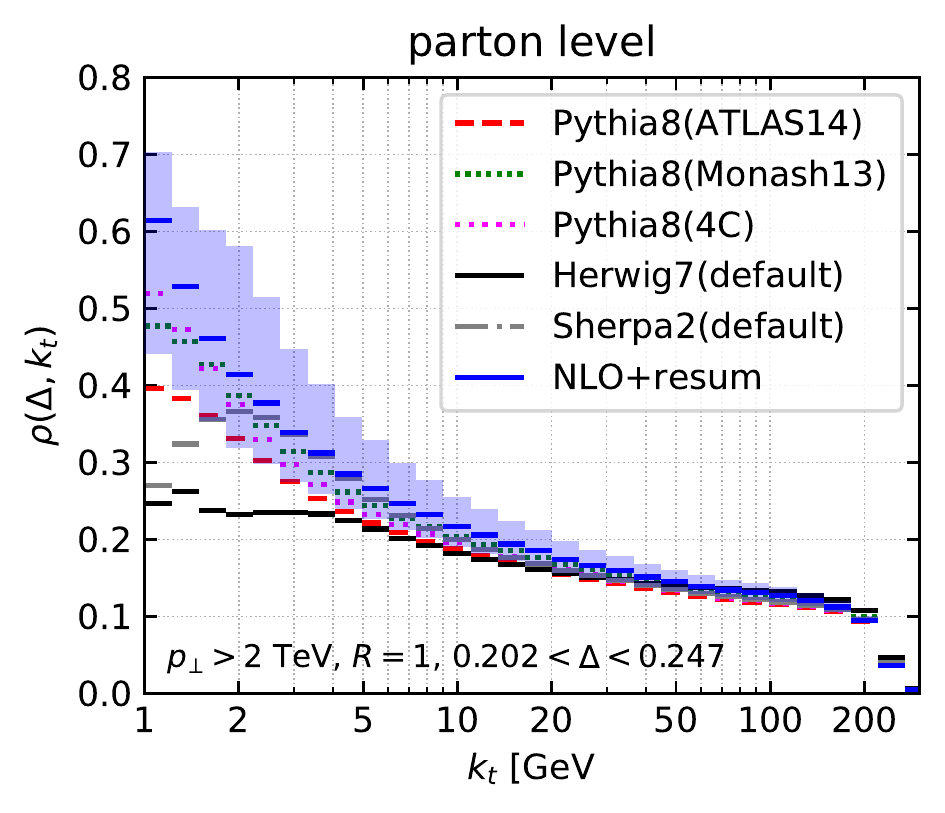}
    \caption{parton level}\label{fig:v-mc-parton}
  \end{subfigure}
  \hfill
  \begin{subfigure}[t]{0.48\textwidth}
    \includegraphics[width=\textwidth,page=2]{figs/v-mc.pdf}
    \caption{hadron+MPI level}\label{fig:v-mc-full}
  \end{subfigure}
  \caption{Comparison between our results for $\rho$ and those
    obtained with the Monte Carlo generators used to estimate
    non-perturbative uncertainties.}\label{fig:v-mc}
\end{figure}

We show in Fig.~\ref{fig:v-mc} a comparison between our analytic
calculations and Monte-Carlo simulations for a slice of the Lund plane
at constant angle.
In Fig.~\ref{fig:v-mc-parton} we compare our perturbative predictions
to parton-level simulations and the (blue) uncertainty band
corresponds to our perturbative scale uncertainty.
In Fig.~\ref{fig:v-mc-full} the comparison is made for the full
prediction, including non-perturbative corrections.\footnote{Obtained
  as discussed in section~\ref{sec:np-effects}, i.e.\ excluding
  Herwig7 from the computation of the average non-perturbative
  corrections to our analytic perturbative results.}

At hadron+MPI level, we see a globally-decent agreement
between our results and those from each Monte Carlo event generator.
At parton-level however, the Herwig7 results are systematically much
smaller that our analytic results for $k_t$ below $\sim 10$~GeV. This
is the main reason for excluding the Herwig7 Monte Carlo when
computing the average non-perturbative correction. 

\section{Non-perturbative corrections for the ATLAS setup}
\label{sec:atlas-np}

\begin{figure}
  \begin{subfigure}[t]{0.32\textwidth}
    \centering
    \includegraphics[width=\textwidth,page=1]{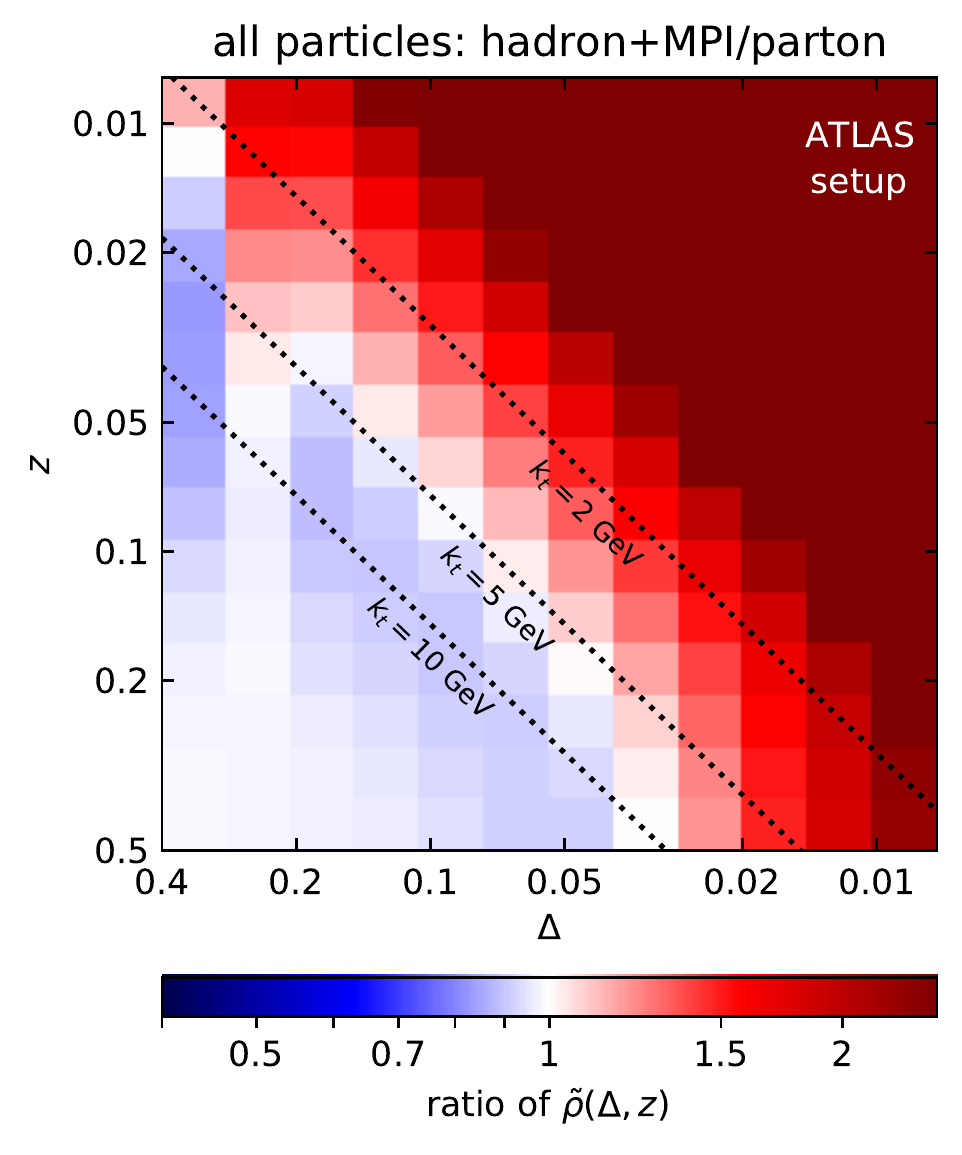}
    \caption{Hadronisation and MPI corrections (all particles)}\label{fig:np-atlas-allpart}
  \end{subfigure}
  \hfill
  \begin{subfigure}[t]{0.32\textwidth}
    \centering
    \includegraphics[width=\textwidth,page=2]{figs/np-effects-atlas.pdf}
    \caption{Effects of the charge track selection.}\label{fig:np-atlas-chg}
  \end{subfigure}
  \hfill
  \begin{subfigure}[t]{0.32\textwidth}
    \centering
    \includegraphics[width=\textwidth,page=3]{figs/np-effects-atlas.pdf}
    \caption{Final set of non-perturbative corrections.}\label{fig:np-atlas-full}
  \end{subfigure}
  \begin{subfigure}[t]{0.32\textwidth}
    \centering
    \includegraphics[width=\textwidth,page=1]{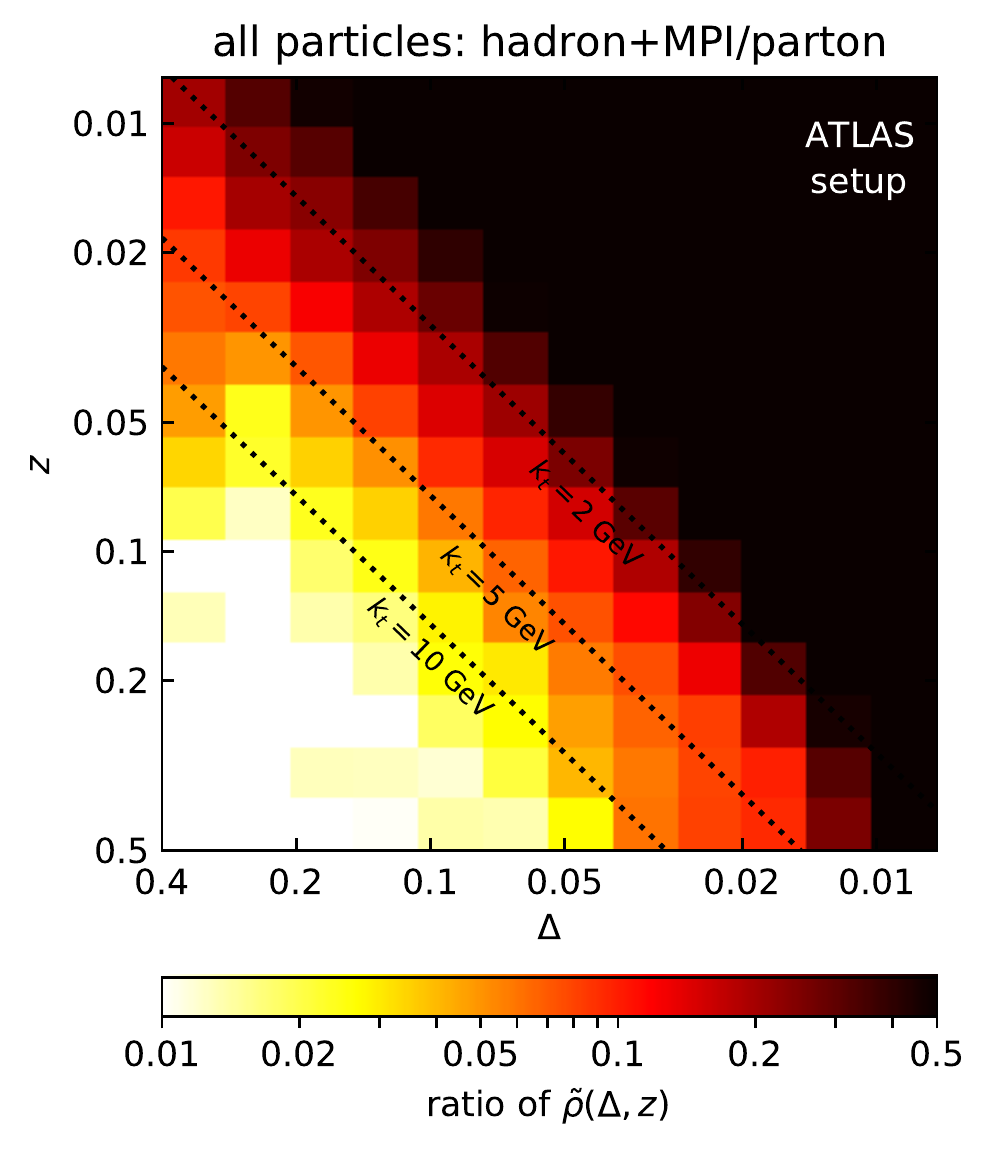}
    \caption{Uncertainties on hadronisation and MPI corrections.}\label{fig:np-atlas-unc-allpart}
  \end{subfigure}
  \hfill
  \begin{subfigure}[t]{0.32\textwidth}
    \centering
    \includegraphics[width=\textwidth,page=2]{figs/np-uncertainties-atlas.pdf}
    \caption{Uncertainties for the charged track selection.}\label{fig:np-atlas-unc-chg}
  \end{subfigure}
  \hfill
  \begin{subfigure}[t]{0.32\textwidth}
    \centering
    \includegraphics[width=\textwidth,page=3]{figs/np-uncertainties-atlas.pdf}
    \caption{Total uncertainties on non-perturbative corrections.}\label{fig:np-atlas-unc-full}
  \end{subfigure}
  \caption{Non-perturbative effects on $\tilde\rho(\Delta,z)$ for the
    ATLAS setup. The top row shows the average corrections and the
    lower row shows the associated uncertainties.}\label{fig:np-atlas}
\end{figure}

The set of non-perturbative corrections included in our calculation of
$\tilde\rho(\Delta, z)$ for the ``\mbox{ATLAS} setup'' (cf.\
section~\ref{sec:primary-lund}) differs from those included in our
default ``high-$p_\perp$ setup.''
The main differences are (i) the use of charged tracks instead of all
particles, (ii) a slightly different clustering procedure
using a radius of $0.4$ for the C/A reclustering, and (iii) the
selection of tracks above 500 MeV and within a distance to the jet
axis calculated using pseudo-rapidity instead of rapidity.

In Fig.~\ref{fig:np-atlas}, we split the full non-perturbative
correction into two separate factors: the corrections due to
hadronisation and multi-parton interactions computed on all particles,
Fig.~\ref{fig:np-atlas-allpart}, and the extra corrections associated
with the use of charged tracks, the 500~MeV transverse-momentum cut
and the track selection based on pseudo-rapidity,
Fig.~\ref{fig:np-atlas-chg}.\footnote{ The effect of reclustering the
  jet constituents with a finite jet radius $R=0.4$ is included
  together with the hadronisation and MPI corrections. We have checked
  that this effect itself is very small, below $0.1\%$.}
The final set of corrections is shown in Fig.~\ref{fig:np-atlas-full}.
We clearly see that the use of just charged tracks with momenta
greater than $500\MeV$ induces a strong reduction of
$\tilde\rho(\Delta,z)$ beyond the perturbative domain 
($k_t< 2$~GeV) and a positive correction for $k_t>2$~GeV.
This effect partially cancels the original effect of hadronisation in
the region where hadronisation depleted the Lund plane density, i.e.\
$k_t\gtrsim 5$~GeV. 
The corresponding uncertainties on the non-perturbative corrections are
shown in Fig.~\ref{fig:np-atlas}(d)--(f).
While the additional corrections associated with the selection of
charged tracks add a little to the uncertainty at large $z$,
the final pattern of uncertainties is largely unmodified compared to
that obtained solely from the hadronisation and MPI uncertainties.

\bibliographystyle{JHEP}
\bibliography{lund}

\providecommand{\href}[2]{#2}\begingroup\raggedright\begin{thebibliography}{10}

\bibitem{Marzani:2019hun}
S.~Marzani, G.~Soyez, and M.~Spannowsky, {\it {Looking inside jets: an
  introduction to jet substructure and boosted-object phenomenology}},
  \href{http://arxiv.org/abs/1901.10342}{{\tt arXiv:1901.10342}}. [Lect. Notes
  Phys.958,pp.(2019)].

\bibitem{Larkoski:2017jix}
A.~J. Larkoski, I.~Moult, and B.~Nachman, {\it {Jet Substructure at the Large
  Hadron Collider: A Review of Recent Advances in Theory and Machine
  Learning}},  {\em Phys. Rept.} {\bf 841} (2020) 1--63,
  [\href{http://arxiv.org/abs/1709.04464}{{\tt arXiv:1709.04464}}].

\bibitem{Asquith:2018igt}
R.~Kogler et~al., {\it {Jet Substructure at the Large Hadron Collider:
  Experimental Review}},  {\em Rev. Mod. Phys.} {\bf 91} (2019), no.~4 045003,
  [\href{http://arxiv.org/abs/1803.06991}{{\tt arXiv:1803.06991}}].

\bibitem{Sirunyan:2019der}
{\bf CMS} Collaboration, A.~M. Sirunyan et~al., {\it {Search for anomalous
  electroweak production of vector boson pairs in association with two jets in
  proton-proton collisions at 13 TeV}},  {\em Phys. Lett. B} {\bf 798} (2019)
  134985, [\href{http://arxiv.org/abs/1905.07445}{{\tt arXiv:1905.07445}}].

\bibitem{Aad:2019wdr}
{\bf ATLAS} Collaboration, G.~Aad et~al., {\it {Measurement of the jet mass in
  high transverse momentum $Z(\rightarrow b\overline{b})\gamma$ production at
  $\sqrt{s}= 13$ TeV using the ATLAS detector}},
  \href{http://arxiv.org/abs/1907.07093}{{\tt arXiv:1907.07093}}.

\bibitem{Aaboud:2019aii}
{\bf ATLAS} Collaboration, M.~Aaboud et~al., {\it {Measurement of
  jet-substructure observables in top quark, $W$ boson and light jet production
  in proton-proton collisions at $\sqrt{s}=13$ TeV with the ATLAS detector}},
  {\em JHEP} {\bf 08} (2019) 033, [\href{http://arxiv.org/abs/1903.02942}{{\tt
  arXiv:1903.02942}}].

\bibitem{Aaboud:2018ngk}
{\bf ATLAS} Collaboration, M.~Aaboud et~al., {\it {Search for chargino and
  neutralino production in final states with a Higgs boson and missing
  transverse momentum at $\sqrt{s} = 13$ TeV with the ATLAS detector}},  {\em
  Phys. Rev. D} {\bf 100} (2019), no.~1 012006,
  [\href{http://arxiv.org/abs/1812.09432}{{\tt arXiv:1812.09432}}].

\bibitem{Sirunyan:2019jbg}
{\bf CMS} Collaboration, A.~M. Sirunyan et~al., {\it {A multi-dimensional
  search for new heavy resonances decaying to boosted WW, WZ, or ZZ boson pairs
  in the dijet final state at 13 TeV}},  {\em Eur. Phys. J. C} {\bf 80} (2020),
  no.~3 237, [\href{http://arxiv.org/abs/1906.05977}{{\tt arXiv:1906.05977}}].

\bibitem{Sirunyan:2019vgt}
{\bf CMS} Collaboration, A.~M. Sirunyan et~al., {\it {Combination of CMS
  searches for heavy resonances decaying to pairs of bosons or leptons}},  {\em
  Phys. Lett. B} {\bf 798} (2019) 134952,
  [\href{http://arxiv.org/abs/1906.00057}{{\tt arXiv:1906.00057}}].

\bibitem{Gras:2017jty}
P.~Gras, S.~Höche, D.~Kar, A.~Larkoski, L.~Lönnblad, S.~Plätzer,
  A.~Siódmok, P.~Skands, G.~Soyez, and J.~Thaler, {\it {Systematics of
  quark/gluon tagging}},  {\em JHEP} {\bf 07} (2017) 091,
  [\href{http://arxiv.org/abs/1704.03878}{{\tt arXiv:1704.03878}}].

\bibitem{Frye:2017yrw}
C.~Frye, A.~J. Larkoski, J.~Thaler, and K.~Zhou, {\it {Casimir Meets Poisson:
  Improved Quark/Gluon Discrimination with Counting Observables}},  {\em JHEP}
  {\bf 09} (2017) 083, [\href{http://arxiv.org/abs/1704.06266}{{\tt
  arXiv:1704.06266}}].

\bibitem{Metodiev:2018ftz}
E.~M. Metodiev and J.~Thaler, {\it {Jet Topics: Disentangling Quarks and Gluons
  at Colliders}},  {\em Phys. Rev. Lett.} {\bf 120} (2018), no.~24 241602,
  [\href{http://arxiv.org/abs/1802.00008}{{\tt arXiv:1802.00008}}].

\bibitem{Larkoski:2019nwj}
A.~J. Larkoski and E.~M. Metodiev, {\it {A Theory of Quark vs. Gluon
  Discrimination}},  {\em JHEP} {\bf 10} (2019) 014,
  [\href{http://arxiv.org/abs/1906.01639}{{\tt arXiv:1906.01639}}].

\bibitem{Andrews:2018jcm}
H.~A. Andrews et~al., {\it {Novel tools and observables for jet physics in
  heavy-ion collisions}},  {\em J. Phys. G} {\bf 47} (2020), no.~6 065102,
  [\href{http://arxiv.org/abs/1808.03689}{{\tt arXiv:1808.03689}}].

\bibitem{Sirunyan:2017bsd}
{\bf CMS} Collaboration, A.~M. Sirunyan et~al., {\it {Measurement of the
  Splitting Function in $pp$ and Pb-Pb Collisions at $\sqrt{s_{_{\mathrm{NN}}}}
  =$ 5.02 TeV}},  {\em Phys. Rev. Lett.} {\bf 120} (2018), no.~14 142302,
  [\href{http://arxiv.org/abs/1708.09429}{{\tt arXiv:1708.09429}}].

\bibitem{Sirunyan:2018gct}
{\bf CMS} Collaboration, A.~M. Sirunyan et~al., {\it {Measurement of the
  groomed jet mass in PbPb and pp collisions at $ \sqrt{s_{\mathrm{NN}}}=5.02 $
  TeV}},  {\em JHEP} {\bf 10} (2018) 161,
  [\href{http://arxiv.org/abs/1805.05145}{{\tt arXiv:1805.05145}}].

\bibitem{Acharya:2019djg}
{\bf ALICE} Collaboration, S.~Acharya et~al., {\it {Exploration of jet
  substructure using iterative declustering in pp and Pb--Pb collisions at LHC
  energies}},  {\em Phys. Lett. B} {\bf 802} (2020) 135227,
  [\href{http://arxiv.org/abs/1905.02512}{{\tt arXiv:1905.02512}}].

\bibitem{Mehtar-Tani:2016aco}
Y.~Mehtar-Tani and K.~Tywoniuk, {\it {Groomed jets in heavy-ion collisions:
  sensitivity to medium-induced bremsstrahlung}},  {\em JHEP} {\bf 04} (2017)
  125, [\href{http://arxiv.org/abs/1610.08930}{{\tt arXiv:1610.08930}}].

\bibitem{Chien:2016led}
Y.-T. Chien and I.~Vitev, {\it {Probing the Hardest Branching within Jets in
  Heavy-Ion Collisions}},  {\em Phys. Rev. Lett.} {\bf 119} (2017), no.~11
  112301, [\href{http://arxiv.org/abs/1608.07283}{{\tt arXiv:1608.07283}}].

\bibitem{Chang:2017gkt}
N.-B. Chang, S.~Cao, and G.-Y. Qin, {\it {Probing medium-induced jet splitting
  and energy loss in heavy-ion collisions}},  {\em Phys. Lett. B} {\bf 781}
  (2018) 423--432, [\href{http://arxiv.org/abs/1707.03767}{{\tt
  arXiv:1707.03767}}].

\bibitem{Milhano:2017nzm}
G.~Milhano, U.~A. Wiedemann, and K.~C. Zapp, {\it {Sensitivity of jet
  substructure to jet-induced medium response}},  {\em Phys. Lett. B} {\bf 779}
  (2018) 409--413, [\href{http://arxiv.org/abs/1707.04142}{{\tt
  arXiv:1707.04142}}].

\bibitem{Caucal:2019uvr}
P.~Caucal, E.~Iancu, and G.~Soyez, {\it {Deciphering the $z_g$ distribution in
  ultrarelativistic heavy ion collisions}},  {\em JHEP} {\bf 10} (2019) 273,
  [\href{http://arxiv.org/abs/1907.04866}{{\tt arXiv:1907.04866}}].

\bibitem{Casalderrey-Solana:2019ubu}
J.~Casalderrey-Solana, G.~Milhano, D.~Pablos, and K.~Rajagopal, {\it
  {Modification of Jet Substructure in Heavy Ion Collisions as a Probe of the
  Resolution Length of Quark-Gluon Plasma}},  {\em JHEP} {\bf 01} (2020) 044,
  [\href{http://arxiv.org/abs/1907.11248}{{\tt arXiv:1907.11248}}].

\bibitem{Butterworth:2008iy}
J.~M. Butterworth, A.~R. Davison, M.~Rubin, and G.~P. Salam, {\it {Jet
  substructure as a new Higgs search channel at the LHC}},  {\em Phys. Rev.
  Lett.} {\bf 100} (2008) 242001, [\href{http://arxiv.org/abs/0802.2470}{{\tt
  arXiv:0802.2470}}].

\bibitem{Krohn:2009th}
D.~Krohn, J.~Thaler, and L.-T. Wang, {\it {Jet Trimming}},  {\em JHEP} {\bf 02}
  (2010) 084, [\href{http://arxiv.org/abs/0912.1342}{{\tt arXiv:0912.1342}}].

\bibitem{Thaler:2010tr}
J.~Thaler and K.~Van~Tilburg, {\it {Identifying Boosted Objects with
  N-subjettiness}},  {\em JHEP} {\bf 03} (2011) 015,
  [\href{http://arxiv.org/abs/1011.2268}{{\tt arXiv:1011.2268}}].

\bibitem{Larkoski:2013eya}
A.~J. Larkoski, G.~P. Salam, and J.~Thaler, {\it {Energy Correlation Functions
  for Jet Substructure}},  {\em JHEP} {\bf 06} (2013) 108,
  [\href{http://arxiv.org/abs/1305.0007}{{\tt arXiv:1305.0007}}].

\bibitem{Dasgupta:2013ihk}
M.~Dasgupta, A.~Fregoso, S.~Marzani, and G.~P. Salam, {\it {Towards an
  understanding of jet substructure}},  {\em JHEP} {\bf 09} (2013) 029,
  [\href{http://arxiv.org/abs/1307.0007}{{\tt arXiv:1307.0007}}].

\bibitem{Larkoski:2014wba}
A.~J. Larkoski, S.~Marzani, G.~Soyez, and J.~Thaler, {\it {Soft Drop}},  {\em
  JHEP} {\bf 05} (2014) 146, [\href{http://arxiv.org/abs/1402.2657}{{\tt
  arXiv:1402.2657}}].

\bibitem{Larkoski:2014pca}
A.~J. Larkoski, J.~Thaler, and W.~J. Waalewijn, {\it {Gaining (Mutual)
  Information about Quark/Gluon Discrimination}},  {\em JHEP} {\bf 11} (2014)
  129, [\href{http://arxiv.org/abs/1408.3122}{{\tt arXiv:1408.3122}}].

\bibitem{Salam:2016yht}
G.~P. Salam, L.~Schunk, and G.~Soyez, {\it {Dichroic subjettiness ratios to
  distinguish colour flows in boosted boson tagging}},  {\em JHEP} {\bf 03}
  (2017) 022, [\href{http://arxiv.org/abs/1612.03917}{{\tt arXiv:1612.03917}}].

\bibitem{Komiske:2017aww}
P.~T. Komiske, E.~M. Metodiev, and J.~Thaler, {\it {Energy flow polynomials: A
  complete linear basis for jet substructure}},  {\em JHEP} {\bf 04} (2018)
  013, [\href{http://arxiv.org/abs/1712.07124}{{\tt arXiv:1712.07124}}].

\bibitem{Dreyer:2018nbf}
F.~A. Dreyer, G.~P. Salam, and G.~Soyez, {\it {The Lund Jet Plane}},  {\em
  JHEP} {\bf 12} (2018) 064, [\href{http://arxiv.org/abs/1807.04758}{{\tt
  arXiv:1807.04758}}].

\bibitem{Cogan:2014oua}
J.~Cogan, M.~Kagan, E.~Strauss, and A.~Schwarztman, {\it {Jet-Images: Computer
  Vision Inspired Techniques for Jet Tagging}},  {\em JHEP} {\bf 02} (2015)
  118, [\href{http://arxiv.org/abs/1407.5675}{{\tt arXiv:1407.5675}}].

\bibitem{deOliveira:2015xxd}
L.~de~Oliveira, M.~Kagan, L.~Mackey, B.~Nachman, and A.~Schwartzman, {\it
  {Jet-images — deep learning edition}},  {\em JHEP} {\bf 07} (2016) 069,
  [\href{http://arxiv.org/abs/1511.05190}{{\tt arXiv:1511.05190}}].

\bibitem{Komiske:2016rsd}
P.~T. Komiske, E.~M. Metodiev, and M.~D. Schwartz, {\it {Deep learning in
  color: towards automated quark/gluon jet discrimination}},  {\em JHEP} {\bf
  01} (2017) 110, [\href{http://arxiv.org/abs/1612.01551}{{\tt
  arXiv:1612.01551}}].

\bibitem{Louppe:2017ipp}
G.~Louppe, K.~Cho, C.~Becot, and K.~Cranmer, {\it {QCD-Aware Recursive Neural
  Networks for Jet Physics}},  {\em JHEP} {\bf 01} (2019) 057,
  [\href{http://arxiv.org/abs/1702.00748}{{\tt arXiv:1702.00748}}].

\bibitem{Egan:2017ojy}
S.~Egan, W.~Fedorko, A.~Lister, J.~Pearkes, and C.~Gay, {\it {Long Short-Term
  Memory (LSTM) networks with jet constituents for boosted top tagging at the
  LHC}},  \href{http://arxiv.org/abs/1711.09059}{{\tt arXiv:1711.09059}}.

\bibitem{Andreassen:2018apy}
A.~Andreassen, I.~Feige, C.~Frye, and M.~D. Schwartz, {\it {JUNIPR: a Framework
  for Unsupervised Machine Learning in Particle Physics}},  {\em Eur. Phys. J.
  C} {\bf 79} (2019), no.~2 102, [\href{http://arxiv.org/abs/1804.09720}{{\tt
  arXiv:1804.09720}}].

\bibitem{Datta:2017lxt}
K.~Datta and A.~J. Larkoski, {\it {Novel Jet Observables from Machine
  Learning}},  {\em JHEP} {\bf 03} (2018) 086,
  [\href{http://arxiv.org/abs/1710.01305}{{\tt arXiv:1710.01305}}].

\bibitem{Komiske:2018cqr}
P.~T. Komiske, E.~M. Metodiev, and J.~Thaler, {\it {Energy Flow Networks: Deep
  Sets for Particle Jets}},  {\em JHEP} {\bf 01} (2019) 121,
  [\href{http://arxiv.org/abs/1810.05165}{{\tt arXiv:1810.05165}}].

\bibitem{CMS:2019gpd}
{\bf CMS} Collaboration, {\it {Machine learning-based identification of highly
  Lorentz-boosted hadronically decaying particles at the CMS experiment}},
  Tech. Rep. CMS-PAS-JME-18-002, July, 2019.

\bibitem{Kasieczka:2019dbj}
A.~Butter et~al., {\it {The Machine Learning Landscape of Top Taggers}},  {\em
  SciPost Phys.} {\bf 7} (2019) 014,
  [\href{http://arxiv.org/abs/1902.09914}{{\tt arXiv:1902.09914}}].

\bibitem{Kasieczka:2018lwf}
G.~Kasieczka, N.~Kiefer, T.~Plehn, and J.~M. Thompson, {\it {Quark-Gluon
  Tagging: Machine Learning vs Detector}},  {\em SciPost Phys.} {\bf 6} (2019),
  no.~6 069, [\href{http://arxiv.org/abs/1812.09223}{{\tt arXiv:1812.09223}}].

\bibitem{Qu:2019gqs}
H.~Qu and L.~Gouskos, {\it {ParticleNet: Jet Tagging via Particle Clouds}},
  {\em Phys. Rev.} {\bf D101} (2020), no.~5 056019,
  [\href{http://arxiv.org/abs/1902.08570}{{\tt arXiv:1902.08570}}].

\bibitem{Andersson:1988gp}
B.~Andersson, G.~Gustafson, L.~Lonnblad, and U.~Pettersson, {\it {Coherence
  Effects in Deep Inelastic Scattering}},  {\em Z. Phys.} {\bf C43} (1989) 625.

\bibitem{Dasgupta:2020fwr}
M.~Dasgupta, F.~A. Dreyer, K.~Hamilton, P.~F. Monni, G.~P. Salam, and G.~Soyez,
  {\it {Parton showers beyond leading logarithmic accuracy}},
  \href{http://arxiv.org/abs/2002.11114}{{\tt arXiv:2002.11114}}.

\bibitem{Aad:2020zcn}
{\bf ATLAS} Collaboration, G.~Aad et~al., {\it {Measurement of the Lund jet
  plane using charged particles in 13 TeV proton-proton collisions with the
  ATLAS detector}},  \href{http://arxiv.org/abs/2004.03540}{{\tt
  arXiv:2004.03540}}.

\bibitem{Cunqueiro:2018jbh}
L.~Cunqueiro and M.~P{\l}osko\'n, {\it {Searching for the dead cone effects
  with iterative declustering of heavy-flavor jets}},  {\em Phys. Rev. D} {\bf
  99} (2019), no.~7 074027, [\href{http://arxiv.org/abs/1812.00102}{{\tt
  arXiv:1812.00102}}].

\bibitem{Zardoshti:2020cwl}
{\bf ALICE} Collaboration, N.~Zardoshti, {\it {First Direct Observation of the
  Dead-Cone Effect}},  in {\em {28th International Conference on
  Ultrarelativistic Nucleus-Nucleus Collisions (Quark Matter 2019) Wuhan,
  China, November 4-9, 2019}}, 2020.
\newblock \href{http://arxiv.org/abs/2004.05968}{{\tt arXiv:2004.05968}}.

\bibitem{Nagy:2003tz}
Z.~Nagy, {\it {Next-to-leading order calculation of three jet observables in
  hadron hadron collision}},  {\em Phys. Rev.} {\bf D68} (2003) 094002,
  [\href{http://arxiv.org/abs/hep-ph/0307268}{{\tt hep-ph/0307268}}].

\bibitem{Cacciari:2008gp}
M.~Cacciari, G.~P. Salam, and G.~Soyez, {\it {The Anti-k(t) jet clustering
  algorithm}},  {\em JHEP} {\bf 04} (2008) 063,
  [\href{http://arxiv.org/abs/0802.1189}{{\tt arXiv:0802.1189}}].

\bibitem{Dokshitzer:1997in}
Y.~L. Dokshitzer, G.~D. Leder, S.~Moretti, and B.~R. Webber, {\it {Better jet
  clustering algorithms}},  {\em JHEP} {\bf 08} (1997) 001,
  [\href{http://arxiv.org/abs/hep-ph/9707323}{{\tt hep-ph/9707323}}].

\bibitem{Wobisch:1998wt}
M.~Wobisch and T.~Wengler, {\it {Hadronization corrections to jet
  cross-sections in deep inelastic scattering}},  in {\em {Monte Carlo
  generators for HERA physics. Proceedings, Workshop, Hamburg, Germany,
  1998-1999}}, pp.~270--279, 1998.
\newblock \href{http://arxiv.org/abs/hep-ph/9907280}{{\tt hep-ph/9907280}}.

\bibitem{Amoroso:2020lgh}
S.~Amoroso et~al., {\it {Les Houches 2019: Physics at TeV Colliders: Standard
  Model Working Group Report}},  in {\em {11th Les Houches Workshop on Physics
  at TeV Colliders}: {PhysTeV Les Houches}}, 3, 2020.
\newblock \href{http://arxiv.org/abs/2003.01700}{{\tt arXiv:2003.01700}}.

\bibitem{Schegelsky:2010xi}
V.~A. Schegelsky, M.~G. Ryskin, A.~D. Martin, and V.~A. Khoze, {\it {A note on
  rapidity distributions at the LHC}},
  \href{http://arxiv.org/abs/1010.2051}{{\tt arXiv:1010.2051}}.

\bibitem{Gallicchio:2018elx}
J.~Gallicchio and Y.-T. Chien, {\it {Quit Using Pseudorapidity, Transverse
  Energy, and Massless Constituents}},
  \href{http://arxiv.org/abs/1802.05356}{{\tt arXiv:1802.05356}}.

\bibitem{Cacciari:2005hq}
M.~Cacciari and G.~P. Salam, {\it {Dispelling the $N^{3}$ myth for the $k_t$
  jet-finder}},  {\em Phys. Lett. B} {\bf 641} (2006) 57--61,
  [\href{http://arxiv.org/abs/hep-ph/0512210}{{\tt hep-ph/0512210}}].

\bibitem{Cacciari:2011ma}
M.~Cacciari, G.~P. Salam, and G.~Soyez, {\it {FastJet User Manual}},  {\em Eur.
  Phys. J. C} {\bf 72} (2012) 1896, [\href{http://arxiv.org/abs/1111.6097}{{\tt
  arXiv:1111.6097}}].

\bibitem{fastjet-contrib}
M.~Cacciari, G.~P. Salam, and G.~Soyez, {\em FastJet contrib}, 2014 (accessed,
  March 7, 2020).
\newblock \url{https://fastjet.hepforge.org/contrib/}.

\bibitem{Catani:1990rr}
S.~Catani, B.~R. Webber, and G.~Marchesini, {\it {QCD coherent branching and
  semi-inclusive processes at large $x$}},  {\em Nucl. Phys.} {\bf B349} (1991)
  635--654.

\bibitem{Dasgupta:2014yra}
M.~Dasgupta, F.~Dreyer, G.~P. Salam, and G.~Soyez, {\it {Small-radius jets to
  all orders in QCD}},  {\em JHEP} {\bf 04} (2015) 039,
  [\href{http://arxiv.org/abs/1411.5182}{{\tt arXiv:1411.5182}}].

\bibitem{Ellis:1986bv}
R.~Ellis, G.~Marchesini, and B.~Webber, {\it {Soft Radiation in Parton Parton
  Scattering}},  {\em Nucl. Phys. B} {\bf 286} (1987) 643. [Erratum:
  Nucl.Phys.B 294, 1180 (1987)].

\bibitem{Kang:2019prh}
Z.-B. Kang, K.~Lee, X.~Liu, D.~Neill, and F.~Ringer, {\it {The soft drop
  groomed jet radius at NLL}},  {\em JHEP} {\bf 02} (2020) 054,
  [\href{http://arxiv.org/abs/1908.01783}{{\tt arXiv:1908.01783}}].

\bibitem{Dasgupta:2002bw}
M.~Dasgupta and G.~P. Salam, {\it {Accounting for coherence in interjet E(t)
  flow: A Case study}},  {\em JHEP} {\bf 03} (2002) 017,
  [\href{http://arxiv.org/abs/hep-ph/0203009}{{\tt hep-ph/0203009}}].

\bibitem{Dasgupta:2001sh}
M.~Dasgupta and G.~P. Salam, {\it {Resummation of nonglobal QCD observables}},
  {\em Phys. Lett.} {\bf B512} (2001) 323--330,
  [\href{http://arxiv.org/abs/hep-ph/0104277}{{\tt hep-ph/0104277}}].

\bibitem{Banfi:2002hw}
A.~Banfi, G.~Marchesini, and G.~Smye, {\it {Away from jet energy flow}},  {\em
  JHEP} {\bf 08} (2002) 006, [\href{http://arxiv.org/abs/hep-ph/0206076}{{\tt
  hep-ph/0206076}}].

\bibitem{Appleby:2003sj}
R.~B. Appleby and M.~H. Seymour, {\it {The Resummation of interjet energy flow
  for gaps between jets processes at HERA}},  {\em JHEP} {\bf 09} (2003) 056,
  [\href{http://arxiv.org/abs/hep-ph/0308086}{{\tt hep-ph/0308086}}].

\bibitem{Banfi:2005gj}
A.~Banfi and M.~Dasgupta, {\it {Problems in resumming interjet energy flows
  with $k_t$ clustering}},  {\em Phys. Lett.} {\bf B628} (2005) 49--56,
  [\href{http://arxiv.org/abs/hep-ph/0508159}{{\tt hep-ph/0508159}}].

\bibitem{Hatta:2013iba}
Y.~Hatta and T.~Ueda, {\it {Resummation of non-global logarithms at finite
  $N_c$}},  {\em Nucl. Phys.} {\bf B874} (2013) 808--820,
  [\href{http://arxiv.org/abs/1304.6930}{{\tt arXiv:1304.6930}}].

\bibitem{Hagiwara:2015bia}
Y.~Hagiwara, Y.~Hatta, and T.~Ueda, {\it {Hemisphere jet mass distribution at
  finite $N_c$}},  {\em Phys. Lett.} {\bf B756} (2016) 254--258,
  [\href{http://arxiv.org/abs/1507.07641}{{\tt arXiv:1507.07641}}].

\bibitem{Larkoski:2014uqa}
A.~J. Larkoski, D.~Neill, and J.~Thaler, {\it {Jet Shapes with the Broadening
  Axis}},  {\em JHEP} {\bf 04} (2014) 017,
  [\href{http://arxiv.org/abs/1401.2158}{{\tt arXiv:1401.2158}}].

\bibitem{Cacciari:2003fi}
M.~Cacciari, S.~Frixione, M.~Mangano, P.~Nason, and G.~Ridolfi, {\it {The t
  anti-t cross-section at 1.8-TeV and 1.96-TeV: A Study of the systematics due
  to parton densities and scale dependence}},  {\em JHEP} {\bf 04} (2004) 068,
  [\href{http://arxiv.org/abs/hep-ph/0303085}{{\tt hep-ph/0303085}}].

\bibitem{Sjostrand:2014zea}
T.~Sjöstrand, S.~Ask, J.~R. Christiansen, R.~Corke, N.~Desai, P.~Ilten,
  S.~Mrenna, S.~Prestel, C.~O. Rasmussen, and P.~Z. Skands, {\it {An
  Introduction to PYTHIA 8.2}},  {\em Comput. Phys. Commun.} {\bf 191} (2015)
  159--177, [\href{http://arxiv.org/abs/1410.3012}{{\tt arXiv:1410.3012}}].

\bibitem{Skands:2014pea}
P.~Skands, S.~Carrazza, and J.~Rojo, {\it {Tuning PYTHIA 8.1: the Monash 2013
  Tune}},  {\em Eur. Phys. J.} {\bf C74} (2014), no.~8 3024,
  [\href{http://arxiv.org/abs/1404.5630}{{\tt arXiv:1404.5630}}].

\bibitem{Corke:2010yf}
R.~Corke and T.~Sjostrand, {\it {Interleaved Parton Showers and Tuning
  Prospects}},  {\em JHEP} {\bf 03} (2011) 032,
  [\href{http://arxiv.org/abs/1011.1759}{{\tt arXiv:1011.1759}}].

\bibitem{ATL-PHYS-PUB-2014-021}
{\it {ATLAS Pythia 8 tunes to 7 TeV datas}},  Tech. Rep. ATL-PHYS-PUB-2014-021,
  CERN, Geneva, Nov, 2014.

\bibitem{Carrazza:2013axa}
S.~Carrazza, S.~Forte, and J.~Rojo, {\it {Parton Distributions and Event
  Generators}},  in {\em {Proceedings, 43rd International Symposium on
  Multiparticle Dynamics (ISMD 13)}}, pp.~89--96, 2013.
\newblock \href{http://arxiv.org/abs/1311.5887}{{\tt arXiv:1311.5887}}.

\bibitem{Corcella:2002jc}
G.~Corcella, I.~G. Knowles, G.~Marchesini, S.~Moretti, K.~Odagiri,
  P.~Richardson, M.~H. Seymour, and B.~R. Webber, {\it {HERWIG 6.5 release
  note}},  \href{http://arxiv.org/abs/hep-ph/0210213}{{\tt hep-ph/0210213}}.

\bibitem{Bellm:2015jjp}
J.~Bellm et~al., {\it {Herwig 7.0/Herwig++ 3.0 release note}},  {\em Eur. Phys.
  J.} {\bf C76} (2016), no.~4 196, [\href{http://arxiv.org/abs/1512.01178}{{\tt
  arXiv:1512.01178}}].

\bibitem{Bellm:2019zci}
J.~Bellm et~al., {\it {Herwig 7.2 Release Note}},
  \href{http://arxiv.org/abs/1912.06509}{{\tt arXiv:1912.06509}}.

\bibitem{Gleisberg:2008ta}
T.~Gleisberg, S.~Hoeche, F.~Krauss, M.~Schonherr, S.~Schumann, F.~Siegert, and
  J.~Winter, {\it {Event generation with SHERPA 1.1}},  {\em JHEP} {\bf 02}
  (2009) 007, [\href{http://arxiv.org/abs/0811.4622}{{\tt arXiv:0811.4622}}].

\bibitem{ATLAS:2019sol}
{\bf ATLAS} Collaboration, T.~A. collaboration, {\it {Measurement of the Lund
  Jet Plane using charged particles with the ATLAS detector from 13 TeV
  proton--proton collisions}}, .

\bibitem{Bendavid:2018nar}
{\em {Les Houches 2017: Physics at TeV Colliders Standard Model Working Group
  Report}}, 3, 2018.

\bibitem{Marzani:2019evv}
S.~Marzani, D.~Reichelt, S.~Schumann, G.~Soyez, and V.~Theeuwes, {\it {Fitting
  the Strong Coupling Constant with Soft-Drop Thrust}},  {\em JHEP} {\bf 11}
  (2019) 179, [\href{http://arxiv.org/abs/1906.10504}{{\tt arXiv:1906.10504}}].

\bibitem{Dokshitzer:1995zt}
Y.~L. Dokshitzer and B.~Webber, {\it {Calculation of power corrections to
  hadronic event shapes}},  {\em Phys. Lett. B} {\bf 352} (1995) 451--455,
  [\href{http://arxiv.org/abs/hep-ph/9504219}{{\tt hep-ph/9504219}}].

\bibitem{Forshaw:2020wrq}
J.~R. Forshaw, J.~Holguin, and S.~Plätzer, {\it {Building a consistent parton
  shower}},  \href{http://arxiv.org/abs/2003.06400}{{\tt arXiv:2003.06400}}.

\bibitem{Forshaw:2019ver}
J.~R. Forshaw, J.~Holguin, and S.~Plätzer, {\it {Parton branching at amplitude
  level}},  {\em JHEP} {\bf 08} (2019) 145,
  [\href{http://arxiv.org/abs/1905.08686}{{\tt arXiv:1905.08686}}].

\bibitem{Nagy:2019pjp}
Z.~Nagy and D.~E. Soper, {\it {Parton showers with more exact color
  evolution}},  {\em Phys. Rev. D} {\bf 99} (2019), no.~5 054009,
  [\href{http://arxiv.org/abs/1902.02105}{{\tt arXiv:1902.02105}}].

\end{thebibliography}\endgroup

\end{document}